\documentclass[12pt, a4paper]{article}

\usepackage{amsmath,amssymb,bm,epsfig,afterpage}
\usepackage{cite}
\usepackage{here}
\usepackage{color}
\usepackage{colortbl}
\usepackage{xcolor}
\usepackage{tabularx}
\usepackage{subcaption}
\usepackage{bbm}
\usepackage{graphicx}
\usepackage{listings}
\usepackage{longtable}
\usepackage[utf8]{inputenc}
\usepackage{cancel}
\usepackage{slashed}

\makeatletter
    
    \@addtoreset{equation}{section}
  \makeatother

\allowdisplaybreaks[3]

\newcommand{\beq}{\begin{equation}}
\newcommand{\eeq}{\end{equation}}

\newcommand{\ka}{\kappa}

\newcommand{\MeV}{\mathrm{MeV}}

\newcommand{\TeV}{\mathrm{TeV}}

\newcommand{\la}{\lambda}
\newcommand{\eps}{\epsilon}
\newcommand{\tla}{\widetilde{\lambda}}
\newcommand{\ty}{\widetilde{y}}

\newcommand{\Mcal}{\mathcal{M}}
\newcommand{\Lcal}{\mathcal{L}}
\newcommand{\Ocal}{\mathcal{O}}
\newcommand{\Hcal}{\mathcal{H}}
\newcommand{\Ucal}{\mathcal{U}}
\newcommand{\Ncal}{\mathcal{N}}
\newcommand{\Dcal}{\mathcal{D}}
\newcommand{\Ecal}{\mathcal{E}}
\newcommand{\Pcal}{\mathcal{P}}
\newcommand{\Fcal}{\mathcal{F}}
\newcommand{\tU}{\widetilde{U}}

\newcommand{\nv}{\mathbf{n}}
\newcommand{\Nv}{\mathbf{N}}

\newcommand{\SM}{\mathrm{SM}}
\newcommand{\PS}{\mathrm{PS}}
\newcommand{\BL}{{B\mathrm{-}L}}
\newcommand{\Pb}{P_{\overline{6}}}

\newcommand{\ol}[1]{\overline{#1}}
\newcommand{\wt}[1]{\widetilde{#1}}
\newcommand{\vev}[1]{\langle{#1}\rangle}
\newcommand{\abs}[1]{\left|{#1}\right|}
\newcommand{\order}[1]{\mathcal{O}\left({#1}\right)}
\newcommand{\br}[2]{\mathrm{BR} \left({#1}\to{#2}\right)}
\newcommand{\id}[1]{\mathbf{1}_{{#1}} }
\newcommand{\zd}[1]{0_{{#1}}}
{\newcommand{\rep}[1]{\mathbf{#1}}
\newcommand{\bra}[1]{\langle{#1}| }
\newcommand{\ket}[1]{|{#1}\rangle }

\newcommand{\Vcal}{\mathcal{V}}
\newcommand{\Wcal}{\mathcal{W}}
\newcommand{\Zcal}{\mathcal{Z}}

\newcommand{\tD}{\widetilde{D}}

\newcommand{\Deladj}{\Delta_8}

\newcommand{\RK}{b\to s\mu\mu}
\newcommand{\RD}{b\to c\tau\nu}

\usepackage[colorlinks=true, linkcolor=black, citecolor=black,
urlcolor=black]{hyperref}

\setcounter{footnote}{0}

\begin{document}

\begin{titlepage}

\begin{flushright}
 {\tt
CTPU-PTC-21-11  
}
\end{flushright}

\vspace{1.2cm}
\begin{center}
{\Large
{\bf
TeV-scale vector leptoquark from Pati-Salam unification 
with vectorlike families
}
}
\vskip 2cm

Syuhei Iguro$^1$, Junichiro Kawamura$^{2,3}$, Shohei Okawa$^4$ and Yuji Omura$^5$

\vskip 0.5cm

{\it $^1$
Department of Physics, Nagoya University, Nagoya 464-8602, Japan}\\[3pt]

{\it $^2$
Center for Theoretical Physics of the Universe, Institute for Basic Science (IBS),
Daejeon 34051, Korea
}\\[3pt]

{\it $^3$Department of Physics,
Keio University, Yokohama 223-8522,  Japan} \\[3pt]

{\it $^4$ Department of Physics and Astronomy, University of Victoria, \\
Victoria, BC V8P 5C2, Canada}

{\it $^5$
Department of Physics, Kindai University, Higashi-Osaka, Osaka 577-8502, Japan}\\[3pt]

\vskip 1.5cm

\begin{abstract}
In this paper, we show an explicit way to realize a TeV-scale vector leptoquark 
from the Pati-Salam (PS) unification with extra vectorlike families. 
The leptoquark mass is constrained to be heavier than PeV-scale 
by the measurement of a flavor violating kaon decay, $K_L \to \mu e$, in conventional models.   
This strong constraint can be avoided by introducing vectorlike families 
consistently with the quark and lepton masses and CKM and PMNS matrices.
The other flavor violating processes are also suppressed.
In this model, the vector leptoquark can be sufficiently light to explain 
the recent $\RK$ anomaly, 
while the $\RD$ anomaly is difficult to be explained due to the strong constraints 
from the $Z^\prime$ boson and vectorlike quark searches at the LHC. 
When the $\RK$ anomaly is explained,  
we show that $\order{0.2}$~\% tuning is required in the fermion matrix,  
the future experiments in $\mu\to e\gamma$ and $\mu$-$e$ conversions 
will cover the most available parameter space,  
and sizable neutral meson mixings, induced by the extra Higgs doublets, are unavoidable. 
\end{abstract}

\end{center}
\end{titlepage}

\tableofcontents
\clearpage

\section{Introduction}

Leptoquark is a hypothetical boson that carries both the baryon number and lepton number~\cite{Buchmuller:1986zs}.
Recently, phenomenology of TeV-scale leptoquarks has been discussed
particularly to explain the anomalies in the semi-leptonic $B$ decays~\cite{Sakaki:2013bfa,Bauer:2015knc,Fajfer:2015ycq,Becirevic:2016yqi,Bhattacharya:2016mcc,Li:2016vvp,Barbieri:2016las,Sahoo:2016pet,Assad:2017iib,Buttazzo:2017ixm,DiLuzio:2017vat,Becirevic:2017jtw,Angelescu:2018tyl,Iguro:2018vqb,Cornella:2019hct}.
An attractive extension which predicts leptoquarks is the Pati-Salam (PS) unification~\cite{Pati:1974yy}.
In this paper, we construct an explicit PS model that leads a vector leptoquark at TeV scale.

In the PS model, quarks and leptons are unified into two multiplets under
the PS gauge symmetry, $G_\PS = SU(4)_C\times SU(2)_L\times SU(2)_R$.
The hypercharge is quantized since all gauge symmetries are non-Abelian.
A scale of the  $G_\PS$ breaking can be much lower than that of the Grand Unification Theories (GUT)
such as $SU(5)$ GUT due to the absence of bosons with di-quark couplings, which induces the proton decay.
A vector leptoquarks arises
as the massive degree of freedom associated with the $G_\PS$ breaking, $SU(4)_C \to SU(3)_C\times U(1)_\BL$,
and hence its mass will be around the PS breaking scale.

The Yukawa couplings for different types of charged fermions in the Standard Model (SM)
are predicted to be unified into one Yukawa coupling at the high-scale where the PS symmetry is restored.
In a conventional scenario~\cite{Poh:2017xvg}, the Yukawa unification is achieved by taking
renormalization group (RG) effects and higher-dimensional operators.
This scenario, however, requires that the $G_\PS$ breaking scale is around the conventional GUT scale,
$\order{10^{16}~\mathrm{GeV}}$,
in order to obtain sizable corrections from the RG effects and higher dimensional operators,
which are usually suppressed by the Planck scale.
It is proposed in Ref.~\cite{Calibbi:2017qbu}
that the Yukawa unification is achieved by introducing vectorlike fermions
even with the TeV-scale $G_\PS$ breaking~\footnote{
Similar attempts in models with fermions whose representations are different from the SM ones are recently studied in Ref.~\cite{Dolan:2020doe}.
},
and thus the TeV-scale vector leptoquark can be realized in such a model.
We also introduce vectorlike fermions in our setup.

One of the recent hot topics in high-energy physics is that
experimental results are deviated from the SM predictions in the observables concerned with the semi-leptonic $B$ decays
in association with $\mu$ or $\tau$ lepton \cite{Aaij:2013qta,Aaij:2015oid,Aaij:2017vbb,Aaij:2014ora,Aaij:2019wad,Abdesselam:2019wac,Abdesselam:2019dgh,Aaij:2020nrf,Lees:2012xj,Lees:2013uzd,Aaij:2015yra,Huschle:2015rga,Sato:2016svk,Hirose:2016wfn,Aaij:2017deq}.
The LHCb collaboration reports deviations from the SM predictions in the measurements
of the semi-leptonic $B$ decays $B \to K^{(*)} \ell \ell $ ($\ell=e, \, \mu$).
The LHCb collaboration studies the
ratio of the branching ratio of $B \to K^{(*)} e e $ to that of $B \to K^{(*)} \mu \mu$
and the result tells that the branching ratio of $B \to K^{(*)} \mu \mu$ is slightly
smaller than the SM prediction.
We call this discrepancy the $\RK$ anomaly. 
The collaboration investigates the
observables related to the angular distribution of the $B \to K^{*} \mu \mu$ decay as well,
and the one observable, namely $P^\prime_5$, deviates from the SM prediction.
The recent observation for the angular observables in charged $B$ meson decay
also supports the result~\cite{Aaij:2020ruw}.
In addition, we can also find the discrepancy in $B \to D^{(*)} \tau \nu $.
The decay mode has been studied in the BaBar, the Belle and the LHCb experiments.
The BaBar collaboration has announced that the experimental result on the lepton universality,
where the branching ratio of $B \to D^{(*)} \tau \nu $ is compared with
$B \to D^{(*)} \ell \nu $, is largely deviated from the SM prediction. 
We call this discrepancy the $\RD$ anomaly. 
The Belle and the LHCb have also studied the decay and have reported their results;
the discrepancy becomes milder but the world average is still about $3\,$-$\,4 \sigma$ away from the
SM value\cite{Bernlochner:2017jka,Bigi:2017jbd,Bordone:2019vic,Amhis:2019ckw,Iguro:2020cpg}.

Motivated by those problems, a lot of new physics possibilities have been proposed.
The leptoquarks provide tree-level explanations of
both $\RK$ and $\RD$ anomalies.
Various types of leptoquarks, scalar or vector, representations under the SM gauge group,
$G_{\SM}=SU(3)_C \times SU(2)_L \times U(1)_Y$,
are discussed in e.g. Refs.~\cite{Bhattacharya:2014wla,Fajfer:2015ycq,Barbieri:2015yvd,Das:2016vkr,Becirevic:2016yqi,Sahoo:2016pet,Bhattacharya:2016mcc}.
It is shown that both of the anomalies can be explained by a vector leptoquark
with $(\rep{3}, \rep{1}, 2/3)$ under $G_\SM$, 
which is realized in 
extensions of the minimal PS model~\cite{DiLuzio:2017vat,Greljo:2018tuh,Cornella:2019hct,DiLuzio:2018zxy,Bordone:2017bld,Bordone:2018nbg,Fuentes-Martin:2020bnh,Fuentes-Martin:2020hvc,Guadagnoli:2020tlx}.  
Besides the leptoquarks, the $\RK$ anomaly is explained
by $Z^\prime$ boson at the tree-level~\cite{Altmannshofer:2014cfa,Crivellin:2015mga,Raby:2017igl, King:2017anf, Falkowski:2018dsl,Kawamura:2019rth,Kawamura:2019hxp}
or a 1-loop box diagrams mediated by extra fermions~\cite{Gripaios:2015gra,Arnan:2016cpy,Grinstein:2018fgb,Arnan:2019uhr, Chiang:2017zkh,Cline:2017qqu,Kawamura:2017ecz,Barman:2018jhz,Cerdeno:2019vpd,Arcadi:2021glq}.
However, the explanation of $\RD$ anomaly with a charged Higgs boson~\cite{Crivellin:2012ye,Tanaka:2012nw,Celis:2012dk,Crivellin:2013wna,Crivellin:2015hha,Chen:2017eby,Iguro:2017ysu,Iguro:2018qzf} is constrained by the collider search\cite{Iguro:2018fni} and $B_c\to \tau \nu$ decay \cite{Alonso:2016oyd,Akeroyd:2017mhr}.

In this paper,
we shall study the phenomenology of the model
with TeV-scale $G_\PS$ breaking and vectorlike fermions.
The typical leptoquark is strongly constrained by the measurement of $K_L \to \mu e$, so that
its mass should be heavier than PeV scale.
It is, however, pointed out in Ref.~\cite{Calibbi:2017qbu} that the bound is relaxed to be less than TeV scale
if the mixing between the chiral and vectorlike fermions has a certain structure.
We explicitly construct a model that realizes such a structure of the fermion mass matrices.
We also point out that the parameters of the model should be tuned at $\order{0.1\%}$ level
to be consistent with the realistic fermion masses and mixing as well as flavor constraints from $K_L \to \mu e$ etc., if the $\RK$ anomaly is solved by the leptoquark.
Interestingly, most of the parameter space will be covered by future experiments
searching for the $\mu$-$e$ conversion process and $\mu \to e\gamma$.
We also show that $\RD$ anomaly is hardly explained in this model
due to the constraints from the $Z^\prime$ boson search at the LHC.
Besides, we study flavor violations via the extra Higgs bosons
which are inevitably induced to explain the realistic Yukawa couplings.

This paper is organized as follows.
We introduce our PS model in Sec.~\ref{sec-model},
and then the phenomenology is discussed in Sec.~\ref{sec-pheno}.
Section~\ref{sec-summary} is devoted to summary.
The details of the model and numerical analysis are shown in Appendix~\ref{sec-detail} and~\ref{sec-analysis},
respectively.

\section{Pati-Salam model with vectorlike fermions}
\label{sec-model}

\begin{table}[t]
\centering
\caption{ \label{table-matct}
The matter content in the Pati-Salam model.
Each type of fermion has three flavors.
}
\begin{tabular}{c|cccc}\hline
fields & spin   & ~~$SU(4)$~~ & ~~$SU(2)_L$~~  & ~~$SU(2)_R$~~      \\ \hline\hline
  $L$  & $1/2$ & ${\bf 4}$         &${\bf2}$    &   ${\bf1}$       \\
   $R$  & $1/2$&  ${\bf 4}$        &${\bf1}$&   ${\bf2}$               \\  \hline
     $F_L$ & $1/2$ &  ${\bf 4}$        &${\bf 2}$ &   ${\bf1}$         \\
          $F_R$ & $1/2$ &  ${\bf 4}$        &${\bf 2}$ &   ${\bf1}$         \\
         $f_L$ & $1/2$ &  ${\bf 4}$        &${\bf 1}$ &   ${\bf2}$        \\
         $f_R$ & $1/2$ &  ${\bf 4}$        &${\bf 1}$ &   ${\bf2}$       \\  \hline
     $\Delta$ & $0$ & ${\bf 15}$         &${\bf1}$ &   ${\bf1}$          \\
         $\Sigma$ & $0$ & ${\bf \ol{10}}$         &${\bf1}$ &   ${\bf3}$        \\
   $\Phi$& $0$  &  ${\bf 1}$        &${\bf2}$ &   $\ol{\bf2}$                    \\
  $S$& $0$  &  ${\bf 1}$        &${\bf1}$ &   ${\bf1}$                \\
\hline
\end{tabular}
\end{table}

We shall consider a model with the PS gauge symmetry,
$G_{\PS} = SU(4)_C \times SU(2)_L\times SU(2)_R$.
In the minimal setup with $G_{\PS}$,
there are three generations of chiral fermions $L$, $R$ and a bi-doublet Higgs field $\Phi$.
The Yukawa couplings of the chiral fermions are given by
\begin{align}
\label{eq-yukSM}
-\Lcal^\mathrm{Ch}_Y=  \ol{L}  y_1 \Phi R+\ol{L}  y_2 \eps^T  \Phi^* \eps R +h.c.,
\end{align}
where $\eps := i\sigma_2$ acts onto the $SU(2)_L$ and $SU(2)_R$ indices.
Here, $y_1$ and $y_2$ are the $3\times 3$ Yukawa matrices in the flavor space.
The linear combination of the two terms leads the splittings of the Yukawa couplings
between the up-type and down-type quarks, as well as the charged leptons and neutrinos.
At the tree-level, however,
the Yukawa couplings to the down-type quarks and charged leptons are coincident,
and thus there is no mass splitting between them.
It is often considered that the mass differences are realized
by RG effects after the PS gauge symmetry breaking
and/or by incorporating higher-dimensional operators involving symmetry breaking
vacuum expectation values (VEVs).
The symmetry breaking scale may need to be around the conventional GUT scale $\sim 10^{16}$ GeV,
in order to realize sufficiently sizable corrections from the RG effects and/or higher-dimensional operators.
Thus, these effects will be too small to explain the realistic fermion masses
if the $G_{\PS}$ breaking scale is around TeV.

In this paper, we show an explicit model with the TeV-scale vector leptoquark.
The matter content and the charge assignment are summarized in Table~\ref{table-matct}.
The PS gauge symmetry is spontaneously broken at the TeV scale in our setup.
We introduce extra vectorlike fermions, an $SU(4)_C$ adjoint scalar $\Delta$,
and another scalar $\Sigma$, ($\mathbf{\ol{10}}$, $\mathbf{1}$, $\mathbf{3}$).
The mass splittings between the charged leptons and down quarks are generated
by the VEV of $\Delta$.
The VEV of $\Sigma$ induces the Majorana masses for the right-handed neutrinos.
In this model,
the non-zero VEV of $\Delta$ breaks $SU(4)_C$ to $SU(3)_C\times U(1)_\BL$
and
that of $\Sigma$ breaks $SU(2)_R\times U(1)_\BL$ to $U(1)_Y$.

The vectorlike fermions are denoted by $F_L$, $F_R$, $f_L$ and $f_R$.
 Each type of vectorlike fermion has three flavors as the chiral fermions.
As shown in Table~\ref{table-matct}, the charge assignment of $L(R)$ is the same as that of $F_L (f_R)$.
In our work, we simply assume that there is an underlying theory or some symmetry, and they can be distinguished from each other.
There are three-generations of vectorlike $SU(2)_L$ and $SU(2)_R$ doublet fermions.
This is the minimal setup such that the leptoquark couplings to the SM fermions are vanishing in a certain limit,
as shown in later.
The vectorlike mass terms and Yukawa couplings are given by
\begin{align}
\label{eq-yukVL}
-\Lcal^\mathrm{VL}_Y =&\
\ol{F}_L M_L F_R+\ol{f}_LM_R f_R
+\ol{f}_L\,m_R R +\ol{L}\, m_L F_R  \\ \notag
& +  \ol{F}_L \ka_L \Delta  F_R+\ol{f}_L  \ka_R \Delta f_R
+\ol{f}_L\, \eps_R\Delta  R +\ol{L}\, \eps_L\Delta F_R
\notag   \\
&+\ol{F}_L\, \la_1\,\Phi \,R+\ol{F}_L\, \la_2\, \eps^T\Phi^*\eps\,R+\ol{L} \wt{\la}_1\, \Phi \, f_R
+\ol{L}\, \wt{\la}_2\, \eps^T \Phi^* \eps \,  f_R
\notag \\ \notag
&+\ol{F}_L \, \wt{y}_1\, \Phi\, f_R+\ol{F}_L \, \wt{y}_2\,  \eps^T \Phi^* \eps\, f_R
+\ol{f}_L\, \wt{y}_1^{\prime}\,\Phi^\dagger\, F_R
+\ol{f}_L \, \wt{y}_2^{\prime} \,  \eps^T \Phi^T \eps \,  F_R
 +h.c.,
\end{align}
where all the masses and Yukawa couplings are $3\times 3$ matrices in the flavor space.
Here, the first line is the tree-level vectorlike mass terms. 
The second line is the Yukawa couplings with the $SU(4)_C$ adjoint $\Delta$
and the last two lines are the Yukawa couplings with the bi-doublet $\Phi$.
The mass splittings between quarks and leptons are induced by the Yukawa couplings involving the adjoint scalar $\Delta$.
In the next subsection, we study the fermion mass matrices originated from
the spontaneous PS symmetry breaking.

In addition, there are Yukawa couplings involving $\Sigma$, which induce Majorana masses after the symmetry breaking.
As discussed in the next subsection, we simply assume that the Yukawa couplings between $R$ and $f_R$ is vanishing
and the effective Majorana mass terms consist of only $f_R$~\footnote{We note that this situation can be realized by
assigning extra symmetry to distinguish $f_R$ from $R$.}.

\subsection{Fermion masses}

After the $G_\PS$ breaking, the fermion multiplets are decomposed as
\begin{align}
& L =
\begin{pmatrix}
e_L  &   n_L  \\
d_L  &  u_L
\end{pmatrix},
\quad
F_L =
\begin{pmatrix}
  E_L & N_L  \\
D_L  &  U_L
\end{pmatrix},
\quad
f_L =
\begin{pmatrix}
\Ecal_L  &  \Ncal_L  \\
 \Dcal_L &  \Ucal_L
\end{pmatrix},  \\
& R =
\begin{pmatrix}
e_R  &   n_R  \\
d_R  &   u_R
\end{pmatrix},
\quad
F_R =
\begin{pmatrix}
E_R &  N_R  \\
D_R  &  U_R
\end{pmatrix},
\quad
f_R =
\begin{pmatrix}
\Ecal_R  & \Ncal_R \\
 \Dcal_R &  \Ucal_R
\end{pmatrix},
\end{align}
where the rows are the $SU(4)_C$ space
and the columns are the $SU(2)_{L(R)}$ space for the $SU(2)_{L(R)}$ doublets.
We parametrize the Dirac mass matrices as
\begin{align}
\label{eq-massgauge}
& -{\cal L}^\mathrm{Ch}_Y- {\cal L}^\mathrm{VL}_Y
\\  \notag
& =
\begin{pmatrix}
\ol{u}_L \\ \ol{U}_L \\ \ol{\Ucal}_L
 \end{pmatrix}^T
\begin{pmatrix}
 y_u v_H & \tla_u v_H             & m_{Q_L} \\
\lambda_u v_H & \ty_u v_H & M_{Q_L} \\
m_{Q_R} & M_{Q_R} & \ty_u^{\prime} v_H
\end{pmatrix}
\begin{pmatrix}
u_R  \\ \Ucal_R \\ U_R
\end{pmatrix}
+ \begin{pmatrix}
\ol{d}_L \\ \ol{D}_L \\ \ol{\Dcal}_L
 \end{pmatrix}^T
\begin{pmatrix}
 y_d v_H & \tla_d v_H             & m_{Q_L} \\
\lambda_d v_H & \ty_d v_H & M_{Q_L} \\
m_{Q_R} & M_{Q_R} & \ty_d^{\prime} v_H
\end{pmatrix}
\begin{pmatrix}
d_R  \\ \Dcal_R \\ D_R
\end{pmatrix}      \\ \notag
& +
 \begin{pmatrix}
\ol{e}_L \\ \ol{E}_L \\ \ol{\Ecal}_L
 \end{pmatrix}^T
\begin{pmatrix}
 y_d v_H & \tla_d v_H             & m_{\ell_L} \\
\lambda_d v_H & \ty_{d} v_H & M_{\ell_L} \\
m_{\ell_R} & M_{\ell_R} & \ty_d^{\prime} v_H
\end{pmatrix}
\begin{pmatrix}
e_R  \\ \Ecal_R \\ E_R
\end{pmatrix}
+
\begin{pmatrix}
\ol{n}_L \\ \ol{N}_L \\ \ol{\Ncal}_L
 \end{pmatrix}^T
\begin{pmatrix}
 y_u v_H & \tla_u v_H             & m_{\ell_L} \\
\lambda_u v_H & \ty_{u} v_H & M_{\ell_L} \\
m_{\ell_R} & M_{\ell_R} & \ty_u^{\prime} v_H
\end{pmatrix}
\begin{pmatrix}
n_R  \\ \Ncal_R \\ N_R
\end{pmatrix}
 +h.c. \\
&=
\ol{\bf u}_L \Mcal_u {\bf u}_R +
\ol{\bf d}_L \Mcal_d  {\bf d}_R +
\ol{\bf e}_L \Mcal_e {\bf e}_R +
\ol{\bf n}_L \Mcal_n {\bf n}_R +
h.c.,
\label{matrix}
\end{align}
where $ \lambda_{u,d}$, $ \tla_{u,d}$, $y_{u,d}$ $\tla_{u,d}$ are the linear combinations of
$ \lambda_{1,2}$, $ \tla_{1,2}$, $y_{1,2}$ and $\ty_{1,2}$, respectively~\footnote{
Their explicit relations are shown in Appendix~\ref{sec-intscal}.
}.
The VEVs of the bi-doublet and adjoint are defined as
\begin{align}
 \vev{\Phi} = v_H \times \mathrm{diag}\left(s_\beta,~c_\beta \right),
\quad
\vev{\Delta} = \frac{v_\Delta}{2\sqrt{3}} \times \mathrm{diag}\left(3,-1,-1,-1\right),
\end{align}
where $s_\beta^2 + c_\beta^2 = 1$ and $v_H \simeq 174$~GeV.
The mass matrices $M_{{\cal F}}$ and $m_{{\cal F}}$ (${\cal F}=Q_{L,R}, \, \ell_{L,R}$) are defined as
\begin{align}
\label{eq-MVL}
 M_{\ell_{L,R}}=&\  M_{L,R} + \frac{3}{2\sqrt 3} \kappa_{L,R} \,  v_\Delta,
\quad
M_{Q_{L,R}}=  M_{L,R} - \frac{1}{2\sqrt 3} \kappa_{L,R}\,   v_\Delta, \\ \notag
m_{\ell_{L,R}}=&\  m_{L,R} + \frac{3}{2\sqrt 3} \epsilon_{L,R} \,  v_\Delta,
\quad
m_{Q_{L,R}}=  m_{L,R} - \frac{1}{2\sqrt 3} \epsilon_{L,R}\,   v_\Delta.
\end{align}
Here, $\Mcal_{u}$, $\Mcal_{d}$, $\Mcal_{e}$ and $\Mcal_{n}$
respectively represent $9 \times 9$ mass matrices for the 9 generations of fermions
$\mathbf{u}_{L,R}$, $\mathbf{d}_{L,R}$, $\mathbf{e}_{L,R}$ and $\mathbf{n}_{L,R}$.
The Yukawa couplings with the SM Higgs bosons are the same
for the down quarks and charged leptons, as well as the up quarks and neutrinos,
while the vectorlike masses and Yukawa couplings with $\Delta$ are common
in each of the quarks and leptons.
Note that the orderings of the $SU(2)_L$ singlet and doublet states are flipped
for which the electroweak (EW) gauge couplings are simplified.
We define the mass basis for the charged fermions as
\begin{align}
\hat{\bf u}_{L,R} =&\
\left(U^u_{L,R}\right)^\dag \mathbf{u}_{L,R},
\quad
\hat{\bf d}_{L,R} =
\left(U^d_{L,R}\right)^\dag  \mathbf{d}_{L,R},
\quad
\hat{\bf e}_{L,R} =
\left(U^e_{L,R}\right)^\dag  \mathbf{e}_{L,R}. 
\end{align}
The unitary matrices diagonalize the mass matrices as
\begin{align}
\left( U^f_L \right)^\dag
\Mcal_f
U^f_R
= \mathrm{diag}\left(m^f_1, m^f_2, \cdots, m^f_9  \right),
\end{align}
where $f=e,d,u$.
The singular values are in ascending ordered, i.e. $m^f_{a} \le m^f_{a+1}$ $(a=1,\cdots,8)$.
The SM fermion masses are given by $m^f_1$, $m^f_2$ and $m^f_3$:
\begin{align}
 (m^e_1, m^e_2,m^e_3 )  = (m_e, m_\mu, m_\tau),
\
 (m^d_1,m^d_2, m^d_3) = (m_d, m_s, m_b),
\
 (m^u_1, m^u_2, m^u_3) = (m_u, m_c, m_t).
\notag
 \end{align}

The neutrino masses are explained by the type-I seesaw mechanism
with Majorana masses for the right-handed neutrinos.
In this model, the Yukawa couplings with $\Sigma$ are given by
\begin{align}
\label{eq-ysig}
- \Lcal_{\Sigma}^y =  \frac{1}{2}  {\ol{f}^c_R}\, h\, \eps^T  \Sigma   f_R + h.c.,
\end{align}
where $h$ is the $3\times 3$ symmetric Yukawa matrix.
The scalar $\Sigma$ is represented as
\begin{align}
 \Sigma_{\alpha\beta} = \Sigma^{k_R}_{\alpha\beta} \tau^{k_R} = \frac{1}{2}
\begin{pmatrix}
\Sigma^3_{\alpha\beta} & \sqrt{2} \Sigma^+_{\alpha\beta} \\ \sqrt{2} \Sigma^-_{\alpha\beta} & - \Sigma^3_{\alpha\beta}
\end{pmatrix},
\quad
\Sigma^{\pm}_{\alpha\beta} = \frac{\Sigma^1_{\alpha\beta} \pm i \Sigma^2_{\alpha\beta}}{\sqrt{2}},
\end{align}
where $k_R=1,2,3$ is $SU(2)_R$ index and $\alpha,\beta =1,2,3,4$ are the $SU(4)_C$ indices.
We shall assume that $\Sigma$ obtains the VEV as
\begin{align}
 \vev{\Sigma_{11}^{+}} = \frac{v_\Sigma}{\sqrt{2}} \ne 0,\quad
\mathrm{others} = 0,
\end{align}
where $\alpha,\beta = 1$ is the leptonic direction in the $SU(4)_C$ space.
After the symmetry breaking, the Majorana mass term is given by
\begin{align}
\label{eq-MajR}
- \Lcal_{h,\Sigma} \supset \frac{1}{2}\;   \ol{\Ncal}^{\;c}_{R} M_R \Ncal_R + h.c.,
\quad
M_{R} = \frac{h}{\sqrt{2}} v_\Sigma.
\end{align}
The full neutrino mass term is given by
\begin{align}
- \Lcal_N = \frac{1}{2} \ol{\rep{N}}_L \Mcal_N \rep{N}_R :=
\begin{pmatrix}
\ol{\rep{n}}_L & \ol{\rep{n}}_R^c
\end{pmatrix}
\begin{pmatrix}
 \zd{9} & \Mcal_n \\ \Mcal_n^T & \Mcal_R
\end{pmatrix}
\begin{pmatrix}
 \rep{n}_L^c \\ \rep{n}_R
\end{pmatrix},
\quad
\Mcal_R =
\begin{pmatrix}
 \zd{3} & \zd{3}  & \zd{3} \\
 \zd{3} & M_R & \zd{3} \\
 \zd{3} & \zd{3}     & \zd{3}
\end{pmatrix}, 
\end{align}
where the Dirac mass matrix is defined in Eq.~\eqref{eq-massgauge}~\footnote{
In this paper, $\zd{n}$ is a $n\times n$ zero matrix. Similarly, $\id{n}$ is an $n\times n$ identity matrix.
}.
Note that $\rep{N}_R^c = \rep{N}_L$.
The mass eigenstate is defined as
\begin{align}
 \hat{\rep{N}}_R = U_N^\dag \rep{N}_R,\quad
 \hat{\rep{N}}_L = U_N^T \rep{N}_L,\quad
\left[U_N^T \Mcal_N U_N\right]_{xy} = m_{\nu_x} \delta_{xy},
\end{align}
with $x,y=1,2,3,\cdots,18$. Here, $U_N$ is an $18\times 18$ unitary matrix.
We define the $9\times 18$ matrices,
\begin{align}
 U^n_L := \Pcal_L U_N^*,\quad U^n_R := \Pcal_R U_N,
\end{align}
where the projection matrices for the neutrino flavors are defined as
\begin{align}
\Pcal_L :=
\begin{pmatrix}
 \id{9} & \zd{9}
\end{pmatrix},
\quad
\Pcal_R :=
\begin{pmatrix}
 \zd{9} & \id{9}
\end{pmatrix}.
\end{align}

\subsection{Leptoquark couplings}
\label{LQcoupling}
The spontaneous PS gauge symmetry breaking generates the mass of
the vector leptoquark, that is part of the $SU(4)_C$ gauge field.
The gauge couplings with the vector leptoquark, $X^\mu$, are given by
\begin{align}
\label{eq-Xcoup}
\Lcal_X =&\  \frac{g_4}{\sqrt{2}} X^\mu
                    \Bigl(\ol{\rep{d}}_L \gamma_\mu \rep{e}_L +
                    \ol{\rep{u}}_L \gamma_\mu  \rep{n}_L + \ol{\rep{d}}_R \gamma_\mu  \rep{e}_R
                    +      \ol{\rep{u}}_R \gamma_\mu X^\mu \rep{n}_R
                    \Bigr)  + h.c.   \\  \notag
           =&\  X^\mu \Bigl(\hat{\ol{\rep{d}}}_L \hat{{g}}^X_{d_L}  \gamma_\mu  \hat{\rep{e}}_L +
                    \hat{\ol{\rep{u}}}_L \hat{{g}}^X_{u_L} \gamma_\mu  \hat{\rep{N}}_L
            +   \hat{\ol{\rep{d}}}_R \hat{{g}}^X_{d_R} \gamma_\mu   \hat{\rep{e}}_R +
                           \hat{\ol{\rep{u}}}_R \hat{{g}}^X_{u_R} \gamma_\mu  \hat{\rep{N}}_R\Bigr) + h.c..
\end{align}
Here, the coupling matrices in the mass basis are given by
\begin{align}
\label{eq-gXs}
& \hat{g}^X_{d_{L}} = \frac{g_4}{\sqrt{2}}\left(U^d_L\right)^\dag U^e_L,
&\quad &
 \hat{g}^X_{d_{R}} =  \frac{{g_4}}{\sqrt{2}} \left(U^d_R\right)^\dag U^e_R, \\ \notag
&
\hat{g}^X_{u_{L}} =  \frac{{g_4}}{\sqrt{2}} \left(U^u_L\right)^\dag U^n_L
&\quad &
 \hat{g}^X_{u_{R}} = \frac{{g_4}}{\sqrt{2}}  \left(U^u_R\right)^\dag U^n_R. 
\end{align}
The couplings of the fermions to the other gauge bosons and scalars are shown in Appendix~\ref{sec-detail}.

All the SM fermion masses and mixings can be explained consistently with the PS relations
by the mass splittings via the Yukawa couplings with the bi-doublets $\Phi$ and adjoint $\Delta$
even if the $G_\PS$ breaking scale is at the TeV scale.
The leptoquark, however, couples to the SM charged leptons and down-type quarks
if the SM fermions dominantly come from the chiral fermions, $L$ and $R$.
When the leptoquark has sizable couplings to the light flavor fermions,
particularly electron and down-type quarks simultaneously,
it is known that the limit from $K_L \to \mu e$ gives the most stringent bound
on the leptoquark mass, that is the $G_\PS$ breaking scale~\cite{Hung:1981pd,Valencia:1994cj}~
\footnote{We note that the flavor constraints from $K \to \pi \mu e$ are milder than the one obtained from $K_L \to \mu e$~\cite{Borsato:2018tcz}.
}.
If $\hat{g}^X_{d_L} = \hat{g}^X_{d_R} =  (g_4/\sqrt{2}) \cdot \id{9}$,
the branching fraction of $K_L \to \mu e$ is estimated as~\footnote{
See Eq.~\eqref{eq-KLemFull} for the full formula.}.
\begin{align}
\label{Ktomue}
\br{K_L}{\mu e} \sim 1.4\times 10^{-11} \times \left(\frac{1~\mathrm{PeV}}{m_X}\right)^4
                                           \left( \frac{g_4}{1.0}\right)^4,
\end{align}
while the current upper bound  is $4.7\times 10^{-12}$.   
Thus the leptoquark should be heavier than 1 PeV in this case,
that is too heavy to explain the current flavor anomalies.
The couplings to light flavor fermions could be suppressed
by introducing vectorlike fermions~\cite{Calibbi:2017qbu} as in our model.
In the following, we will show explicit mass matrices which are consistent
with the TeV-scale vector leptoquark,
SM fermion mass and mixing.
We also see that most of flavor violating processes as well as $K_L \to \mu e$
are sufficiently suppressed.

\subsection{How to suppress $K_L \to \mu e$}

We propose an explicit way to avoid the strong constraint from the $K_L\to\mu e$ at the tree level,
by tuning the vectorlike masses such that
\begin{align}
\label{eq-tuning}
m_{Q_R} \ll M_{Q_R},\quad M_{Q_L} \ll m_{Q_L}, \quad
M_{\ell_R} \ll m_{\ell_R},\quad m_{\ell_L} \ll M_{\ell_L}.
\end{align}
The mass matrices of the down quarks and charged leptons are schematically given by
\begin{align}
\label{eq-texture}
 \mathcal{M}_d \sim
\begin{pmatrix}
\zd{3}      & \Hat{m}_e & m_{Q_L} \\
\Hat{m}_d & \zd{3}      & \zd{3} \\
\zd{3}      & M_{Q_R}& \zd{3} 
\end{pmatrix},
\quad
\mathcal{M}_e \sim
\begin{pmatrix}
\zd{3}      & \Hat{m}_e & \zd{3} \\
\Hat{m}_d & \zd{3}      & M_{\ell_L} \\
m_{\ell_R}   & \zd{3} & \zd{3} 
\end{pmatrix},
\end{align}
where $\Hat m_d$ and $\Hat m_e$ are the $3\times 3$ mass matrices proportional to $v_H$.
The same structure will arise in the up quark and neutrino sector due to the hierarchy in Eq.~\eqref{eq-tuning}.
With this texture, $\Hat m_d$ and $\Hat m_e$ are approximately the mass matrices for the down-type quarks
and charged leptons in the SM generations since the mixing with the other blocks are suppressed.
The down quarks are originated from $(F_L, R)$, while the charged leptons are originated from $(L, f_R)$.
Thus, the masses and mixing of the SM fermions can be explained separately,
and the leptoquark couplings to the SM down-type fermions 
are more suppressed as the hierarchical structure in Eq.~\eqref{eq-tuning} is more strict.

\subsection{Parametrization}

In our notation, it is convenient to express the mass matrices by
a $6\times 6$ block on the upper-left and a $3\times 3$ block on the bottom-right.
Without loss of generality~\footnote{
Here, the mass matrices are expressed by the singular value decomposition,
as usually applied to the Dirac mass matrix.
},
we can parametrize the Dirac mass matrices in the gauge basis as 
\begin{align}
\label{eq-parametrization}
 \Mcal_d =&\
\begin{pmatrix}
 \tD_d & V_{Q_L} \tD_{Q_L} W_{Q_R}^\dag  \\ W_{Q_L} \tD_{Q_R} V_{Q_R}^\dag & \Delta_{d}
\end{pmatrix},
\quad
 \Mcal_e =
\begin{pmatrix}
 \tD_d & V_{\ell_L} \tD_{\ell_L} W_{\ell_R}^\dag  \\ W_{\ell_L} \tD_{\ell_R} V_{\ell_R}^\dag & \Delta_{d}
\end{pmatrix}, \\ \notag
 \Mcal_u =&\
\begin{pmatrix}
v_{u_L} \tD_u v_{u_R}^\dag  & V_{Q_L} \tD_{Q_L} W_{Q_R}^\dag  \\
W_{Q_L} \tD_{Q_R} V_{Q_R}^\dag & w_{u_L} \Delta_{u} w_{u_R}^\dag
\end{pmatrix},
\quad
 \Mcal_n =
\begin{pmatrix}
 v_{u_L} \tD_u v_{u_R}^\dag  & V_{\ell_L} \tD_{\ell_L} W_{\ell_R}^\dag \\
W_{\ell_L} \tD_{\ell_R} V_{\ell_R}^\dag & w_{u_L} \Delta_{u} w_{u_R}^\dag
\end{pmatrix},
\end{align}
where
\begin{align}
 \tD_d  =
\begin{pmatrix}
  \zd{3} & D_e \\
  D_d  & \zd{3}
\end{pmatrix},
\quad
 \tD_u  =
\begin{pmatrix}
  \zd{3}  & D_n \\
  D_u  & \zd{3}
\end{pmatrix},
\end{align}
and
\begin{align}
 \tD_{Q_R} =
\begin{pmatrix}
\zd{3} &  D_{Q_R}
\end{pmatrix},
\quad
 \tD_{Q _L} =
\begin{pmatrix}
D_{Q_L} \\ \zd{3}
\end{pmatrix},
\quad
 \tD_{\ell_R} =
\begin{pmatrix}
 D_{\ell_R} & \zd{3}
\end{pmatrix},
\quad
 \tD_{\ell_L} =
\begin{pmatrix}
\zd{3} \\ D_{\ell_L}
\end{pmatrix}.
\end{align}
Here, $D_{f}$ ($f=u,d,e,n$) and $D_{F_{L,R}}$ ($F=Q,\ell$) are $3\times 3$ diagonal matrices.
There are four $6\times 6$ unitary matrices $V_{Q_{L,R}}$, $V_{\ell_{L,R}}$
and eight $3\times 3$ unitary matrices $W_{Q_{L,R}}$, $W_{\ell_{L,R}}$
and  $v_{u_{L,R}}$, $w_{u_{L,R}}$.
By definition, the mass matrices are unchanged under
\begin{align}
\label{eq-redundancy}
 V_{\ell_{L,R}} \to &\
\begin{pmatrix}
 u_{\ell_{L,R}} & \zd{3} \\ \zd{3} & \id{3}
\end{pmatrix}
V_{\ell_{L,R}},
\quad
V_{Q_{L,R}} \to
\begin{pmatrix}
 u_{Q_{L,R}} & \zd{3} \\ \zd{3} & \id{3}
\end{pmatrix}
V_{Q_{L,R}},
\end{align}
where $u_{F_{L,R}}$ are arbitrary $3\times 3$ unitary matrices.
We start from a basis in which $D_d$, $D_e$ and $\Delta_d$ are diagonalized,
which can be done without changing any couplings with the gauge bosons or scalars. 
See Appendix~\ref{sec-diago} for more details.
We further assume that the Majorana matrix $M_R$ is proportional to an identity matrix
in this basis for simplicity.

This parametrization is defined such that the SM-like fermion masses and couplings are realized
when all the unitary matrices are identity matrices except $v_{u_L}$ which should be
\begin{align}
v_{u_L} =
\begin{pmatrix}
U_{\mathrm{PMNS}}  & \zd{3} \\ \zd{3} &  V_{\mathrm{CKM}}^\dag
\end{pmatrix},
\end{align}
so that the CKM and PMNS matrices are realized.
In this canonical case, the (Dirac) mass matrices are diagonalized by the following unitary matrices, 
\begin{align}
& U^{0,d}_{L} = U^{0,e}_{R} = U^{0,n}_R =
\begin{pmatrix}
 \zd{3} & \id{3}  & \zd{3} \\ \id{3} & \zd{3} & \zd{3}  \\ \zd{3}  & \zd{3} & \id{3}
\end{pmatrix},
\quad
U^{0,u}_{L} =
\begin{pmatrix}
 \zd{3} & \id{3}  & \zd{3} \\  V_{\mathrm{CKM}}^\dag  & \zd{3} & \zd{3}  \\ \zd{3}  & \zd{3} & \id{3}
\end{pmatrix},  \\ \notag
& U^{0,u}_{R} = U^{0,d}_{R} = U^{0,e}_{L} =
\begin{pmatrix}
 \id{3} & \zd{3} & \zd{3} \\ \zd{3} & \zd{3} & \id{3}  \\ \zd{3}  & \id{3} & \zd{3}
\end{pmatrix},
\quad
U^{0,n}_{L} =
\begin{pmatrix}
 U_{\mathrm{PMNS}} & \zd{3} & \zd{3} \\ \zd{3} & \zd{3} & \id{3}  \\ \zd{3}  & \id{3} & \zd{3}
\end{pmatrix},
\end{align}
up to $\order{v_H^2/v_\Delta}$ corrections.
For the full neutrino matrix, the diagonalization matrix is given by
\begin{align}
 U_N^0 \sim
\begin{pmatrix}
  \left(U^{0,n}_L\right)^* & \zd{9} \\   \zd{9} & U^{0,n}_R
\end{pmatrix}. 
\end{align}
Then, the leptoquark couplings are approximately given by
\begin{align}
 \hat{g}_{d_L}^X \sim  
\left(\hat{g}_{d_R}^{X}\right)^T \sim 
\Pcal_R  \left( \hat{g}_{u_R}^{X}  \right)^T
\sim \frac{g_4}{\sqrt{2}}
\begin{pmatrix}
 \zd{3}&\zd{3}&\id{3} \\
 \id{3}&\zd{3}&\zd{3} \\
 \zd{3}&\id{3}&\zd{3} \\
\end{pmatrix},
\quad
\hat{g}_{u_L}^X \Pcal_L^T
\sim \frac{g_4}{\sqrt{2}}
\begin{pmatrix}
 \zd{3}&\zd{3}& V_{\mathrm{CKM}} \\
 U_{\mathrm{PMNS}} &\zd{3}&\zd{3} \\
 \zd{3}&\id{3}&\zd{3} \\
\end{pmatrix}.
\end{align}
Hence, the leptoquark couplings to the SM fermions,
which corresponds to the most upper-left block in the coupling matrix,
e.g. $[\hat{g}^X_{d_L}]_{ij}$ ($i,j \le 3$), are vanishing.
Thus, in this canonical case, there is no flavor violation at the tree-level,
although there might be flavor violation, such as $\mu\to e\gamma$,
from the loop-effects involving the vectorlike states.
In order to explain the flavor anomalies, the leptoquark should couple to the SM fermions with a certain pattern.
In the following, we will turn on the mixing angles in the unitary matrices
which are chosen to be identity in the canonical limit.
The diagonalization for a general case with the parametrization in Eq.~\eqref{eq-parametrization}
is discussed in Appendix~\ref{sec-diago}.

For simplicity, we assume that the singular values for vectorlike fermions
and $\Delta_{u,d}$ are universal~\footnote{
Precisely, we introduce $\order{0.1~\mathrm{GeV}}$ corrections to the vectorlike fermion masses
to avoid numerical instabilities due to the degeneracy as can be seen in Appendix~\ref{sec-analysis}.
}, i.e.
\begin{align}
 D_{Q_L} = d_{Q_L} \id{3},\quad
 D_{Q_R} = d_{Q_R} \id{3},\quad
 D_{\ell_L} = d_{\ell_L} \id{3},\quad
 D_{\ell_R} = d_{\ell_R} \id{3},
\end{align}
and
\begin{align}
 \Delta_d = \delta_d \id{3}, \quad  \Delta_u = \delta_u \id{3}.
\end{align}
We also assume that the Majorana mass matrix is given by
\begin{align}
\Mcal_R = m_N
\begin{pmatrix}
 \zd{3} & \zd{3} & \zd{3} \\ \zd{3} & \id{3} & \zd{3} \\ \zd{3} & \zd{3} & \zd{3}
\end{pmatrix}.
\end{align}
Further, we take $W_{Q_{L,R}}$, $W_{\ell_{L,R}}$ and $w_{u_{L,R}}$ identity matrices,
since these are not significant for the SM fermion couplings.

When the $V_{Q_{L,R}}$ and $V_{\ell_{L,R}}$ are not identity matrices,
the diagonalization matrices for the down-type fermions in Eq.~\eqref{eq-parametrization}
are approximately given by
\begin{align}
 U_{e_L} \sim&\
\begin{pmatrix}
 0_{6\times 3} & V_{\ell_L} \\ W_{\ell_L} & 0_{3\times 6}
\end{pmatrix}
P_{1}
,
\quad &
 U_{e_R} \sim&\
\begin{pmatrix}
V_{\ell_R} & 0_{6\times 3}  \\ 0_{3\times 6} & W_{\ell_R}
\end{pmatrix}
P_{1} ,  \\
 U_{d_L} \sim&\
\begin{pmatrix}
V_{Q_L} & 0_{6\times 3} \\  0_{3\times6} & W_{Q_L}
\end{pmatrix}
P_{1} ,
\quad &
 U_{d_R} \sim&\
\begin{pmatrix}
0_{6\times 3} &  V_{Q_R} \\ W_{Q_R}  & 0_{3\times 6}
\end{pmatrix}
P_{1} ,
\end{align}
where
\begin{align}
 P_1 :=
\begin{pmatrix}
\zd{3}&\id{3} &\zd{3}\\
\id{3} & \zd{3} &\zd{3}\\
\zd{3}& \zd{3} &\id{3}
\end{pmatrix}, 
\end{align}
is introduced so that the singular values are increasingly ordered.
The derivation and those for the up-type fermions are shown in Appendix~\ref{sec-diago}.
With this parametrization, the leptoquark couplings to the SM fermions are approximately given by
\begin{align}
\label{eq-XcoupApp}
\left[\hat{g}^X_{d_L}\right]_{ij}  \sim \left[\hat{g}^X_{u_L}\right]_{ij}
\sim \frac{g_4}{\sqrt{2}} \left[ V_{Q_L}^\dag V_{\ell_L} \right]_{3+i,j},\quad
\left[\hat{g}^X_{d_R}\right]_{ij} \sim \left[\hat{g}^X_{u_R}\right]_{ij}  \sim \frac{g_4}{\sqrt{2}} \left[ V_{Q_R}^\dag V_{\ell_R} \right]_{i,j+3},
\end{align}
with $i,j=1,2,3$.
Hence, the leptoquark couplings to the SM families are induced
from the mixing between the first three and the second three states.
We parametrize the unitary matrices, $V_{F_{X}}$, $F=\ell,Q$, $X=L,R$,
as~\footnote{Here, we assumed the matrices are real for simplicity.}
\begin{align}
\label{eq-VFX}
 V_{F_X} := 
R^{11}_{F_X} R^{12}_{F_X} R^{13}_{F_X}
R^{21}_{F_X} R^{22}_{F_X} R^{23}_{F_X}
R^{31}_{F_X} R^{32}_{F_X} R^{33}_{F_X},
\end{align}
where the rotation matrix $R^{ij}_{F_X}$ mixes the $i$-th and ($j$+3)-th elements, i.e.
\begin{align}
 R^{ij}_{F_X} =
\begin{pmatrix}
  1           & \cdots & 0                 & \cdots & 0                  & \cdots & 0 \\
   \vdots  &             & \vdots          &           & \vdots         &             & \vdots \\
   0          & \cdots & c^{ij}_{F_X} &  \cdots & s^{ij}_{F_X} & \cdots & 0 \\
   \vdots  &             & \vdots          &           & \vdots         &             & \vdots \\
   0          & \cdots & -s^{ij}_{F_X} &  \cdots& c^{ij}_{F_X} & \cdots  & 0 \\
   \vdots  &             & \vdots          &           & \vdots         &              &\vdots  \\
   0          & \cdots & 0&  \cdots& 0 & \cdots  & 1 \\
\end{pmatrix},
\end{align}
where $(c^{ij}_{F_X})^2+(s^{ij}_{F_X})^2 = 1$.
In our analysis, we assume that $V_{F_X}$'s are real
and we will not consider the mixing inside the first three and second three states.

In our numerical analysis of Section~\ref{sec-pheno}, the vectorlike fermion masses
$d_{Q_{L,R}}$, $d_{\ell_{L,R}}$, $\delta_{u,d}$ and the unitary matrices $V_{F_X}$ are input parameters.
The rest of parameters,  $\tilde{D}_d$, $\tilde{D}_u$ and $v_{u_L}$, $v_{u_R}$
are fitted to explain the observed fermion masses and mixing for a given set of input parameters.

\subsection{Fine-tuning}
The texture in Eq.~\eqref{eq-texture}, or the hierarchy in Eq.~\eqref{eq-tuning}
requires fine-tunings between the PS symmetric mass parameters 
in the first line of Eq.~\eqref{eq-yukVL}
and the mass terms originated from the $SU(4)_C$ adjoint $\vev{\Delta}$. 
To quantify the degree of tunings, we define the tuning measure $\Delta_{\mathrm{FT}}$ as
\begin{align}
 \Delta_{\mathrm{FT}}:=
\underset{A,B}{\mathrm{min}}
\left( \left[\Delta_{d_L}\right]_A, \left[ \Delta_{d_R}\right]_B  \right),\
\label{eq:tuning}
\end{align}
where
\begin{align}
\left[\Delta_{d_L} \right]_{A} :=
 \frac{\mathrm{min}\left(\abs{\left[\Mcal_d\right]_{A,6+\tilde{A}}}, \abs{\left[\Mcal_e\right]_{A,6+\tilde{A}}} \right)}
        {\mathrm{max}\left(\abs{\left[\Mcal_d\right]_{A,6+\tilde{A}}}, \abs{\left[\Mcal_e\right]_{A,6+\tilde{A}}} \right)},
\quad
\left[\Delta_{d_R} \right]_{B} :=
 \frac{\mathrm{min}\left(\abs{\left[\Mcal_d\right]_{6+B,\tilde{B}}}, \abs{ \left[\Mcal_e\right]_{6+B,\tilde{B}}} \right)}
        {\mathrm{max}\left(\abs{\left[\Mcal_d\right]_{6+B,\tilde{B}}}, \abs{\left[\Mcal_e\right]_{6+B,\tilde{B}}} \right)}.
\end{align}
Here, $A,B = 1,2,3,\cdots,6$ and $\tilde{A}$, $\tilde{B}$ are respectively residues of $A$, $B$ divided by $3$, so that it measures the degree of cancellations in the diagonal elements of the vectorlike masses.
For instance, 10~$\%$ tuning is required if $\Delta_{\mathrm{FT}}$ is 0.1.

\section{Phenomenology}
\label{sec-pheno}

We study phenomenology in this setup. We have seen the way to suppress
$K_L \to \mu e$. The suppression, however, may require severe fine-tuning.
If the fine-tuning is relaxed, other flavor violating processes would become sizable.
First of all, we discuss the possibility that
the TeV-scale leptoquark explains the anomalies in the semi-leptonic $B$ meson decays.
Then, we investigate the other flavor violating processes paying attention to
the degree of tuning.

\subsection{Vector leptoquark explanation of the anomalies}

Interestingly, the TeV-scale leptoquark may be able to explain all anomalies in the semi-leptonic $B$ meson decays,
$B\to K^{(*)}\mu\mu$ and $B\to D^{(*)}\tau\nu$.
It has been pointed out in the literature that the vector leptoquark from the PS model can explain both anomalies simultaneously,
but we show that the $\RD$ anomaly is hardly solved in our model
due to the correlation between the leptoquark and $Z^\prime$ boson masses
~\footnote{The correlation between leptoquark and $Z^\prime$ has been also studied in a more generic way \cite{Baker:2019sli}.}.

\subsubsection{$\RK$ anomaly}
The $\RK$ anomaly may be a signal of new physics.
The effective Hamiltonian is given by
\begin{align}
\Hcal_{\mathrm{eff}}^{\RK}
= -\frac{4G_F}{\sqrt{2}} \frac{\alpha}{4\pi} \sum_{a=9,10} \left( C_a \Ocal_a + C_a^\prime \Ocal_a^\prime \right),
\end{align}
where
\begin{align}
 \Ocal_9 =&\ \left(\ol{s} \gamma^\rho P_L b \right) \left( \ol{\mu} \gamma_\rho \mu \right),
\quad
 \Ocal_{10} = \left(\ol{s} \gamma^\rho P_L b \right) \left( \ol{\mu} \gamma_\rho \gamma_5 \mu \right), \\
 \Ocal_9^\prime
=&\ \left(\ol{s} \gamma^\rho P_R b \right) \left( \ol{\mu} \gamma_\rho \mu \right),
\quad
 \Ocal_{10}^\prime
= \left(\ol{s} \gamma^\rho P_R b \right) \left( \ol{\mu} \gamma_\rho \gamma_5 \mu \right).
\end{align}

The leptoquark contributions to the Wilson coefficients, $C_9$ and $C_{10}$, are given by
\begin{align}
\Delta C_9 = -\Delta C_{10} =&\
   - \frac{\sqrt{2}}{4G_F} \frac{4\pi}{\alpha} \frac{1}{V_{tb}^* V_{ts}}\frac{1}{2m_X^2}
                     \left[ \hat{g}^X_{d_L} \right]_{32}^*  \left[\hat{g}^X_{d_L} \right]_{22}   \\ \notag
   \sim &\ - 0.51 \times \left(\frac{5~\mathrm{TeV}}{m_X}\right)^2
  \left( \frac{\left[ \hat{g}^X_{d_L} \right]_{32}^*  \left[\hat{g}^X_{d_L} \right]_{22}}{0.02}  \right).
 \end{align}
The experimental results favor $-0.59 \lesssim \Delta C_9 \lesssim-0.41$
at $1\sigma$ level~\cite{Alguero:2019ptt},
and thus $m_X \sim \order{10~\mathrm{TeV}}$ is a suitable size to explain the $\RK$ anomaly.
The $Z^\prime$ boson would contribute to the $\RK$ anomaly,
but the flavor violating effects from the $Z^\prime$ boson is expected to be very suppressed
in our model as discussed in Appendix~\ref{sec-intgauge}.

With the parametrization of Eq.~\eqref{eq-VFX},
the relevant couplings are given by
\begin{align}
\label{eq-gXRK}
 \left[\hat{g}^X_{d_L}\right]_{22} = &\
\frac{g_4}{\sqrt{2}}  c_{\ell_L}^{23}
\left( s_{Q_L}^{22} c_{\ell_L}^{22} - c_{Q_L}^{22} s_{\ell_L}^{22} \right),  \notag \\
  \left[\hat{g}^X_{d_L}\right]_{32} = &\
\frac{g_4}{\sqrt{2}} \left\{  s_{Q_L}^{23} c_{\ell_L}^{23}
 \left( c_{Q_L}^{22} c_{\ell_L}^{22} + s_{Q_L}^{22} s_{\ell_L}^{22} \right)
   - s_{\ell_L}^{23}  c_{Q_L}^{23} \right\},
\end{align}
if the angles only in $R^{22}_{Q_L}$,  $R^{23}_{Q_L}$,  $R^{22}_{\ell_L}$ and $R^{23}_{\ell_L}$
are turned on.
We fix the angles at
\begin{align}
\label{eq-angleRK}
 s^{23}_{Q_L} = - s^{23}_{\ell_L} = \frac{1}{\sqrt{2}},
\quad
 s^{22}_{Q_L} = - s^{22}_{\ell_L} = 0.04\times \frac{1}{\sqrt{2}},
\end{align}
so that the $\RK$ anomaly is explained with $m_X \sim 5$ TeV
and $|[\hat{g}_{d_L}^X]_{22}/[\hat{g}_{d_L}^X]_{23}| \sim 0.04$.
As discussed in later, $[\hat{g}_{d_L}^X]_{22}$ should be small to suppress $K_L\to\mu e$.
Note that this model, in general, predicts the leptoquark contributions to the other lepton flavors,
as well as $C_{9,10}^\prime$.
In our analysis,  we will consider the parameter space where these are negligibly small.

\subsubsection{$\RD$ anomaly}

The effective Hamiltonian relevant to $\RD$ within the our model is given by
\begin{align}
 \Hcal_{\mathrm{eff}}^{\RD} = \frac{4G_F}{\sqrt{2}} V_{cb} C_{V_1}
    \left(\ol{c} \gamma^\mu P_L b \right) \left(\ol{\tau} P_L \nu_\tau\right).
\end{align}
The leptoquark contribution to the Wilson coefficients $C_{V_1}$
is given by
\begin{align}
\Delta  C_{V_1} =&\ \frac{\sqrt{2}}{4G_F}\frac{1}{ V_{cb}} \frac{1}{m_X^2}
                         \left[ \hat{g}^X_{d_L} \right]_{33}^*\left[ \hat{g}^X_{u_L} \right]_{2\nu} \\ \notag
   \sim&\   0.092 \times \left(\frac{1.4~\mathrm{TeV}}{m_X}\right)^2
  \left( \frac{\left[ \hat{g}^X_{d_L} \right]_{33}^* \left[\hat{g}^X_{u_L} \right]_{2{\nu}}}{0.25}  \right),
\end{align}
where $[\hat{g}^X_{u_L}]_{2{\nu}} := \sum_{i=1,2,3} [\hat{g}^X_{u_L}]_{2i}$.
The experimental results favor $0.052 \lesssim \Delta C_{V_1} \lesssim 0.124$ at $2\sigma$ level \cite{Iguro:2020keo}.
Note that
$\left[ \hat{g}^X_{d_L} \right]_{33}^*  \left[\hat{g}^X_{d_L} \right]_{2{\nu}} \sim g_4^2/4 \sim 0.25$
is the maximal value as far as the $SU(4)_C$ gauge coupling constant is $g_4 \sim 1.0$
that is consistent with the strong coupling constant at the TeV scale~\cite{Antusch:2013jca}.
Thus, $\Delta C_{V_1} \sim 0.09$ (0.057) could be explained if $m_X \sim 1.4$ (1.8) TeV.

The LHC result~\cite{Aad:2019fac} searching for di-lepton resonance severely constraints
a $Z^\prime$ boson mass if the $Z^\prime$ boson decays to a pair of electrons or muons.
In our model, the $Z^\prime$ boson couples to the fermions in the similar way as the $Z$ boson.
Therefore the resonant production cross section via the Drell-Yan process can be large as long as $Z^\prime$ boson is light, 
and there are sizable branching fractions to di-leptons, 
as we see in Eq.~\eqref{eq-ZprimeC} of Appendix~\ref{sec-intgauge}. 
If all the vectorlike fermions are heavier than the half of the $Z^\prime$ mass, the limit is about 5 TeV.
The limit is relaxed to about 4.5 TeV if $Z^\prime$ boson can decay to vectorlike fermions,
since the branching fractions to di-leptons are suppressed by $1/3$~\footnote{
The larger decay width will also relax the limit.
}.
From Eqs.~\eqref{eq-gmatrix} and~\eqref{eq-mZZpapp} in Appendix~\ref{sec-intgauge},
the $Z^\prime$ boson mass is bounded by the leptoquark mass,
\begin{align}
 m_{Z^\prime} < \sqrt{\frac{2g_R^2 + 3 g_4^2}{g_4^2}}\;  m_X
\sim 3.5~\mathrm{TeV}\times \left(\frac{m_X}{1.8~\mathrm{TeV}} \right), 
\end{align}
where the upper bound is saturated when $v_\Delta = 0$.
Hence, the $Z^\prime$ boson is too light to be consistent with the current limit.
Furthermore, the vectorlike fermions will have masses of $\order{v_\Delta}$
unless the Yukawa couplings are non-perturbatively large.
The LHC limit on a single vectorlike quark is about 1.2 TeV
when a vectorlike quark decays to a SM boson and a quark in the third generation~\cite{Aaboud:2018pii}.
Thus, the vector leptoquark explanation of the $\RD$ anomaly
is excluded by the $Z^\prime$ and vectorlike quark searches.
We note that it would be possible that the $\RD$ anomaly is explained in the scalar leptoquarks in PS models
as studied in e.g. Ref.~\cite{Heeck:2018ntp}.

\subsection{Flavor physics}

We shall discuss the flavor physics induced by the leptoquark and the extra Higgs bosons
when $m_X = 5$ TeV and the sizable leptoquark couplings are given
by Eqs.~\eqref{eq-gXRK} and~\eqref{eq-angleRK},
so that the $\RK$ anomaly is explained.
The formulas and values of constants used
in our numerical analysis are shown in Appendix~\ref{sec-analysis}.

In this paper, we study our predictions at the tree level in flavor physics, except for
$\mu \to e \gamma$. 
We do not include tree-level contributions of the $SU(4)_C$ adjoint $\Delta$
and $SU(2)_R$ triplet $\Sigma$.
As shown in Appendix~\ref{sec-intscal},
the adjoint $\Delta$ does not couple to two SM fermions up to $\order{v_H/v_\Delta}$.
Hence tree-level processes induced by those scalars are very suppressed.
The loop effects would be as large as those from the leptoquark
if the Yukawa couplings are $\order{1}$. The loop corrections involving both
scalars and leptoquark possibly enhance flavor violating processes, such as $K_L \to \mu e$,
even if all $s^{ij}_{F_X}$ are vanishing. This study is work in progress and will be shown near future.    
For $\Sigma$,
it is shown that the scalar leptoquarks from it
can explain both $\RK$ and $\RD$ anomalies~\cite{Heeck:2018ntp}
as well as inducing various flavor violations
if the corresponding Yukawa couplings have a certain structure.
The phenomenology of those scalar fields are interesting subject,
but this is beyond the scope of this paper.
We simply assume that $v_\Delta$ is so large that the corresponding Yukawa
couplings are negligible and the Yukawa coupling $h\propto \id{3}$,
and thus we expect that flavor violating effects from those scalar fields are negligibly small.

\subsubsection{$\mu\mathrm{-}e$ flavor violation versus fine-tuning}

As discussed in Sec.\,\ref{LQcoupling}, the flavor violating decay of $K$ meson,
$K_L \to \mu e$, gives the most stringent bound to the leptoquark mass
if it is coupled to the first generation fermions.
The branching fraction is approximately given by
\begin{align}
 \br{K_L}{\mu e} \sim&\
\frac{\tau_{K_L}}{128 \pi m_X^4}  \frac{m_{K}^5 f_K^2}{m_s^2}  \left(1-\frac{m_\mu^2}{m_K^2}\right)^2
                           \abs{\left[\hat{g}_{d_L}^X\right]_{22} \left[\hat{g}_{d_R}^X\right]_{11} }^2      \\ \notag
     \sim&\ 2.3\times 10^{-12}\times  \left(\frac{5~\mathrm{TeV}}{m_X}\right)^4
            \left( \frac{\abs{\left[\hat{g}_{d_L}^X\right]_{22} \left[\hat{g}_{d_R}^X\right]_{11} }}{10^{-5}}\right)^2.
\end{align}
Here we assume that $[\hat{g}^X_{d_L}]_{22}$ is much
larger than the other couplings except for those involving the third generation.
The experimental upper bound on the branching ratio is $4.7 \times 10^{-12}$~\cite{Zyla:2020zbs}.

The leptoquark coupling will be also strongly constrained
by the measurements of the $\mu$-$e$ conversion for an aluminum (gold) target
in the future (current) experiments.
The conversion per capture rate for an aluminum target is approximately given by
\begin{align}
 \br{\mu}{e}^{\mathrm{Al}} & \sim
\frac{16 m_\mu^5 S_p^2 m_p^2}{m_X^4 \Gamma_{\mathrm{capt}}}
\abs{\frac{f_{S_p}^s}{m_s} \left[\hat{g}_{d_R} \right]_{21}^* \left[\hat{g}_{d_L}\right]_{22}
 + \frac{2}{27} \frac{f_{G}^p}{m_b} \left[\hat{g}_{d_R} \right]_{31}^* \left[\hat{g}_{d_L}\right]_{32} }^2  \\ \notag
 \sim&\ 1.5\times 10^{-14} \times
\left(\frac{5~\mathrm{TeV}}{m_X}\right)^2
\abs{ 0.45 \left( \frac{ \left[ \hat{g}_{d_R} \right]_{21}^* \left[\hat{g}_{d_L}\right]_{22}}{10^{-5}} \right)
         + 0.016 \left( \frac{ \left[\hat{g}_{d_R} \right]_{31}^* \left[\hat{g}_{d_L}\right]_{32}}{10^{-5}} \right)}^2,
\end{align}
where the mass and form factors for the neutron are set to be those for the proton for simplicity
in the second line.
This rate is the same order of magnitude for the gold target.
The future (current) limit on the conversion rate per capture rate is~\cite{Bertl:2006up,Natori:2014yba,Kuno:2013mha,Abrams:2012er}
\begin{align}
 \br{\mu}{e}^{\mathrm{Al(Au)}} = \frac{\Gamma_{\mathrm{conv}}}{\Gamma_{\mathrm{capt}}}
       < 6\times 10^{-17}~\left(7\times 10^{-13}\right).
\end{align}
We note that even if the $K_L \to \mu e$ is sufficiently suppressed,
the $\mu$-$e$ conversion rate can be larger than the future sensitivity because of the contribution from the other coupling parameters.

There may be constraints from $\mu \to e \gamma$ induced
by the loop effects mediated by the leptoquarks and vectorlike down quarks.
The branching fraction is given by~\cite{Lavoura:2003xp},
\begin{align}
 \br{\mu}{e\gamma} = \tau_\mu \frac{\alpha_e m_\mu^3}{1024\pi^4 m_X^4}
           \left( \abs{C_L^{\mu e}}^2 + \abs{C_R^{\mu e}}^2 \right),
\end{align}
where $\tau_\mu$ denotes the muon lifetime and 
\begin{align}
 C_L^{\mu e} =  \sum_{A=1}^9   \left[
             m_\mu  \left[\hat{g}^X_{d_L} \right]_{A1}^* \left[\hat{g}^X_{d_L} \right]_{A2}
                        F \left(\frac{m_{d_A}^2}{m_X^2} \right)
                           +
          m_{d_A} \left[\hat{g}^X_{d_L} \right]_{A1}^* \left[\hat{g}^X_{d_R} \right]_{A2}
                               G \left( \frac{m_{d_A}^2}{m_X^2} \right)
                          \right].
\end{align}
$C_R^{\mu e}$ is obtained by formally replacing $L \leftrightarrow R$.
The loop functions are defined as
\begin{align}
F(t) =&\ -\frac{4-16t+39t^2-28t^3+t^4 + 6t^2(1+2t)\log t}{4(t-1)^4},  \\
G(t) =&\ \frac{-4+27t-24t^2 +t^3 + 6t(1+2t) \log t}{2(t-1)^3}.
\end{align}
The second term will be the dominant due to the chiral enhancement
by the Yukawa coupling with Higgs bosons
if $\delta_d \le \order{v_H}$ in Eq.~\eqref{eq-parametrization} is larger than the muon mass.
Since the muon couples to one vectorlike down quark with $\order{1}$ mixing,
the branching fraction is estimated as
\begin{align}
 \br{\mu}{e\gamma} \sim 1.2\times 10^{-13} \times
\left(\frac{5.0~\TeV}{m_X} \right)^4 \left( \frac{g^X_L g^X_R}{0.005}   \right)^2
           \left( \frac{\delta_d}{100~\MeV}\right)^2,
\end{align}
where $g^X_{L(R)}$ is a typical size of leptoquark couplings to mu/electron in the left (right)-current.
Thus the Higgs boson couplings with the vectorlike down-type fermions should be so suppressed
that the current limit $\br{\mu}{e\gamma} < 4.2\times 10^{-13}$~\cite{TheMEG:2016wtm}
is satisfied.
This process can probe different parameter space
from the $K_L \to \mu e$ decay and $\mu$-$e$ conversion which are induced at the tree-level,
since it directly constrains the leptoquark couplings to the SM charged leptons and vectorlike quarks.
We will see that there are parameter spaces that can be probed only by $\mu\to e\gamma$
even if $\delta_d = 10^{-4} v_H$ and the chiral enhancement effect is suppressed.

With the severe constraint from $\mu\to e\gamma$,
the muon anomalous magnetic moment $\Delta a_\mu \sim \order{10^{-9}}$
is difficult to be explained by the 1-loop effects of the leptoquark and vectorlike quark.
It is roughly estimated as
\begin{align}
\Delta a_\mu \lesssim  \frac{m_\mu \delta_d}{16\pi^2 m_X^2} N_c N_{\mathrm{VL}}
  \sim 5 \times 10^{-11} \times \left(\frac{5~\TeV}{m_X}\right)^2 \left(\frac{\delta_d}{100~\MeV}\right),
\end{align}
where $N_c = 3$ and $N_{\mathrm{VL}} = 6$.
Thus it is, at least, two orders of magnitude smaller than the currently preferred value
as far as $\delta_d \lesssim m_\mu$ to suppress $\mu\to e\gamma$.

In our model,
the stringent constraints on the $\mu$-$e$ flavor violation
can be avoided due to the texture of Eq.~\eqref{eq-texture}
which is achieved by tuning the vectorlike mass terms
such that there are cancellations between
the mass parameters and the adjoint VEV $\vev{\Delta}$, see Eq.~\eqref {eq-MVL}.
To quantify how accurately this cancellation should be hold,
we turn on the angles universally except for those relevant to the $\RK$ anomaly,
\begin{align}
  s_{Q_L} := s^{\hat{i}\hat{j}}_{q_L}, \quad
  s_{Q_R} := s^{ij}_{q_R},
\quad
 s_{\ell_R} := s^{\hat{i}\hat{j}}_{\ell_L}, \quad
 s_{\ell_R} := s^{ij}_{\ell_R}.
\end{align}
where $i,j = 1,2,3$ run over all the combinations
and $\hat{i},\hat{j} = 1,2,3$ also run over all the combinations except $(\hat{i},\hat{j}) = (2,2), (2,3)$,
to keep $\Delta C_9$ unchanged approximately.
In our numerical analysis,
we set $\delta_u = \delta_d  =10^{-4}$, 
so that the chiral enhancement effect to $\mu \to e\gamma$ is negligible and obtain conservative limits.

\begin{figure}[t]
\centering
\includegraphics[height=100mm]{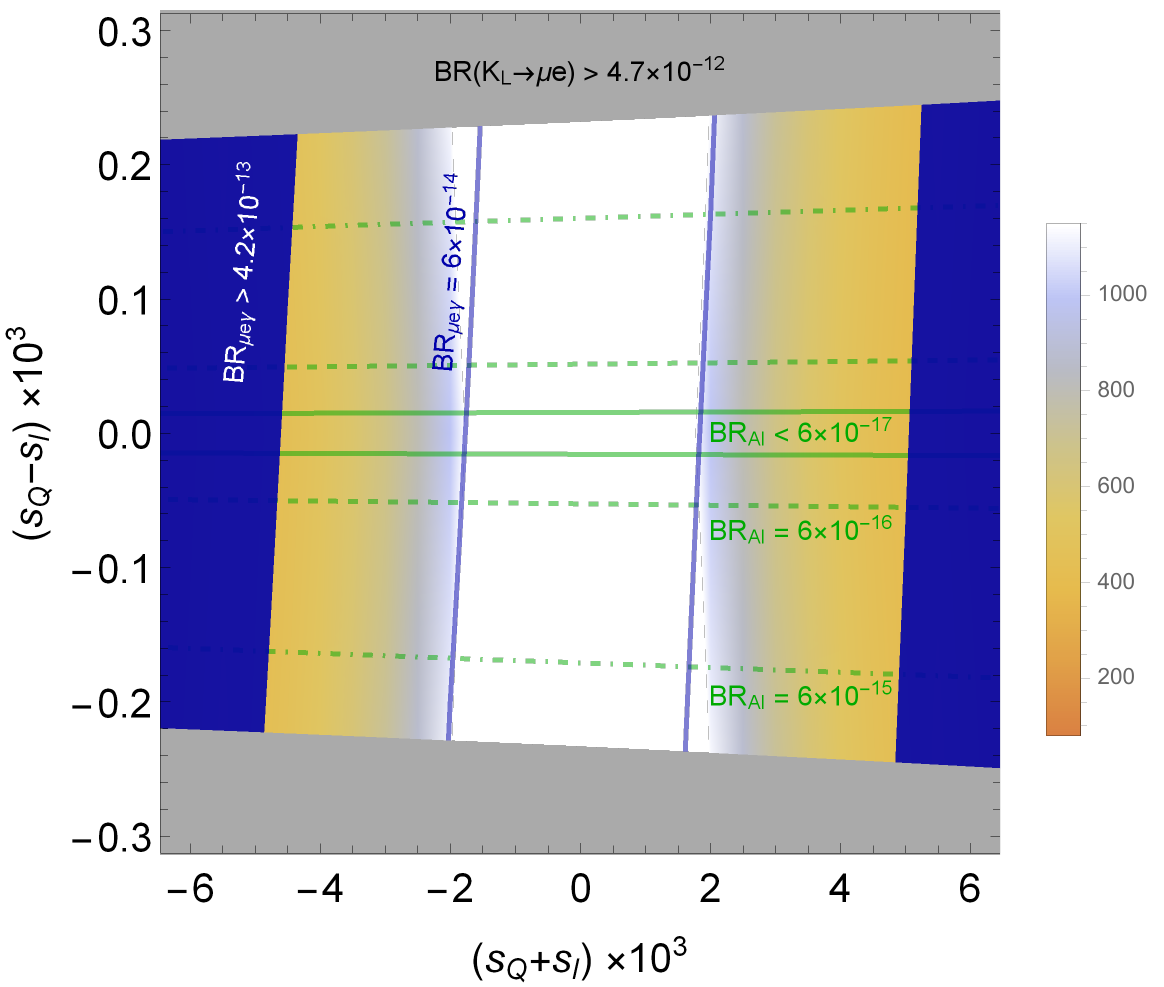}
\caption{\label{fig-tuning}
The degree of tuning $\Delta_{\mathrm{FT}}^{-1}$,
constraints and values of the $\mu$-$e$ violation observables.
See the text for details.
}
\end{figure}
Figure~\ref{fig-tuning} shows the degree of tuning $\Delta_{\mathrm{FT}}$ and
constraints from the $\mu$-$e$ violations on the $(s_Q + s_\ell)$ vs $(s_Q-s_\ell)$ plane,
where
\begin{align}
 s_Q := s_{Q_L} = s_{Q_R}
\quad
 s_\ell := s_{\ell_L} = s_{\ell_R}.
\end{align}
The density plot shows $\Delta_{\mathrm{FT}}^{-1}$.
The gray (blue) region is excluded
by the current limit on $\br{K_L}{\mu e}$ ($\br{\mu}{e\gamma}$). 
The blue lines are the future sensitivity of $\br{\mu}{e\gamma}=6\times 10^{-14}$
at the MEG experiment\cite{Baldini:2018nnn}.
The green lines show $\br{\mu}{e}^{\mathrm{Al}}$
and it is below the future sensitivity of $\br{\mu}{e}^{\mathrm{Al}} = 6\times 10^{-17}$\cite{Kuno:2013mha}
between the solid green lines.
There is no region excluded by the current limit on $\br{\mu}{e}^{\mathrm{Au}}$~\cite{Bertl:2006up}.
Since the leptoquark couplings to the SM fermions are induced by $\abs{s_Q - s_\ell}$,
the tree-level $\mu$-$e$ violations are enhanced in the upper and lower regions.
On the other hand, $\br{\mu}{e\gamma}$ is enhanced as $\abs{s_Q+s_\ell}$ increases
since it comes from the coupling with the vectorlike quarks.
The limit from $\br{\mu}{e\gamma}$ will be tightened
if $\delta_d$ is larger.
Note that the $\RK$ anomaly is explained in the whole region in this plot.

The degree of tuning $\Delta_{\mathrm{FT}}^{-1} \gtrsim 500$ ($1000$)
at $\abs{s_Q+s_\ell} \gtrsim 0.005$ $(0.002)$.
One may concern about the tuning between $s_Q$ and $s_\ell$,
but the cancellation is mild since $(s_Q - s_\ell)/(s_Q + s_\ell) \gtrsim \order{0.01}$
even if $\br{\mu}{e}^{\mathrm{Al}} < 6\times 10^{-17}$.
Therefore, we conclude that
the degree of tuning to explain the $\RK$ anomaly consistently with the current limit
is $\Delta_{\mathrm{FT}}^{-1} \sim 500$ which corresponds to $\order{0.2\%}$ tuning.
Most of the parameter space, outside of the two green solid lines, 
will be tested by the future experiments of the $\mu$-$e$ conversion.

\begin{figure}[t]
\centering
\includegraphics[height=100mm]{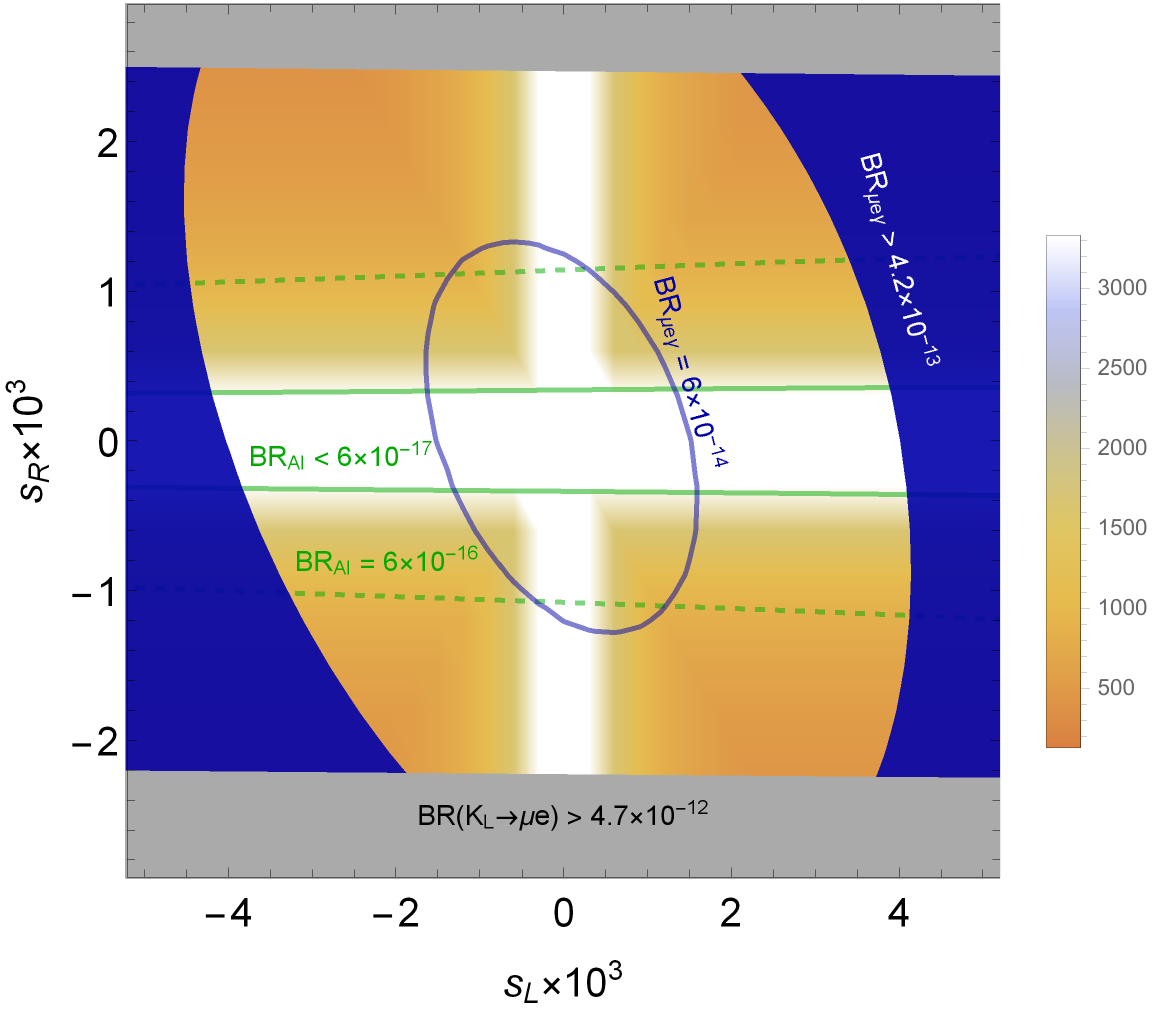}
\caption{\label{fig-tuningLR}
The same plot as Fig.~\ref{fig-tuning} on $(s_L, s_R)$ plane.
}
\end{figure}
Figure~\ref{fig-tuningLR} shows the same plot as Fig.~\ref{fig-tuning},
but on $(s_L, s_R)$ plane, where
\begin{align}
\label{eq-sLsR}
s_L:=  s_{Q_L} = -  s_{\ell_L}/1.1,
\quad
s_R :=  s_{Q_R} = - s_{\ell_R}/1.1.
\end{align}
We keep $s_{Q_L} \sim s_{\ell_L}$ and $s_{Q_R} \sim s_{\ell_R}$
to avoid the large $\mu$-$e$ violation.
The tree-level $\mu$-$e$ flavor violations are sensitive to $s_R$,
since $[\hat{g}^X_{d_L}]_{s\mu}$ is sizable to explain the $\RK$ anomaly.
The future measurement of the $\mu\to e$ conversion will cover $\Delta_{\mathrm{FT}}^{-1} \lesssim 3000$
in the case of Eq.~\eqref{eq-sLsR}.
The radiative decay, $\mu\to e\gamma$ is equally sensitive to $s_L$ and $s_R$,
and the region surrounded by the blue line will be covered by the MEG experiment.

\subsubsection{Flavor violation via Higgs bosons}

This model predicts sizable tree-level flavor violating couplings involving Higgs bosons as well.
Even in the canonical limit, the heavy Higgs boson couplings to the down-type SM quarks are given by
\begin{align}
\label{eq-hFLVdom}
 \left[\hat{Y}_d^H\right]_{ij} \sim&\ -i \left[\hat{Y}^A_d\right]_{ij}
\sim \frac{1}{\sqrt{2}} \left[\hat{Y}^{H^+}_d \right]_{ij}
\\ \notag
\sim&\
-\frac{1}{\sqrt{2}v_H} \left(
\frac{2\tan\beta}{1-\tan^2\beta} \left[D_d \right]_{ij}
+
\frac{1+\tan^2\beta}{1-\tan^2\beta} \left[ V_{\mathrm{CKM}}^\dag D_u  \right]_{ij}
  \right),
\end{align}
where $i,j = 1,2,3$.
Here we take the decoupling limit of the Higgs bosons~\footnote{
See Appendix~\ref{sec-intscal} for more details of the Higgs couplings.
}. In this limit, all the scalar masses of the heavy Higgs doublet are degenerate.
The second term in the parenthesis inevitably induces flavor violations whose typical values are estimated as
\begin{align}
v_H^{-1} \abs{V^\dag_{\mathrm{CKM}}  {D_u}}
\sim
\begin{pmatrix}
5\times 10^{-5} & 0.0006 & 0.007 \\  1\times 10^{-6} & 0.003 & 0.03 \\ 2\times 10^{-8} & 0.0001 & 0.78
\end{pmatrix}.
\end{align}
As we see, flavor violating couplings, that are off-diagonal elements,
are sizable and hence flavor violating processes will be large if the Higgs bosons are light,
although the chirality structure may suppress them because
only the upper-right elements are sizable.

In the charged lepton sector, similarly, there are flavor violations, that are estimated as $U_{\mathrm{PMNS}} D_n/v_H$. The effects are, however, expected to be negligible
because the Yukawa coupling to the neutrinos is estimated as $D_n/v_H \sim \order{10^{-5}}$
to explain the neutrino masses with the $\order{10~\mathrm{TeV}}$ Majorana masses.
Therefore the neutral meson mixing is the most sensitive process to the flavor violation via the Higgs bosons.
The effective interactions relevant to our model are given by~\footnote{
We use the basis of the operators used in e.g. Ref.~\cite{Buras:2001ra}.
}
\begin{align}
 \Hcal_{\mathrm{eff}}^{\Delta F=2} = \sum_{I,A} C_I^A Q_I^A,
\end{align}
where $(I,A) = (1,\mathrm{VLL}), (2,\mathrm{LR}), (1,\mathrm{SLL}), (1,\mathrm{SRR})$.
The four-Fermi operators are defined as
\begin{align}
 Q_{1}^{\mathrm{VLL}} =&\ \left(\ol{F}^a \gamma_\mu P_L f^a \right)
                                      \left(\ol{F}^b \gamma^\mu P_L f^b  \right),
\quad &
 Q_{2}^{\mathrm{LR}} =&\ \left(\ol{F}^a  P_L f^a  \right) \left(\ol{F}^b P_R f^b  \right), \\
 Q_{1}^{\mathrm{SLL}} =&\ \left(\ol{F}^a  P_L f^a  \right) \left(\ol{F}^b P_L f^b  \right),
\quad &
 Q_{1}^{\mathrm{SRR}} =&\ \left(\ol{F}^a  P_R f^a  \right) \left(\ol{F}^b P_R f^b \right), 
\end{align}
where $a, b=1,2,3$ are the color indices.
Here, $(F,f) = (s,d), (b,d), (b,s)$ for $K$-$\ol{K}$,
$B_d$-$\ol{B}_d$ and $B_s$-$\ol{B}_s$ mixing, respectively.
We define the ratios of off-diagonal matrix element of our model to that of the SM as
\begin{align}
 C_M :=&\ \frac{\langle{M}| \Hcal_{\mathrm{eff}}^{\Delta F=2} |\ol{M}\rangle}
                    {\langle{M}| \Hcal^{\mathrm{SM}}_{\mathrm{eff}} |\ol{M}\rangle }
          = \frac{C^{M}_{\mathrm{SM}} \Ocal^{VLL}_1
 + C_{2}^{LR} \Ocal^{LR}_2 +  \left(C_1^{SLL}+C_1^{SRR} \right)\Ocal^{SLL}_1
  }
     {C^{M}_{\mathrm{SM}} \Ocal^{VLL}_1}.
\end{align}
Here, $\Ocal^A_I := \bra{M} Q^A_I \ket{\ol{M}}/(2 m_M)$ with $m_M$ is the meson mass. 
We use the values of $\Ocal^A_I$ shown in Ref.~\cite{Kawamura:2019hxp}.  
The SM contribution $C^M_{\mathrm{SM}}$ is shown in e.g. Ref.~\cite{Buras:1998raa}.
In our model, the Wilson coefficients are given by
\begin{align}
C_2^{LR}=&\   -
     \sum_{S=h,H,A} \frac{y_L^S y_R^S }{2m_{S}^2},
\quad
C_1^{SLL}=
     - \sum_{S=h,H,A} \frac{y_L^S y_L^S }{2m_{S}^2},
\quad
C_1^{SRR}=
     - \sum_{S=h,H,A} \frac{y_R^S y_R^S }{2m_{S}^2},
\end{align}
where
\begin{align}
 y^S_L =  \left[\hat{Y}^{S}_{d} \right]_{ji}^*, \quad
 y^S_R =  \left[\hat{Y}^{S}_{d} \right]_{ij}, 
\end{align}
with $(i,j) = (2,1), (3,1), (3,2)$ for $M = K, B_d, B_s$, respectively.
The flavor violation from the adjoint field is negligible as discussed in Appendix~\ref{sec-intscal}.
We neglect loop corrections from the leptoquarks in our analysis.
The current constraints at 95\% C.L. given by the UT-Fit~\cite{Bona:2005vz,Bona:2007vi} are
\begin{align}
0.87 \le \mathrm{Im}\; C_{K} \le 1.39
,\quad
0.83 \le \abs{C_{B_d}} \le 1.29,  \quad
 0.942 \le \abs{C_{B_s}} \le 1.288.
\end{align}
As pointed out in Ref.~\cite{DiLuzio:2019jyq}, 
the uncertainties are reduced 
in the ratio of mass differences of the $B_d$ and $B_s$ meson mixing, 
\begin{align}
 \frac{\Delta M_d}{\Delta M_s} = 0.0298^{+0.0005}_{-0.0009}.   
\end{align}  
We define the parameter, 
\begin{align}
 C_{B_d/B_s} := \frac{1}{0.0298} \times 
                      \abs{\frac{\bra{B_d} \mathcal{H}_\mathrm{eff}^{\Delta F=2} \ket{\ol{B}_d}}
                      {\bra{B_s} \mathcal{H}_\mathrm{eff}^{\Delta F=2} \ket{\ol{B}_s}} }. 
\end{align}

\begin{figure}[t]
\centering
\begin{minipage}[c]{0.48\hsize}
\centering
\includegraphics[height=70mm]{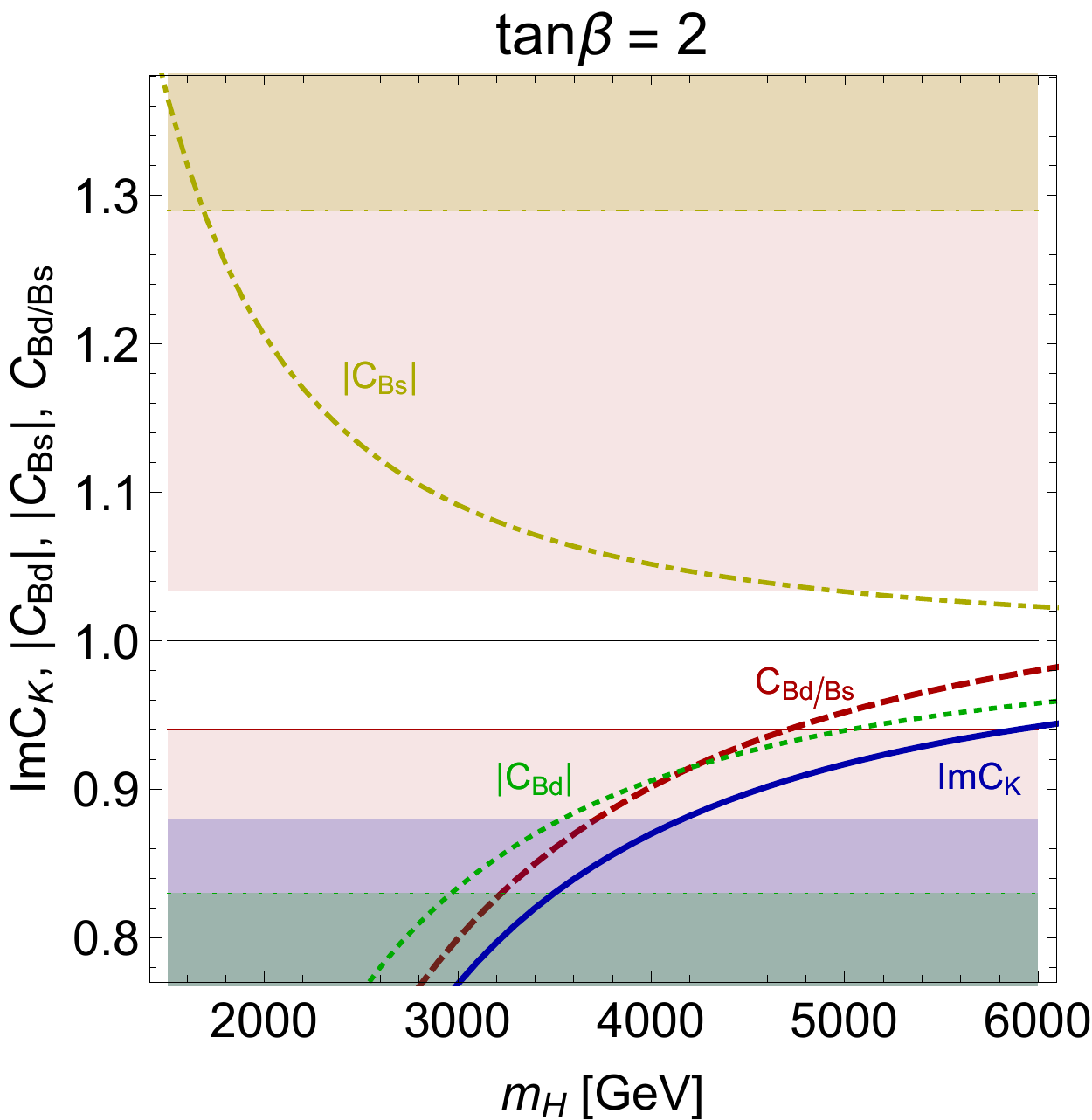}
\end{minipage}
\begin{minipage}[c]{0.48\hsize}
\centering
\includegraphics[height=70mm]{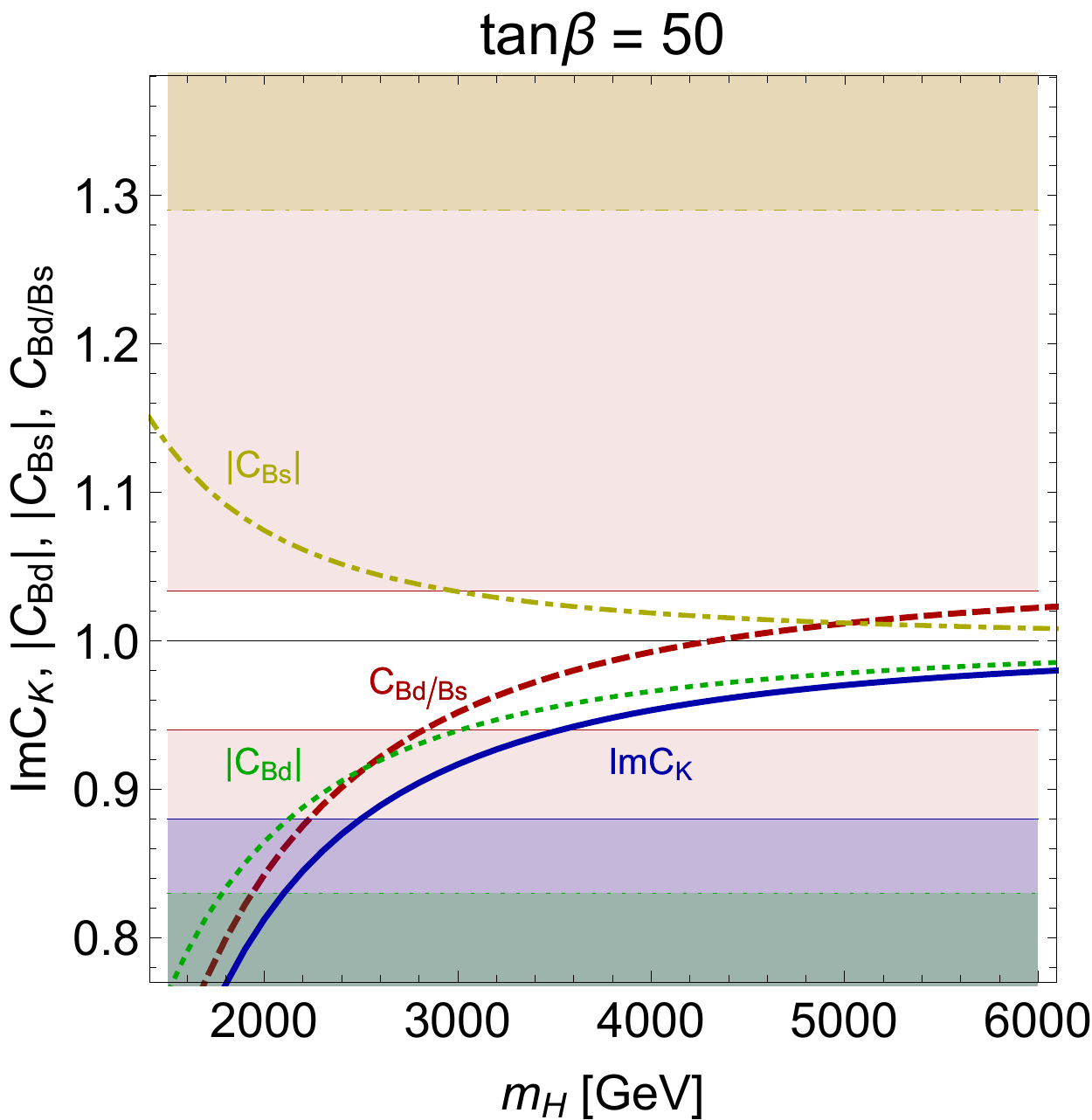}
\end{minipage}
\caption{\label{fig-HiggsFCNC}
The limits on the heavy Higgs boson mass $m_H$ from the observables concerned with the neutral meson mixings:
$|C_{B_d}|$ (green), $|C_{B_s}|$ (yellow), $C_{B_d/B_s}$ (red) and $\mathrm{Im}\; C_{K}$ (blue).
The excluded region for each observable is colored in the same color as the corresponding line.
$\tan \beta$ is fixed at
$\tan\beta = 2.0$ ($50$) in the left (right) panel.}
\end{figure}
Figure~\ref{fig-HiggsFCNC} shows experimental limits from the neutral meson mixing
on the heavy Higgs boson mass for $\tan\beta = 2.0$ $(50)$ on the left (right) panel.
Our predictions for $|C_{B_d}|$, $|C_{B_s}|$, $C_{B_d/B_s}$ and $\mathrm{Im}\; C_{K}$ are depicted by the green, yellow, red, and blue lines, respectively.
The excluded region for each observable is colored in the same color as the corresponding line.
In the red regions, the deviation of $C_{B_d/B_s}$ from the central value is   
more than twice as large as the uncertainties.
In this figure, our parameters satisfy $s_Q + s_\ell = 0.0015$ and $s_Q - s_\ell = 0.6\times 10^{-5}$
that can not be probed by looking for the $\mu$-e flavor violations.
The limits for the cases with larger angles $s_Q$, $s_\ell$ are quite similar to this result
because the dominant effect comes from Eq.~\eqref{eq-hFLVdom} which is independent of the angles,
although there are mild dependence on them.
We see that $C_{B_d/B_s}$ gives the most stringent limits for both $\tan\beta = 2$ and $50$. 
Interestingly, the bound from $C_{B_d/B_s}$ is stronger than the others, since 
the uncertainty is small and our predictions of
$\abs{C_{B_s}}$ and $\abs{C_{B_d}}$ move in the opposite directions. 
The limits from the phases of $C_{B_{d,s}}$ are weaker than those from the absolute values.
The lower bound on the heavy Higgs boson mass is about 4.8 (2.8) TeV for $\tan\beta = 2.0$ $(50)$.
The limit is stronger for smaller $\tan\beta$ since the up-type Yukawa coupling constants are enhanced,
see Eq.~\eqref{eq-hFLVdom}.

\subsection{$n-\ol{n}$ oscillation}

Before closing this section, let us discuss neutron anti-neutron ($n$-$\ol{n}$) oscillation.
In general, gauge unified models predict baryon number violating processes,
such as proton decays and $n$-$\ol{n}$ oscillation,
that provide a useful tool to test the unification models.
In the PS unification,
the gauge bosons from the PS symmetry breaking do not mediate the baryon number violating processes,
since their interactions respect the $B-L$ and $B+L$ symmetries.
However, the scalars responsible for the PS symmetry breaking can generate the baryon number violation,
depending on the representations.
Since the PS breaking scale is relatively low in our model,
the baryon number violation induced by such scalars may provide a stringent constraint.

In our model,
three (non-singlet) scalar fields, namely
$\Phi: (\rep{1},\rep{2},\rep{2})$, $\Delta:(\rep{15},\rep{1},\rep{1})$
and $\Sigma:(\rep{\ol{10}},\rep{1},\rep{3})$,
are introduced.
With these representations,
the stability of proton is ensured even after the PS symmetry breaking
because of the discrete symmetry under the following transformation~\cite{Mohapatra:1980qe},
\begin{align}
\Sigma_{ab} & \to e^{-2i\pi/3} \Sigma_{ab},&\quad
\Sigma_{a1} &\to e^{-i\pi/3} \Sigma_{a1},& \quad
\Sigma_{11} \to \Sigma_{11}, \\
{\bf u}_{L,R} &\to e^{i\pi/3} {\bf u}_{L,R} ,\quad&
{\bf d}_{L,R} &\to e^{i\pi/3} {\bf d}_{L,R},&
\end{align}
where 
$a,b=2,3,4$ are the color indices.
Since this symmetry does not protect processes that change the baryon number by an even integer,
the $n$-$\ol{n}$ oscillation can occur in our model.
In what follows,
we shall estimate the transition amplitude and discuss the compatibility with the TeV-scale leptoquark scenario.

Of three scalars in our model,
only the symmetric representation scalar $\Sigma$ breaks the baryon number symmetry
when it develops the VEV, and hence contributes to the $n-\ol{n}$ transition.
The transition requires three $ \Sigma \ol{f}_R^{c}\; h\; f_R$ vertices,
the VEV of $\Sigma$ and the following quartic coupling of $\Sigma$,
\begin{equation}
V_{n-\ol{n}} = \lambda_\Sigma
\left[ \mathrm{Tr} \left(\Sigma_{\alpha \zeta} \Sigma_{\beta\eta} \right) 
\mathrm{Tr} \left( \Sigma_{\gamma\kappa} \Sigma_{\delta\lambda} \right)
\epsilon^{\alpha\beta\gamma\delta} \epsilon^{\zeta\eta\kappa\lambda} + h.c. \right] ,
\end{equation}
where $\alpha,\beta,\dots,\lambda=1,2,3,4$ are the $SU(4)_C$ indices and
the trace is for the $SU(2)_R$ indices.
After the PS symmetry breaking,
the interactions relevant to the $n-\ol{n}$ oscillation are given by
\begin{equation}
V_{n-\ol{n}} \supset  2 \la_\Sigma v_\Sigma \, \epsilon^{abc} \epsilon^{def}
\Sigma^{2/3}_{ad} \left( \Sigma^{-1/3}_{be} \Sigma^{-1/3}_{cf} + 2 \Sigma^{2/3}_{be} \Sigma^{-4/3}_{cf} \right)
 + h.c.,
\end{equation}
where $a,b,\dots,f=1,2,3$ are the $SU(3)_c$ indices and
we denote the scalars of symmetric representation under $SU(3)_c$ as
$\Sigma^{Y} : (\rep{\ol{6}}, \rep{1}, Y)$, with $Y=-4/3, -1/3, 2/3$, under $G_\SM$.
Their masses are denoted by $m_{\Sigma^{Y}}$.
Integrating out these heavy scalars induces six-quark operators,
\begin{align}
{\cal L}_{\rm eff}^{n-\ol{n}}  = - \la_\Sigma v_\Sigma \, \epsilon^{abc} \epsilon^{def}
& \left[  \frac{h_{dd}^2 h_{uu}}{ m_{\Sigma^{-4/3}}^4 m_{\Sigma^{2/3}}^2 }
     (\ol{d}_{R,a}^c d_{R,d}) (\ol{d}_{R,b}^c d_{R,e}) (\ol{u}_{R,c}^c u_{R,f}) \right. \notag \\
& \quad \left. - \, 2 \frac{h_{dd} h_{ud}^2}{ m_{\Sigma^{-4/3}}^2 m_{\Sigma^{-1/3}}^4 }
     (\ol{d}_{R,a}^c d_{R,d}) (\ol{d}_{R,b}^c u_{R,e}) (\ol{u}_{R,c}^c d_{R,f}) \right],
\end{align}
where $h_{qq^\prime}$ ($q,q^\prime = u,d$)
is the Yukawa coupling of $\Sigma$ to $q$ and $q^\prime$ quarks in the mass basis,
given by
\begin{align}
\label{eq-hqq}
h_{q q^\prime} & = \sum_{i,j=1,2,3} \frac{1}{\sqrt{2}} \left[h\right]_{ij}
   \left[U^{q}_R\right]_{3+i,1}  \left[U^{q^\prime}_R\right]_{3+j,1}.  
\end{align}
Assuming, for simplicity,
\begin{equation}
m_\Sigma : = m_{\Sigma^{-4/3}} = m_{\Sigma^{2/3}} = m_{\Sigma^{-1/3}} , \quad
h_Q := h_{uu} = h_{dd} = h_{ud}.
\end{equation}
The Yukawa coupling of the SM quarks to $\Sigma$ is estimated to be $h_Q \sim s_Q^2$
when the Yukawa coupling in the gauge basis $h \sim \order{1}$,
and hence it is suppressed by the mixing between the chiral and vectorlike quarks
which is at most $0.005$ as read from Fig.~\ref{fig-tuning}.
With the naive dimensional analysis~\cite{Phillips:2014fgb}, we find the transition amplitude to be
\begin{align}
\tau_{n-\ol{n}} &\sim \frac{m_\Sigma^6}{\la_\Sigma v_\Sigma h_Q^3} \frac{1}{\Lambda_{\rm QCD}^6} \\
\notag
    &\sim 1.2 \times 10^{14} \, {\rm sec}
\times
\left(\frac{m_\Sigma}{10\,{\rm TeV}}\right)^6
\left(\frac{10\,{\rm TeV}}{v_\Sigma}\right)
\left(\frac{180\,{\rm MeV}}{\Lambda_{\rm QCD}}\right)^6
\left(\frac{0.005}{s_Q}\right)^6
\left(\frac{1.0}{h}\right)^3
\left(\frac{0.1}{\lambda_\Sigma}\right) .
\end{align}
Here, $\Lambda_{\rm QCD}$ denotes the QCD scale.
It follows that the $n-\ol{n}$ transition is very sensitive to the $\Sigma$ mass.
The current limit is
\begin{equation}
\tau_{n-\ol{n}} \geq 
\begin{cases}
2.7 \times 10^8\,{\rm sec} & \mbox{bound neutron~\cite{Abe:2011ky}} \\
8.6 \times 10^7\,{\rm sec} & \mbox{reactor free neutron~\cite{Phillips:2014fgb}}
\end{cases}
\end{equation}
Therefore, the constraint from the $n-\ol{n}$ oscillation can be avoided
even if the relevant couplings, $h$ and $\la_\Sigma$, are of order unity.

\section{Summary}
\label{sec-summary}

In this paper,
we have proposed an explicit model with the PS gauge symmetry and extra vectorlike fermions
that realizes
\begin{itemize}
\item {a vector leptoquark which accounts for the $\RK$ anomaly}
\item {the realistic mass spectrum of the SM fermions}
\item {sufficiently suppressed $\mu$-$e$ flavor violations due to tuning of the parameters.}
\end{itemize}
The texture in Eq.~\eqref{eq-texture} of the vectorlike fermion masses
is a key idea to suppress the $\mu$-$e$ flavor violation, especially $K_L \to \mu e$ decay.
We have showed in Fig.~\ref{fig-tuning} that $\br{K_L}{e\mu}$ is less than the current limit
if we adopt the $\order{0.2\%}$ tuning of the parmeters.
With this texture, the TeV-scale vector leptoquark is allowed and thus the $\RK$ anomaly can be explained.
We have, on the other hand, pointed out that
the combination of the $Z^\prime$ boson and vectorlike quark searches exclude the light leptoquark to explain the $\RD$ anomaly.

The idea to relax the bound from $K_L\to \mu e$ by introducing vectorlike families 
was proposed in Ref.~\cite{Calibbi:2017qbu}. 
In this paper, 
we have shown an explicit texture of the mass matrices to suppress the $K_L \to\mu e$ 
together with the SM fermion mass and mixing matrices. 
By this explicit construction, 
we found that the $\RD$ anomaly is difficult to be explained, 
the $\order{0.1}$ \% tuning is requried to satisfy the phenomenological conditions 
for the $\order{1}$~TeV leptoquark, 
and the flavor violating couplings of the heavy Higgs boson are unavoidable.

On top of these, 
we conclude that the available parameter space to explain the $\RK$ anomaly is probed by future searches
for $\mu \to e\gamma$ and $\mu \mathrm{Al} \to e \mathrm{Al}$ processes.
The former process is sensitive to the couplings with the SM leptons and vectorlike quarks,
while the later is sensitive to the SM leptons and SM down quarks.
The future experiments cover the parameter space that satisfies $\abs{s_Q+s_\ell} \gtrsim 0.002$ and $\abs{s_Q - s_\ell} \gtrsim 3.0\times 10^{-5}$.
We have also found the fact that the flavor violating copings of the extra Higgs bosons are inevitable.
It is shown in Fig.~\ref{fig-HiggsFCNC} that
the heavy Higgs bosons lighter than $4.8$ ($2.8$) TeV are excluded by the measurement
of $B$ meson mixing when $\tan\beta = 2$ ($50$). 
In our model, the neutron oscillation is also predicted by the couplings involving $\Sigma$.
We estimate the transition amplitude and conclude that our prediction is much below the current experimental bound.

Before closing our discussion, let us comment on loop corrections to the flavor violating processes, 
e.g. $K_L\to \mu e$. In our setup, the tree-level contributions via $\Delta$
and leptoquark exchanging are very suppressed because of the unique structure in the fermion mass matrices. 
This setup leads almost flavor-diagonal couplings, and we concentrate on
the tree-level predictions induced by leptoquark exchanging. 
However, the leptoquark and scalar couplings between quarks/leptons and heavy fermions
could be sizable. The one-loop corrections involving heavy fermions
and scalars may be large in the TeV-scale scenario.
The study on the loop effect is not so simple because of many parameters. 
We need more careful study, taking into account the mass spectrum of all fermions as well.
This study is work in progress and will be shown near future.

\section*{Acknowledgment}
S. I. would like to thank the warm hospitality at KEK where he stayed during the work.
The work of S. I. is supported by the Japan Society for the Promotion of Science (JSPS) Research Fellowships for Young Scientists, No. 19J10980 and the JSPS Core-to-Core Program, No.JPJSCCA20200002.
The work of J.K. 
is supported in part by 
the Institute for Basic Science (IBS-R018-D1), 
the Department of Energy (DOE) under Award No.\ DE-SC0011726,
and the Grant-in-Aid for Scientific Research from the
Ministry of Education, Science, Sports and Culture (MEXT), Japan No.\ 18K13534. 
S. O. is supported in part by NSERC, Canada.
The work of Y. O. is supported by Grant-in-Aid for Scientific research from the MEXT, Japan, No. 19H04614, No. 19H05101, and No. 19K03867.

\appendix
\section{Model details}
\label{sec-detail}

\subsection{Diagonalization of the fermion mass matrices}
\label{sec-diago}

\subsubsection{Dirac mass matrix}
Let us discuss the fermion mass matrices parametrized as in Eq.~\eqref{eq-parametrization}.
The unitary matrices
\begin{align}
 U^0_{\ell_L} =
\begin{pmatrix}
 0_{6\times 3} & V_{\ell_L} \\ W_{\ell_L} & 0_{3\times 6}
\end{pmatrix},
\quad
 U^0_{\ell_R} =
\begin{pmatrix}
V_{\ell_R} & 0_{6\times 3}  \\ 0_{3\times 6} & W_{\ell_R}
\end{pmatrix},
\quad
 U^0_{Q_L} =
\begin{pmatrix}
V_{Q_L} & 0_{6\times 3} \\  0_{3\times6} & W_{Q_L}
\end{pmatrix},
\quad
 U^0_{Q_R} =
\begin{pmatrix}
0_{6\times 3} &  V_{Q_R} \\ W_{Q_R}  & 0_{3\times 6}
\end{pmatrix},
\end{align}
approximately diagonalize the mass matrices as
\begin{align}
\label{eq-appdiag}
\left(U_{\ell_L}^0\right)^\dag \Mcal_e U_{\ell_R}^0 =&\
 \begin{pmatrix}
  D_{\ell_R} & \zd{3} &   \cdot \\
  \cdot & d_e & \zd{3} \\
  \cdot & \cdot        & D_{\ell_L}
 \end{pmatrix},
& \quad &
\left(U_{Q_L}^0\right)^\dag \Mcal_d U_{Q_R}^0 =
 \begin{pmatrix}
  D_{Q_L} & \cdot & \cdot    \\
 \zd{3}              & d_d  & \cdot \\
\cdot &  \zd{3}    & D_{Q_R}\\
 \end{pmatrix}, \\
\left(U_{\ell_L}^0\right)^\dag \Mcal_n U_{\ell_R}^0 = &\
 \begin{pmatrix}
  D_{\ell_R} & \zd{3} &   \cdot \\
 \cdot  & m_n & \zd{3} \\
 \cdot  & \cdot       & D_{\ell_L}
 \end{pmatrix},
 & \quad &
\left(U_{Q_L}^0\right)^\dag \Mcal_u U_{Q_R}^0 =
 \begin{pmatrix}
  D_{Q_L} & \cdot & \cdot   \\
 \zd{3}              & m_u  & \cdot \\
 \cdot &  \zd{3}   & D_{Q_R}\\
 \end{pmatrix},
\end{align}
where $\cdot$ represents $\order{v_H}$ entries.
Here, $d_e$ and $d_d$ are chosen to be diagonal
by using the redundancies of Eq.\eqref{eq-redundancy} in $V_{\ell_{L,R}}$ and $V_{Q_{L,R}}$.
To make the singular values in increasing order
and the SM up-type quark and neutrino masses, $M_u$ and $M_n$, diagonalized,
we introduce
\begin{align}
\label{eq-Utilde}
 \tU_{e_{L,R}} := &\ U_{\ell_{L,R}}^0
\begin{pmatrix}
\zd{3}&\id{3} &\zd{3}\\
\id{3} & \zd{3} &\zd{3}\\
\zd{3}& \zd{3} &\id{3}
\end{pmatrix},
&\quad
 \tU_{n_{L,R}} :=&\  U_{\ell_{L,R}}^0
\begin{pmatrix}
\zd{3}&\id{3} &\zd{3}\\
u_{n_{L,R}}  & \zd{3} &\zd{3}\\
\zd{3}& \zd{3} &\id{3}
\end{pmatrix},
\\
 \tU_{d_{L,R}} := &\ U_{Q_{L,R}}^0
\begin{pmatrix}
\zd{3}&\id{3} &\zd{3}\\
\id{3} & \zd{3} &\zd{3}\\
\zd{3}& \zd{3} &\id{3}
\end{pmatrix},
&\quad
 \tU_{u_{L,R}} :=&\  U_{Q_{L,R}}^0
\begin{pmatrix}
\zd{3}&\id{3} &\zd{3}\\
u_{u_{L,R}} & \zd{3} &\zd{3}\\
\zd{3}& \zd{3} &\id{3}
\end{pmatrix},
\end{align}
where the unitary matrices $u_{u_{L,R}}$ and $u_{n_{L,R}}$ diagonalize $m_u$ and $m_n$ respectively,
\begin{align}
 u_{n_L}^\dag m_n u_{n_R} = d_n, 
\quad
 u_{u_L}^\dag m_u u_{u_R} = d_u. 
\end{align}
The Dirac mass matrices are diagonalized as
\begin{align}
 \left(\tU_{e_L}\right)^\dag \Mcal_e  \tU_{e_R} \sim&\ \mathrm{diag}\left(d_e, D_{\ell_R}, D_{\ell_L}\right), \quad&
\left( \tU_{n_L}\right)^\dag \Mcal_n  \tU_{n_R} \sim &\ \mathrm{diag}\left(d_n, D_{\ell_R}, D_{\ell_L}\right),
\notag \\
\left( \tU_{d_L}\right)^\dag \Mcal_d  \tU_{d_R} \sim &\ \mathrm{diag}\left(d_d, D_{Q_R}, D_{Q_L}\right), \quad&
\left(\tU_{u_L}\right)^\dag \Mcal_u  \tU_{u_R} \sim &\ \mathrm{diag}\left(d_u, D_{Q_R}, D_{Q_L}\right).
\end{align}
The corrections from the off-diagonal blocks to the SM fermion mass matrix
are $\order{d_f^2/v_\Delta}$, $f=e,n,d,u$,
so it may be sub-dominant compared with the leading matrix $\sim d_f$.

\subsubsection{Neutrino masses}
The $18\times 18$ mass matrix is given by
\begin{align}
\notag
- \Lcal_N = \frac{1}{2} \ol{\Nv}_L \Mcal_N \Nv_R
=
\begin{pmatrix}
\ol{\nv}_L & \ol{\nv}_R^c
\end{pmatrix}
\begin{pmatrix}
0 & \Mcal_n \\ \Mcal_n^T &  \Mcal_R
\end{pmatrix}
\begin{pmatrix}
 \nv_L^c \\ \nv_R
\end{pmatrix},
\end{align}
where the Dirac mass matrix $\Mcal_n$ and Majorana mass matrix $\Mcal_n$
are given in Eqs.~\eqref{eq-parametrization} and~\eqref{eq-MajR}, respectively.
The mass basis is defined as
\begin{align}
\notag
\hat{\Nv}_R = U_N^\dag \Nv_R,
\quad
\hat{\Nv}_L = U_N^T  \Nv_L,
\end{align}
where $U_N^T \Mcal_N U_N$ is diagonalized.

We introduce the unitary matrices,
\begin{align}
 \wt{U}_N =
\begin{pmatrix}
 \wt{U}_{n_L}^* & \zd{9} \\ \zd{9} & \wt{U}_{n_R}
\end{pmatrix},
\end{align}
where $\wt{U}_{n_{L,R}}$ are defined in Eq.~\eqref{eq-Utilde}.
After multiplying these matrices, we obtain
\begin{align}
 \wt{\Mcal}_{N} :=&\
\tU_N^T \Mcal_N \tU_N
=
\begin{pmatrix}
 \zd{3} & \zd{3} & \zd{3} & d_n & \cdot & \zd{3} \\
 \zd{3} & \zd{3} & \zd{3} & \zd{3}  & D_{n_R} & \cdot \\
 \zd{3} & \zd{3} & \zd{3} & \cdot  & \cdot & D_{n_L} \\
d_n & \zd{3} & \cdot              & \tilde{M}_R & \tilde{\mu}_N & \zd{3} \\
\cdot  & D_{n_R} & \cdot & \tilde{\mu}_N^T  & \tilde{M}_N & \zd{3} \\
\zd{3} & \cdot & D_{n_L}         & \zd{3} & \zd{3} & \zd{3} \\
\end{pmatrix},
\end{align}
where
\begin{align}
\begin{pmatrix}
 \tilde{M}_R & \tilde{\mu}_N \\ \tilde{\mu}_N^T & \tilde{M}_N
\end{pmatrix}
=
\begin{pmatrix}
 \zd{3} & u_{n_R}^T \\ \id{3} & \zd{3}
\end{pmatrix}
V_{\ell_R}^T
\begin{pmatrix}
 M_R & \zd{3} \\ \zd{3} & \zd{3}
\end{pmatrix}
V_{\ell_R}
\begin{pmatrix}
 \zd{3} & \id{3} \\  u_{n_R} & \zd{3}.
\end{pmatrix}.
\end{align}

In numerical analysis,
we studied the case which $V_{\ell_R} \sim \id{6}$, $u_{n_R} \sim \id{3}$ and $M_R \propto \id{3}$.
We assume this structure in the analytical analysis in this Appendix,
but the non-zero effects are included in our numerical analysis.
The SM Majorana neutrino masses arise after the block-diagonalization by
\begin{align}
\tU^1_N \sim
\begin{pmatrix}
                \id{3} &  \zd{3} & \zd{3} & d_n \tilde{M}_R^{-1}  & \zd{3} & \zd{3} \\
 \zd{3}               & \id{3}  & \zd{3}  & \zd{3} & \zd{3} & \zd{3} \\
 \zd{3}               & \zd{3} & \id{3}  & \zd{3} & \zd{3} & \zd{3} \\
-\tilde{M}_R^{-1} d_n  & \zd{3}   & \zd{3} & \id{3} &  \zd{3} & \zd{3} \\
                \zd{3}  & \zd{3} & \zd{3}  & \zd{3} & \id{3} & \zd{3} \\
                \zd{3}  & \zd{3} & \zd{3}  & \zd{3} & \zd{3} & \id{3} \\
\end{pmatrix}.
\end{align}
The other states have diagonal Dirac mass matrices.
In this case, the active neutrino mass is approximately given by
\begin{align}
 \tilde{M}_\nu \sim - d_n \tilde{M}_R^{-1} d_n,
\end{align}
and this is already diagonalized.
The total unitary matrix is thus given by $U_N \sim \tU_N \tU^1_N$.

\subsection{Gauge interactions}
\label{sec-intgauge}

We shall discuss gauge interactions.
The PS symmetry is broken by non-zero VEVs of $\Delta$ and $\Sigma$.
We name the massive gauge bosons in $SU(4)_C/SU(3)_C\times U(1)_\BL$ as leptoquark $X_\mu$,
and those in $SU(2)_R\times U(1)_\BL/U(1)_Y$ as $Z^\prime_\mu$ and $W_{R\mu}^{\pm}$.
In this subsection for the gauge interactions,
$A,B,\cdots  = 1,2,\cdots,15$ and $\alpha,\beta,\cdots = 1,2,3,4$
are for the $SU(4)_C$ indices of adjoint and fundamental representations, respectively.
The $SU(2)_{X}$, $X=L,R$, indices of adjoint and fundamental representations
are respectively denoted as $k_{X},\cdots = 1,2,3$ and $\alpha_{X},\cdots = 1,2,3$.

\subsubsection{Symmetry breaking and vector boson masses}

The covariant derivative terms of the symmetry breaking scalar fields are given by
\begin{align}
 \Lcal_{\mathrm{kin}}^{\mathrm{SC}}
= \mathrm{Tr}_4 \left( D^\mu \Delta \cdot D_\mu \Delta \right) +
   \sum_{\alpha,\beta} 
     \mathrm{Tr}_2 \left[ \left(D^\mu \Sigma_{\alpha\beta}  \right)^\dagger  D_\mu \Sigma_{\alpha\beta} \right] +
   \mathrm{Tr}_2 \left[ \left(D^\mu \Phi\right)^\dagger D_\mu \Phi \right],
\end{align}
where $\mathrm{Tr}_4$ is the trace for the $SU(4)_C$ indices
and $\mathrm{Tr}_2$ is that for the $SU(2)_L$ and $SU(2)_R$ indices.
The covariant derivatives are given by
\begin{align}
 D_\mu \Delta =&\ \partial_\mu \Delta - ig_4 \left[\Vcal_\mu,\ \Delta\right], \\
 D_\mu \Sigma_{\alpha \beta} =&\ \partial_\mu \Sigma_{\alpha\beta} - ig_R
     \left[W_{R\mu},\ \Sigma_{\alpha\beta} \right]
                +ig_4\Sigma_{\gamma \delta}
 \left(\Vcal^\gamma_{\alpha\mu} \delta^{\delta}_\beta
   + \Vcal^\delta_{\beta \mu} \delta^{\gamma}_\alpha  \right) , \\
 D_\mu \Phi =&\ \partial_\mu \Phi - i g_L W_{L\mu} \Phi + i g_R \Phi W_{R\mu} ,
\end{align}
where the indices of the fields are
\begin{align}
 \Delta        =&\ \Delta^\alpha_\beta = \Delta^A 
 T^A, \quad
\Sigma_{\alpha\beta}  
= \sqrt{1+\delta_{\alpha\beta}}\; \Sigma_{\alpha\beta}^{k_R} \tau^{k_R}_R,
\quad
\Phi = 
\begin{pmatrix}
H_1&  \tilde{H}_2
\end{pmatrix}, \\
\Vcal_\mu =&\    \Vcal_\mu^A T^A, 
\quad
W_{L\mu} = 
W_{L\mu}^{k_L} \tau_L^{k_L}, \quad
W_{R\mu} = 
W_{R\mu}^{k_R} \tau_R^{k_R}, \quad
\end{align}
where $\Sigma^{\alpha\beta} = \Sigma^{\beta\alpha}$.
Here, the scalar fields $\Delta^A$ and $\Sigma^{k_R}_{\alpha\beta}$ are canonically normalized.
The hermitian matrices $T^A$ and $\tau^{k_{L(R)}}_{L(R)}$
are the generators of fundamental representation of $SU(4)_C$ and $SU(2)_{L(R)}$, respectively.
$H_1$ and $\tilde{H}_2 := \eps H^*_2$ are the $SU(2)_L$ doublets.

In this paper, we assume a certain potential gives the following VEVs,
\begin{align}
 \vev{\Delta} =&\ \frac{v_\Delta}{2\sqrt{3}}
\begin{pmatrix}
 3 & 0 \\ 0 & -\id{3}
\end{pmatrix},
\quad
\vev{\Sigma_{\alpha\beta}} = 
\frac{v_\Sigma}{\sqrt{2}}
\begin{pmatrix}
 0 & 1 \\ 0 & 0
\end{pmatrix}
\otimes
\delta_{\alpha1} \delta_{\beta1},
\quad
\vev{\Phi} 
=v_H
\begin{pmatrix}
  c_\beta & 0 \\ 0 & s_\beta
\end{pmatrix}.
\end{align}
where $s_\beta^2 + c_\beta^2 = 1$.
The $SU(4)_C$ gauge boson $\Vcal_\mu$
and $SU(2)_{L,R}$ gauge boson $\Wcal_{L,R\mu}$ are decomposed as
\begin{align}
 \Vcal =&\
\sqrt{\frac{3}{8}}
\begin{pmatrix}
 1 & 0 \\ 0 & -\dfrac{1}{3} \cdot \id{3}
\end{pmatrix}
V^\BL_\mu +
\begin{pmatrix}
0 & X^\dagger/\sqrt{2} \\ X/\sqrt{2} & G 
\end{pmatrix},
\\ \notag
W_{L,R} =&\ \frac{1}{2}
\begin{pmatrix}
 W^3_{L,R} & \sqrt{2} W^+_{L,R} \\ \sqrt{2} W^-_{L,R} & -W^3_{L,R}
\end{pmatrix}, \quad
W_{L,R}^\pm = \frac{W^1_{L,R} \pm i W^2_{L,R}}{\sqrt{2}}  
\end{align}
where the Lorentz index is omitted.
Here, $V^\BL$, $X$, $G$ are ${\BL}$ gauge boson, leptoquarks and gluon, respectively.
The mass terms of the gauge bosons are given by
\begin{align}
 \Lcal^{V}_\mathrm{mass} = m_X^2 X^\dagger_\mu X^\mu
                             +  \Wcal^+_\mu \Mcal_W^2 \Wcal^{-\mu}
                             + \frac{1}{2}  \Zcal_\mu \Mcal_Z^2 \Zcal^\mu,
\end{align}
where the mass and matrices are given by
\begin{align}
\label{eq-gmatrix}
 m_X        = g_4 \sqrt{\frac{4}{3} v_\Delta^2 +  \frac{1}{2} v_\Sigma^2} , \quad
 \Mcal^2_W = \frac{1}{2}
\begin{pmatrix}
g_L^2 v_H^2  & -s_{2\beta} g_L g_R v_H^2  \\
-s_{2\beta} g_L g_R v_H^2 & g_R^2 \left(v_H^2 + v_\Sigma^2 \right)
\end{pmatrix},
\end{align}
\begin{align}
\Mcal_Z^2 =
\frac{1}{2}
 \begin{pmatrix}
 g_L^2 v_H^2       & - g_L g_R v_H^2 & 0 \\
 -g_L g_R v_H^2 &  g_R^2 v_H^2 + 2 g_R^2 v_\Sigma^2 & -\sqrt{6}g_Rg_4 v_\Sigma^2 \\
0 & -\sqrt{6} g_Rg_4 v_\Sigma^2 & 3 g_4^2 v_\Sigma^2 \\
 \end{pmatrix},
\end{align}
with the gauge bosons
\begin{align}
 \Wcal^\pm_{\mu}  :=
\begin{pmatrix}
 W_{L\mu}^\pm \\ W_{R\mu}^\pm
\end{pmatrix},
\quad
\Zcal_\mu
:=
\begin{pmatrix}
 W^3_{L\mu} \\ W^3_{R\mu} \\ V^\BL_{\mu}
\end{pmatrix}.
\end{align}
Here, $s_{2\beta} = 2 c_\beta s_\beta$ and $c_{2\beta} = c_\beta^2 - s_\beta^2$.
The mass basis of the gauge bosons are defined as
\begin{align}
\Wcal^\pm_\mu
= R_W
\begin{pmatrix}
 W^\pm_\mu  \\ W^{\prime \pm}_\mu
\end{pmatrix}, \quad
\Zcal_\mu
= R_Z
\begin{pmatrix}
 A_\mu \\ Z_\mu \\  Z^{\prime}_\mu
\end{pmatrix},
\end{align}
where $R_W$ and $R_Z$ are orthogonal matrices diagonalizing $\Mcal^2_W$ and $\Mcal^2_Z$, respectively.

The rotation matrix $R_W$ and eigenvalues of the mass matrix, $m_W$ and $m_{W^\prime}$
are exactly given by
\begin{align}
 R_W =&\
\begin{pmatrix}
 c_\omega & s_\omega \\ -s_\omega & c_\omega
\end{pmatrix},  \\
m_W^2 =&\ \frac{v_\Sigma^2}{4} \left[ g_R^2 + (g_R^2+g_L^2) \eta - D_W \right],
\quad
m_{W^\prime}^2 = \frac{v_\Sigma^2}{4} \left[ g_R^2 + (g_R^2+g_L^2) \eta + D_W \right],
 \end{align}
where
\begin{align}
 D_W =&\  \sqrt{g_R^4 + 2 g_R^2(g_R^2-g_L^2) \eta +
\left\{\left(g_R^2-g_L^2\right)^2 + 4 s_{2\beta}^2 g_L^2 g_R^2 \right\} \eta^2},  \\
c_\omega =&\ \frac{1}{\sqrt{2}} \sqrt{1+ \frac{g_R^2 + (g_R^2-g_L^2)\eta }{D_W}}, \quad
s_\omega = - \frac{1}{\sqrt{2}} \sqrt{1- \frac{ g_R^2 + (g_R^2-g_L^2)\eta }{D_W}},
\end{align}
and $\eta := v_H^2/v_\Sigma^2$.
The rotation matrix for the neutral bosons are given by
\begin{align}
 R_Z =
\begin{pmatrix}
 c_{12} & s_{12} & 0 \\
 -s_{12} & c_{12} & 0 \\
0 & 0 & 1
\end{pmatrix}
\begin{pmatrix}
 c_{13} & 0 & s_{13}  \\
0  & 1 & 0 \\
 -s_{13} & 0 & c_{13} \\
\end{pmatrix}
\begin{pmatrix}
1 & 0 & 0 \\
0 & c_{23} & s_{23}  \\
0 &  -s_{23} & c_{23} \\
\end{pmatrix}.
\end{align}
Here, the angles are given by
\begin{align}
 c_{12} =&\ \frac{g_R}{\sqrt{g_L^2+g_R^2}},  \quad
 s_{12} =-  \frac{g_L}{\sqrt{g_L^2+g_R^2}},  \quad
 c_{13} = - \sqrt{3(g_L^2+g_R^2)} \frac{g_4}{\tilde{g}^2}, \quad
 s_{13} = \sqrt{2} \frac{g_Lg_R}{\tilde{g}^2},
\end{align}
\begin{align}
 c_{23} = \frac{1}{\sqrt{2}}\sqrt{1-\frac{s-2\tilde{g}^4/ (g_L^2+g_R^2)}{D_Z}},\quad
 s_{23} = \frac{1}{\sqrt{2}}\sqrt{1+\frac{s-2\tilde{g}^4/(g_L^2+g_R^2)}{D_Z}},
\end{align}
where
\begin{align}
\tilde{g}^2 = \sqrt{2g_R^2 g_L^2 + 3g_4^2 (g_L^2+g_R^2)},\quad
 s = 3 g_4^2 + 2g_R^2 + (g_L^2+g_R^2) \eta,\quad   
 D_Z = \sqrt{s^2-4\tilde{g}^4\eta}.
\end{align}
The $Z$ and $Z^\prime$ boson masses are given by
\begin{align}
m_Z^2 = \frac{v_\Sigma^2}{4} \left(s-D_Z \right),\quad
m_{Z^\prime}^2 = \frac{v_\Sigma^2}{4} \left(s+D_Z \right).
\end{align}
The rotation matrices $R_W$, $R_Z$ diagonalize the mass matrices as
\begin{align}
 R_W^T \Mcal_W^2 R_W = \mathrm{diag}(m^2_W, m^2_{W^\prime}),
\quad
 R_Z^T \Mcal_Z^2 R_Z = \mathrm{diag}(0, m^2_Z, m^2_{Z^\prime}).
\end{align}

If the sub-leading terms in $\eta$ are neglected,
the vector boson masses are given by
\begin{align}
m_W^2 =  \frac{g_L^2}{2}v_H^2,\quad m_{W^\prime}^2 = \frac{g_R^2}{2}v_\Sigma^2,
\end{align}
and
\begin{align}
\label{eq-mZZpapp}
 m_Z^2 = \frac{\tilde{g}^4}{3g_4^2+2g_R^2} \frac{v_H^2}{2}  ,\quad
 m_{Z^\prime}^2 =  \left(3g_4^2+2g_R^2\right) \frac{v_\Sigma^2}{2}.
\end{align}

\subsubsection{Gauge interactions with fermions}

With the PS symmetry, the covariant derivatives terms for the fermions,
$\Fcal_L \in (L, F_L, f_L)$ and $\Fcal_R \in (R, f_R, F_R)$, are given by
\begin{align}
 D^\mu \Fcal_L = \partial^\mu \Fcal_L - i g_L W_L^\mu \Fcal_L -ig_4 \Vcal^\mu \Fcal_L,
\quad
 D^\mu \Fcal_R = \partial^\mu R - i g_R W_R^\mu \Fcal_L -ig_4 \Vcal^\mu \Fcal_R,
\end{align}
The gauge couplings to the vector leptoquarks are shown in Eqs.~\eqref{eq-Xcoup} and~\eqref{eq-gXs}.

The gauge couplings with the charged $SU(2)$ bosons, $(W^-_L, W^-_R)$ are given by
\begin{align}
 \Lcal_W =&\ \frac{g_L}{\sqrt{2}} W^-_{L\mu} \left(
                    \ol{\rep{u}}_L \gamma^\mu P_6  \rep{d}_L
                     +  \ol{\rep{n}}_L \gamma^\mu P_6   \rep{e}_L
                           +  \ol{\rep{u}}_R \gamma^\mu \Pb   \rep{d}_R
                               +  \ol{\rep{n}}_R \gamma^\mu \Pb  \rep{e}_R
                                  \right)  \\ \notag
               &\ + \frac{g_R}{\sqrt{2}} W^-_{R\mu} \left(
                    \ol{\rep{u}}_L \gamma^\mu \Pb  \rep{d}_L
                     +  \ol{\rep{n}}_L \gamma^\mu \Pb  \rep{e}_L
                           +  \ol{\rep{u}}_R \gamma^\mu P_6  \rep{d}_R
                               +  \ol{\rep{n}}_R \gamma^\mu P_6 \rep{e}_R
                                  \right)  + h.c. \\ \notag
             =&\  W^-_{L\mu} \left(
                    \ol{\hat{\rep{u}}}_L \hat{g}^{W_L}_{q_L}   \gamma^\mu  \hat{\rep{d}}_L
                     +  \ol{\hat{\rep{N}}}_L \hat{g}^{W_L}_{\ell_L} \gamma^\mu  \hat{\rep{e}}_L
                           +  \ol{\hat{\rep{u}}}_R \hat{g}^{W_L}_{q_R} \gamma^\mu  \hat{\rep{d}}_R
                               +  \ol{\hat{\rep{N}}}_R \hat{g}^{W_L}_{\ell_R} \gamma^\mu  \hat{\rep{e}}_R
                    \right) \\ \notag
            &\  +  W^-_{R\mu} \left(
                    \ol{\hat{\rep{u}}}_L \hat{g}^{W_R}_{q_L}   \gamma^\mu  \hat{\rep{d}}_L
                     +  \ol{\hat{\rep{N}}}_L \hat{g}^{W_R}_{\ell_L} \gamma^\mu  \hat{\rep{e}}_L
                           +  \ol{\hat{\rep{u}}}_R \hat{g}^{W_R}_{q_R} \gamma^\mu  \hat{\rep{d}}_R
                               +  \ol{\hat{\rep{N}}}_R \hat{g}^{W_R}_{\ell_R} \gamma^\mu  \hat{\rep{e}}_R
                    \right) + h.c.,
\end{align}
where the couplings in the mass basis of the fermions are given by
\begin{align}
&\hat{g}^{W_L}_{q_L} = \frac{g_L}{\sqrt{2}} \left(U_L^u\right)^\dag P_6 U_L^d,
&\quad &
 \hat{g}^{W_L}_{\ell_L} = \frac{g_L}{\sqrt{2}} \left(U_L^n\right)^\dag P_6 U_L^e,\\ \notag
& \hat{g}^{W_L}_{q_R} = \frac{g_L}{\sqrt{2}} \left(U_R^u\right)^\dag \Pb U_R^d,
&\quad &
 \hat{g}^{W_L}_{\ell_R} = \frac{g_L}{\sqrt{2}} \left(U^n_R\right)^\dag \Pb U_R^e, \\ \notag
&\hat{g}^{W_R}_{q_L} = \frac{g_R}{\sqrt{2}} \left(U_L^u\right)^\dag \Pb U_L^d,
&\quad &
 \hat{g}^{W_R}_{\ell_L} = \frac{g_R}{\sqrt{2}} \left(U^n_L\right)^\dag \Pb U_L^e,\\ \notag
&
\hat{g}^{W_R}_{q_R} = \frac{g_R}{\sqrt{2}} \left(U_R^u\right)^\dag P_6 U_R^d,
&\quad &
\hat{g}^{W_R}_{\ell_R} = \frac{g_R}{\sqrt{2}} \left(U^n_R\right)^\dag  P_6 U_R^e.
\end{align}
Here, the projection matrices are defined as
\begin{align}
 P_6 :=
\begin{pmatrix}
 \id{6} & 0_{6\times 3} \\ 0_{3\times 6} & \zd{3}
\end{pmatrix},
\quad
\Pb := \id{9} - P_6.
\end{align}
Note that $(W_L, W_R)$ is not a mass base of the gauge bosons.
For instance, the couplings to the left-handed quarks are given by
\begin{align}
\begin{pmatrix}
 \hat{g}_{q_L}^{W} \\  \hat{g}_{q_L}^{W^\prime}
\end{pmatrix}
=
R_W^T
\begin{pmatrix}
 \hat{g}_{q_L}^{W_L} \\  \hat{g}_{q_L}^{W_R}
\end{pmatrix},
\end{align}
and those for the fermions can be obtained in the same way.

The gauge couplings with the neutral gauge bosons $(W_L^3, W_R^3, V^\BL)$ are given by
\begin{align}
 \Lcal_Z = &\
\frac{g_L}{2} W^3_{L\mu} \sum_{\rep{f}= \rep{u},\rep{d},\rep{e},\rep{n}}  I_{\rep{f}}
                     \left( \ol{\rep{f}}_L \gamma^\mu P_6  \rep{f}_L
                             + \ol{\rep{f}}_R \gamma^\mu \Pb \rep{f}_R   \right)   \\ \notag
           &\   + \frac{g_R}{2} W^3_{R\mu} \sum_{\rep{f}= \rep{u},\rep{d},\rep{e},\rep{n}}  I_{\rep{f}} 
                     \left( \ol{\rep{f}}_L \gamma^\mu \Pb  \rep{f}_L
                             + \ol{\rep{f}}_R \gamma^\mu P_6  \rep{f}_R   \right)   \\ \notag
                 &\ - \sqrt{\frac{3}{8}} g_4 V^{\BL}_\mu \sum_{\rep{f}= \rep{u},\rep{d},\rep{e},\rep{n}}  Q^{\rep{f}}_\BL
                     \left( \ol{\rep{f}}_L \gamma^\mu  \rep{f}_L
                             + \ol{\rep{f}}_R \gamma^\mu  \rep{f}_R   \right) \\
  =&\   \sum_{V=W^3_L,W^3_R,V^{\BL}} V_\mu \sum_{\rep{f}= \rep{u},\rep{d},\rep{e},\rep{N}}
                     \left( \ol{\hat{\rep{f}}}_L \hat{g}^{V}_{\rep{f}_L} \gamma^\mu   \hat{\rep{f}}_L
                      + \ol{\hat{\rep{f}}}_R \hat{g}^{V}_{\rep{f}_R} \gamma^\mu   \hat{\rep{f}}_R \right) ,
\end{align}
where $I_{\rep{f}} = +1$ $(-1)$ for $\rep{f} = \rep{e}$, $\rep{d}$ ($\rep{u}$, $\rep{n}$).
$Q_\BL^\rep{f}$ is the $\BL$ number of a fermion $\rep{f}$.
The gauge coupling matrices in the fermion mass basis are given by
\begin{align}
 &\hat{g}^{W^3_L}_{f_L} = \frac{g_L}{{2}} I^f_L \left(U_L^f\right)^\dag P_6 U_L^f,
&\quad &
 \hat{g}^{W^3_L}_{f_R} = \frac{g_L}{{2}} I^f_L \left(U_R^f\right)^\dag \Pb U_R^f,  \\
 &\hat{g}^{W^3_R}_{f_L} = \frac{g_R}{{2}} I^f_R \left(U_L^f\right)^\dag \Pb U_L^f,
&\quad &
 \hat{g}^{W^3_R}_{f_R} = \frac{g_R}{{2}} I^f_R \left(U_R^f\right)^\dag P_6 U_R^f,  \\
 &\hat{g}^{V^\BL}_{f_L} = - \sqrt{\frac{3}{{8}}} g_4 Q_\BL^f  \left(U_L^f\right)^\dag U_L^f,
&\quad &
 \hat{g}^{V^\BL}_{f_R} = - \sqrt{\frac{3}{{8}}} g_4 Q_\BL^f  \left(U_R^f\right)^\dag U_R^f.
\end{align}
The coupling matrices in the mass basis are
\begin{align}
 \begin{pmatrix}
 \hat{g}_{f_L}^{A} \\  \hat{g}_{f_L}^{Z} \\ \hat{g}_{f_L}^{Z^\prime}
\end{pmatrix}
=
R_Z^T
\begin{pmatrix}
 \hat{g}_{f_L}^{W^3_L} \\  \hat{g}_{f_L}^{W^3_R} \\  \hat{g}_{f_L}^{V^\BL}
\end{pmatrix},
\quad
 \begin{pmatrix}
 \hat{g}_{f_R}^{A} \\  \hat{g}_{f_R}^{Z} \\ \hat{g}_{f_R}^{Z^\prime}
\end{pmatrix}
=
R_Z^T
\begin{pmatrix}
 \hat{g}_{f_R}^{W^3_L} \\  \hat{g}_{f_R}^{W^3_R} \\  \hat{g}_{f_R}^{V^\BL}
\end{pmatrix}.
\end{align}
Approximately, $R_W$ and $R_Z$ are given by
\begin{align}
\label{eq-RZWapp}
 R_W =
\begin{pmatrix}
 1 & 0 \\ 0 & 1
\end{pmatrix}
+ \order{\eta},
\quad
R_Z =
\begin{pmatrix}
- \sqrt{3}g_4 g_R/\tilde{g}^2 & - \sqrt{s} g_L / \tilde{g}^2 & 0 \\
 -\sqrt{3} g_4 g_L/\tilde{g}^2 & 3g_4^2 g_R / (\sqrt{s}\tilde{g}^2) & \sqrt{2} g_R/\sqrt{s} \\
 -\sqrt{2} g_L g_R / \tilde{g}^2 & \sqrt{6} g_4 g_R^2 / (\sqrt{s}\tilde{g}^2) & - \sqrt{3} g_4 / \sqrt{s}
\end{pmatrix}
+ \order{\eta}.
\end{align}

At the leading order in $\eta$, $W \sim W_L$ and $W^\prime \sim W_R$.
The coupling matrices to the fermions are approximately given by
\begin{align}
 \hat{g}_{q_L}^{W} \sim&\ \frac{g_L}{\sqrt{2}}
\begin{pmatrix}
 u_{u_L}^\dag & \zd{3} & \zd{3} \\
 \zd{3} & \id{3} & \zd{3} \\
 \zd{3} & \zd{3} & \zd{3} \\
\end{pmatrix},
\quad  &
 \hat{g}_{q_R}^{W} \sim&\ \frac{g_L}{\sqrt{2}}
\begin{pmatrix}
 \zd{3} & \zd{3} & \zd{3} \\
 \zd{3} & \id{3} & \zd{3} \\
 \zd{3} & \zd{3} & \zd{3} \\
\end{pmatrix},
 \\
\Pcal_L \hat{g}_{\ell_L}^{W} \sim&\ \frac{g_L}{\sqrt{2}}
\begin{pmatrix}
 u_{n_L}^\dag & \zd{3} & \zd{3} \\
 \zd{3} & \zd{3} & \zd{3} \\
 \zd{3} & \zd{3} & \id{3} \\
\end{pmatrix},
\quad  &
\Pcal_R  \hat{g}_{\ell_R}^{W} \sim&\ \frac{g_L}{\sqrt{2}}
\begin{pmatrix}
 \zd{3} & \zd{3} & \zd{3} \\
 \zd{3} & \zd{3} & \zd{3} \\
 \zd{3} & \zd{3} & \id{3} \\
\end{pmatrix},
 \\
\hat{g}_{q_L}^{W^\prime} \sim&\ \frac{g_R}{\sqrt{2}}
\begin{pmatrix}
 \zd{3} & \zd{3} & \zd{3} \\
 \zd{3} & \zd{3} & \zd{3} \\
 \zd{3} & \zd{3} & \id{3} \\
\end{pmatrix},
\quad  &
 \hat{g}_{q_R}^{W^\prime} \sim&\ \frac{g_R}{\sqrt{2}}
\begin{pmatrix}
 u_{u_R}^\dag & \zd{3} & \zd{3} \\
 \zd{3} & \zd{3} & \zd{3} \\
 \zd{3} & \zd{3} & \id{3} \\
\end{pmatrix},
\\
 \Pcal_L \hat{g}_{\ell_L}^{W^\prime} \sim&\ \frac{g_R}{\sqrt{2}}
\begin{pmatrix}
 \zd{3} & \zd{3} & \zd{3} \\
 \zd{3} & \id{3} & \zd{3} \\
 \zd{3} & \zd{3} & \zd{3} \\
\end{pmatrix},
\quad  &
\Pcal_R  \hat{g}_{\ell_R}^{W^\prime} \sim&\ \frac{g_R}{\sqrt{2}}
\begin{pmatrix}
 u_{n_R}^\dag & \zd{3} & \zd{3} \\
 \zd{3} & \id{3} & \zd{3} \\
 \zd{3} & \zd{3} & \zd{3} \\
\end{pmatrix}. 
\end{align}
The other blocks in the neutrino couplings are vanishing.
For the SM families, the $W$ boson couples via the left-current,
while the $W^\prime$ boson couples via the right-current.
Their flavor structure depends on independent unitary matrices,
$u_{u_{L,R}}$ and $u_{n_{L,R}}$.
In particular,
$u_{u_L}^\dag$ and $u_{n_L}^\dag$ correspond to the CKM and PMNS matrices, respectively.

The coupling matrices to the neutral bosons $(W_L^3, W_R^3, V^\BL)$ are given by
\begin{align}
& \hat{g}^{W_L^3}_{q_L} \sim  \frac{g_L}{2} I_q  P^Z_{q_L},
\quad
 \hat{g}^{W_R^3}_{q_L} \sim  \frac{g_R}{2} I_q  P^{Z^\prime}_{q_L}
\quad
 \hat{g}^{V^{\BL}}_{q_L} \sim  - g_4 \sqrt{\frac{1}{24}}  \id{9}, \\
& \hat{g}^{W_L^3}_{q_R} \sim  \frac{g_L}{2} I_q   P^Z_{q_R},
\quad
 \hat{g}^{W_R^3}_{q_R} \sim  \frac{g_R}{2} I_q   P^{Z^\prime}_{q_R},
\quad
 \hat{g}^{V^{\BL}}_{q_R} \sim  - g_4 \sqrt{\frac{1}{24}}  \id{9}, \\
& \hat{g}^{W_L^3}_{\ell_L} \sim  \frac{g_L}{2} I_\ell  P^{Z}_{\ell_L},
\quad
 \hat{g}^{W_R^3}_{\ell_L} \sim  \frac{g_R}{2} I_\ell  P^{Z^\prime}_{\ell_L},
\quad
 \hat{g}^{V^{\BL}}_{\ell_L} \sim   g_4 \sqrt{\frac{3}{8}}  \id{9}, \\
& \hat{g}^{W_L^3}_{\ell_R} \sim  \frac{g_L}{2} I_\ell  P^{Z}_{\ell_R},
\quad
 \hat{g}^{W_R^3}_{\ell_R} \sim  \frac{g_R}{2} I_\ell  P^{Z^\prime}_{\ell_R},
\quad
 \hat{g}^{V^{\BL}}_{\ell_R} \sim   g_4 \sqrt{\frac{3}{8}}  \id{9},
\end{align}
where $q = u,d$ and $\ell = e, n$.
The coupling matrices are given by
\begin{align}
&P^Z_{q_L} = P^{Z^\prime}_{\ell_R} = 
\begin{pmatrix}
 \id{3} & \zd{3} & \zd{3} \\  \zd{3} & \id{3} & \zd{3} \\  \zd{3} & \zd{3} & \zd{3} \\
\end{pmatrix}
, \quad
P^Z_{q_R} = P^{Z^\prime}_{\ell_L} = 
\begin{pmatrix}
 \zd{3} & \zd{3} & \zd{3} \\  \zd{3} & \id{3} & \zd{3} \\  \zd{3} & \zd{3} & \zd{3} \\
\end{pmatrix},
\\
&P^Z_{\ell_L} = P^{Z^\prime}_{q_R}=  
\begin{pmatrix}
 \id{3} & \zd{3} & \zd{3} \\  \zd{3} & \zd{3} & \zd{3} \\  \zd{3} & \zd{3} & \id{3} \\
\end{pmatrix}
, \quad
P^Z_{\ell_R} = P^{Z^\prime}_{q_L} =  
\begin{pmatrix}
 \zd{3} & \zd{3} & \zd{3} \\  \zd{3} & \zd{3} & \zd{3} \\  \zd{3} & \zd{3} & \id{3} \\
\end{pmatrix}.
\end{align}
For the neutrinos,
these are the $9\times 9$ upper-left (bottom-right) block in the $18\times 18$ coupling matrices
for the left-handed (right-handed) neutrinos.
Using Eq.~\eqref{eq-RZWapp}, the gauge couplings to the neutral gauge bosons in the mass basis,
$(A, Z, Z^\prime)$ are approximately given by
\begin{align}
 \hat{g}^A_{f_X} \sim e Q^{f}_e \id{9},
\quad
 \hat{g}^Z_{f_X} \sim \frac{g_L}{c_W}\left(-\frac{I_f}{2} P^Z_{f_X} - s_W^2 Q^{f}_e \id{9} \right),
\quad
 \hat{g}^{Z^\prime}_{f_X} \sim \frac{g_{Z^\prime}}{2}
               \left(   s^2_{Z^\prime} P^{Z^\prime}_{f_X} I_f +  c^2_{Z^\prime} Q_\BL^f \id{9} \right),
\label{eq-ZprimeC}
\end{align}
where the electric charge, coupling constant and $Z^\prime$ gauge coupling constant are given by
\begin{align}
Q^f_e = \frac{Q_\BL^f-I_f}{2},\quad
e = \sqrt{3} \frac{g_4 g_L g_R}{\tilde{g}^2}, \quad
g_{Z^\prime} = \sqrt{\frac{2g_R^2 + 3 g_4^2}{2}}.
\end{align}
For the $Z$ and $Z^\prime$ boson couplings, the angles are defined as
\begin{align}
c_W = \frac{g_L \sqrt{3g_4^2+2g_R^2}}{\tilde{g}^2},
\quad
s_W = \frac{\sqrt{3}g_4 g_R}{\tilde{g}^2},
\quad
c_{Z^\prime} = \sqrt{\frac{2 g_R^2}{2g_R^2 + 3g_4^2}},
\quad
s_{Z^\prime} = \sqrt{\frac{3 g_4^2}{2g_R^2 + 3g_4^2}}.
\end{align}
Therefore, the EW gauge couplings coincides with the SM values
when $\order{\eta}$ effects are negligible.

In the above approximate formulas,
we neglected $\order{v_H^2/v_\Delta^2, \eta}$ effects in the diagonalization unitary matrices
for the fermion and gauge boson mass matrices.
In the fermion mass matrix,
the $\order{v_H}$ elements, denoted by $\cdot$ in Eq.~\eqref{eq-appdiag},
in the off-diagonal blocks are neglected.
These off-diagonal entries will induce flavor violating couplings with the EW gauge bosons.
The flavor violating coupling to $f_i f_j$ will be $ \order{ g_L [d_f]_i [d_f]_j / v_\Delta^2}$,
where $[d_f]_i$ is the mass of the SM fermion of $i$-th generation~\footnote{
See Refs.~\cite{Kawamura:2019rth,Kawamura:2019hxp} for the similar analysis
in a model with one vectorlike generation.
}.
Here we assume that all the vectorlike fermion masses are $\order{v_\Delta}$.
If $v_\Delta \sim \order{10~\mathrm{TeV}}$ as considered in this paper,
the induced flavor violating coupling is at most $\order{10^{-6}}$ for top and charm quarks,
and the smaller for the light flavor fermions.
Thus the flavor violation from the EW gauge bosons are too small to be measured by experiments.
The mixing of $Z$-$Z^\prime$ and $W$-$W^\prime$ will affect to EW precision observables,
since these induce exotic right-current interactions.
Again, when $v_\Sigma \sim \order{10~\mathrm{TeV}}$, the effect is $\order{10^{-4}}$,
and thus may be too small to be measured.
Note that $Z^\prime$ and $W^\prime$ can be lighter
while keeping the leptoquark $\order{10~\mathrm{TeV}}$
if $v_\Sigma \ll v_\Delta$ .
This would be an interesting possibility but is beyond the scope of this paper.
In our numerical analysis, flavor violating effects from the extra gauge bosons are neglected.

\subsection{Scalar Interactions}
\label{sec-intscal}
In our model, there are three scalar fields $\Delta$, $\Sigma$ and $\Phi$
introduced to break the PS to SM symmetry.
In this paper, we will not consider the scalar potential explicitly,
and we assume that the scalar potential has the global minimum at the VEVs which we assumed.

The EW symmetry is broken by the VEV of bi-doublet $\Phi$,
\begin{align}
\notag
 \Phi =
\begin{pmatrix}
H_1^0 &  H_2^{+ *}    \\
H_1^+  &    -H_2^{0*}
\end{pmatrix}
=
\begin{pmatrix}
H_1  &  \tilde{H}_2
 \end{pmatrix},
\end{align}
where, $H_{1,2}$ are the $SU(2)_L$ doublets which can be expanded as
\begin{align}
 H_k =&\
\begin{pmatrix}
 H_k^0 \\  H_k^+
\end{pmatrix}
= \frac{1}{\sqrt{2}}
\begin{pmatrix}
   v_k + h_k + i a_k  \\
\sqrt{2} H_k^+
\end{pmatrix},
\quad k = 1,2.
\end{align}

The mass basis of the doublet Higgs bosons are defined as
\begin{align}
\begin{pmatrix}
 h_1 \\ h_2
\end{pmatrix}
= R_\alpha
\begin{pmatrix}
 h \\ H
\end{pmatrix}
,\quad
\begin{pmatrix}
 a_1 \\ a_2
\end{pmatrix}
= R_{\beta_0}
\begin{pmatrix}
 G^0 \\ A
\end{pmatrix},
\quad
\begin{pmatrix}
 H_1^{+} \\ H_2^{+}
\end{pmatrix}
= R_{\beta_+}
\begin{pmatrix}
 G^+ \\ H^+
\end{pmatrix},
\end{align}
where $G^0, G^+$ are the NG bosons.
The rotations matrices are defined with the angles as
\begin{align}
 R_\phi =
\begin{pmatrix}
 \cos \phi &- \sin \phi \\
\sin \phi  & \cos \phi
\end{pmatrix},
\quad
 \phi = \alpha, \, \beta_0, \, \beta_\pm.
\end{align}
In the decoupling limit ($m_A, m_{H^\pm} \gg m_h$),  these angles are aligned as
$ \alpha = \beta_0 = \beta_\pm = \beta$ where $\tan\beta \equiv v_2/v_1$.
The Higgs couplings to the fermions are given by
\begin{align}
 -\mathcal{L}_{H}
 =&\  \sum_{S=h, H, A} S  \, \left(\ol{u}_L Y^S_{\rep{u}} \rep{u}_R +
                  \ol{\rep{d}}_L Y^S_{d} \rep{d}_R +
                  \ol{\rep{e}}_L Y^S_{e} \rep{e}_R +
                  \ol{\rep{n}}_L Y^S_{n} \rep{n}_R
                \right)  \\ \notag
 &\ + H^+ \left(\ol{\rep{u}}_L Y^{H^+}_{d} \rep{d}_R + \ol{\nu}_L Y^{H^+}_{e} \rep{e}_R \right)
       + H^- \left(\ol{\rep{d}}_L Y^{H^-}_{u} \rep{u}_R + \ol{e}_L Y^{H^-}_{n} \rep{n}_R \right)+  h.c. .
\end{align}
The Yukawa matrices $Y^S_f$ are given by linear combinations of
\begin{align}
 Y_a =
\begin{pmatrix}
 y_a  & \tilde{\lambda}_a  & 0_3 \\
\lambda_a & \tilde{y}_a  & 0_3 \\
0_3 & 0_3 & \tilde{y}_a^{\prime}
\end{pmatrix},
\quad
a = 1,2,
\end{align}
where the $3\times 3$ Yukawa matrices are defined in Eqs.~\eqref{eq-yukSM} and~\eqref{eq-yukVL}.
The relation of these Yukawa matrices to the quark Yukawa matrix is given by
\begin{align}
\begin{pmatrix}
 Y_d \\ Y_u
\end{pmatrix}
:=
\frac{d}{d v_H}
\begin{pmatrix}
 \Mcal_d \\ \Mcal_u
\end{pmatrix}
=
\begin{pmatrix}
 c_\beta & -s_\beta \\ -s_\beta & c_\beta
\end{pmatrix}
\begin{pmatrix}
 Y_1 \\ Y_2
\end{pmatrix},
\end{align}
where $Y_d$ and $Y_u$ are the Yukawa matrices aligned to the relevant block of the mass matrix.
The Yukawa matrices for $S=h,H,A,H^\pm$ are given by
\begin{align}
\begin{pmatrix}
Y_d^{h} \\ Y_u^{h}
\end{pmatrix}
=&\ \frac{1}{\sqrt{2}\cos{2\beta}}
\begin{pmatrix}
\cos(\alpha+\beta) &  -\sin(\alpha-\beta) \\ -\sin(\alpha-\beta) & \cos(\alpha+\beta)
\end{pmatrix}
\begin{pmatrix}
Y_d \\ Y_u
\end{pmatrix}, \\
 \begin{pmatrix}
Y_d^{H} \\ Y_u^{H}
\end{pmatrix}
=&\ \frac{-1}{\sqrt{2}\cos{2\beta}}
\begin{pmatrix}
\sin(\alpha+\beta) &  \cos(\alpha-\beta) \\ \cos(\alpha-\beta) & \sin(\alpha+\beta)
\end{pmatrix}
\begin{pmatrix}
Y_d \\ Y_u
\end{pmatrix}, \\
 \begin{pmatrix}
Y_d^{A} \\ Y_u^{A}
\end{pmatrix}
=&\ \frac{-i}{\sqrt{2}\cos{2\beta}}
\begin{pmatrix}
\sin(\beta_0+\beta) &  \cos(\beta_0-\beta) \\ - \cos(\beta_0-\beta) & -\sin(\beta_0+\beta)
\end{pmatrix}
\begin{pmatrix}
Y_d \\ Y_u
\end{pmatrix}, \\
 \begin{pmatrix}
Y_d^{H^+} \\ Y_u^{H^-}
\end{pmatrix}
=&\ \frac{-1}{\cos{2\beta}}
\begin{pmatrix}
\sin(\beta_\pm+\beta) &  \cos(\beta_\pm-\beta) \\  -\cos(\beta_\pm-\beta) & - \sin(\beta_\pm+\beta)
\end{pmatrix}
\begin{pmatrix}
Y_d \\ Y_u
\end{pmatrix}.
\end{align}
In the gauge basis, the lepton Yukawa couplings are the same as those of the quarks,
e.g. $Y^h_d = Y^h_e$.
The Yukawa matrices in the mass basis are given by
\begin{align}
 \hat{Y}^{S^0}_f = \left(U^f_{L}\right)^\dagger Y^{S^0}_f U^f_R,
  \end{align}
where $S^0=h,H,A$ and $f=u,d,e,n$. The charge Higgs couplings are given by
  \begin{align}
&
\hat{Y}^{H^+}_d = \left(U^u_{L}\right)^\dagger Y^{H^+}_d U^d_R,
\quad
\hat{Y}^{H^-}_u = \left(U^d_{L}\right)^\dagger Y^{H^-}_u U^u_R,  \\
&
\hat{Y}^{H^+}_e = \left(U^n_{L}\right)^\dagger Y^{H^+}_e U^e_R,
\quad
\hat{Y}^{H^-}_n = \left(U^e_{L}\right)^\dagger Y^{H^-}_n U^n_R.
\end{align}
In the decoupling limit, the SM Higgs couplings are aligned
with the mass matrix, i.e. $Y^h_{d(u)} \propto Y_{d(u)}$,
while the Yukawa couplings to the heavier Higgs bosons are not.
Therefore the heavy Higgs bosons generically induce flavor violation.

The $SU(4)_C$ adjoint scalar $\Delta$ is expanded as
\begin{align}
 \Delta =&\ \frac{1}{2\sqrt{3}} \left( v_\Delta + \frac{h_\Delta}{\sqrt{2}} \right)
\begin{pmatrix}
3&0  \\ 
0&-\rep{1}_3 \\
\end{pmatrix}
+
\begin{pmatrix}
 0 & 0 \\ 0 & \Deladj
\end{pmatrix}.
\end{align}
Here, $h_{\Delta}$ is a CP-even neutral scalar and $\Deladj$ is a $SU(3)_C$ adjoint scalar field.
The Yukawa couplings involving $\Delta$ and $\Deladj$ are given by
\begin{align}
- \mathcal{L}_{\Delta} =
 \sum_{\rep{f}=\rep{u},\rep{d},\rep{e},\rep{n }}
       \sqrt{\frac{{3}}{8}} h_\Delta Q_{\BL}^{\rep{f}} \ol{\rep{f}}_L  Y_\Delta \rep{f}_R
+\sum_{\rep{f}=\rep{u},\rep{d}}  \ol{\rep{f}}_L  Y_\Delta \, \Deladj \rep{f}_R
+ h.c.,
\end{align}
where the Yukawa coupling matrices in the gauge basis are common
for the fermions,
\begin{align}
 Y_\Delta =
\begin{pmatrix}
 0_3  & 0_{3} & \epsilon_L \\
 0_{3} & 0_{3} & \kappa_L \\
 \epsilon_R & \kappa_R& 0_{3}
\end{pmatrix}
=
\frac{\sqrt{3}}{2 v_\Delta} \left(\Mcal_{e}-\Mcal_{d} \right).
\label{Yukawa-Delta}
\end{align}
In the mass basis, the couplings are given by
\begin{align}
\hat{Y}_\Delta^f = \left(U^f_L\right)^\dagger Y_\Delta U^f_R,  \quad f=u,d,e,n.
\end{align}

The flavor violation is also induced by the $\Delta$ couplings,
although the sizable contributions appear only with vectorlike generations.
For instance, the Yukawa coupling to the charged leptons is approximately given by,
\begin{align}
\label{eq-YeDelta}
\hat{Y}_\Delta^e \sim \left(\wt{U}_{e_L}\right)^\dag Y_\Delta \wt{U}_{e_R}
                =
\begin{pmatrix}
  \zd{3}& \zd{3} & \Lambda_{L_1}^e \\
\Lambda_{R_2}^e & \Lambda_{R_1}^{e} & \zd{3} \\
  \zd{3}& \zd{3} & \Lambda_{L_2}^e \\
\end{pmatrix}
+ \order{\frac{v_H}{v_\Delta}},
\end{align}
where the $3\times 3$ coupling matrices $\Lambda^e_{L,R_{1,2}}$ are given by
\begin{align}
 \begin{pmatrix}
  \Lambda_{R_1}^e & \Lambda_{R_2}^{e}
 \end{pmatrix}
=&\ \tilde{D}_{\ell_R} - W^\dag_{\ell_L} W_{Q_L} \tilde{D}_{Q_R} V_{Q_R}^\dag  V_{\ell_R},  \\
 \begin{pmatrix}
  \Lambda_{L_1}^e \\ \Lambda_{L_2}^{e}
 \end{pmatrix}
=&\ \tilde{D}_{\ell_L} - V^\dag_{\ell_L} V_{Q_L} \tilde{D}_{Q_L} W_{Q_R}^\dag  W_{\ell_R}.
\end{align}
Thus, there is no couplings of $\Delta$ with two SM fermions at the leading order.
It might be possible that loop effects mediated by the vectorlike fermions induce flavor violations,
such as $\mu \to e \gamma$.
For $\mu\to e\gamma$,
the chirality enhanced effect enhanced by $v_H$
will be proportional to $[\hat{Y}^e_\Delta]_{1a} [\hat{Y}^e_\Delta]_{a2}$, $a=1,2,\cdots, 9$,
which are all zero in Eq.~\eqref{eq-YeDelta}.
Hence, we expect that $\Delta$ will not give significant flavor violating effects
no matter how the leptoquark couplings are.

\section{Analysis details}
\label{sec-analysis}

\subsection{Formulas of flavor observables}
\subsubsection{$K_L \to \mu e$}

\begin{table}[t]
 \centering
\caption{\label{tab-valmsn} 
Values of parameters of the mesons taken from PDG and HFLAG2019 \cite{Aoki:2019cca}.
The units of masses and decay constants are GeV, 
that of lifetime is GeV$^{-1}$.
}
\begin{tabular}[t]{cc|cc|cc|cc} \hline
$m_{B_d}$ & 5.280 & $\tau_{B_d}\times 10^{-12}$ & 2.3230 & $f_{B_d}$ & 0.1920 & $B_{B_d}$ & 1.30  \\
$m_{B_s}$ & 5.367 & $\tau_{B_s}\times 10^{-12}$ & 2.2930 & $f_{B_s}$ & 0.2284 & $B_{B_s}$ & 1.35  \\
$m_{B_c}$ & 6.275 & $\tau_{B_c}\times 10^{-12}$ & 0.7703 & $f_{B_c}$ & 0.4340 & - & -  \\
$m_{K}$ & 0.4976 & $\tau_{K_L}\times 10^{-17}$ & 7.7730 & $f_{K}$ & 0.1552 & $B_{K}$ & 0.717  \\
\hline
\end{tabular}
\end{table}

The branching fraction of $K_L \to e_i e_j$ is given by
\small
\begin{align}
\label{eq-KLemFull}
& \br{K_L}{e_i e_j}  \\ \notag
&\ = \frac{\tau_{K_L}}{512\pi m_X^4} (m_{e_i}+m_{e_j})^2 m_{K} f_K^2
\sqrt{\left(1-\frac{(m_{e_i}+m_{e_j})^2}{m_K^2}\right) \left(1-\frac{(m_{e_i}-m_{e_j})^2}{m_K^2}\right)}  \\ \notag
&\times \left[
\left| \left[\hat{g}_{d_L}^X\right]_{2i} \left[\hat{g}_{d_L}^X\right]_{1j}^* +
          \left[\hat{g}_{d_R}^X\right]_{2i} \left[\hat{g}_{d_R}^X\right]_{1j}^*   \right. \right. \\ \notag
  &  \left.\left. \hspace{1.0cm}   -  \frac{2m_K^2}{(m_{e_i}+m_{e_j})(m_d+m_s)}
        \left(
            \left[\hat{g}_{d_R}^X\right]_{2i} \left[\hat{g}_{d_L}^X\right]_{1j}^* +
            \left[\hat{g}_{d_L}^X\right]_{2i} \left[\hat{g}_{d_R}^X\right]_{1j}^*
         \right)
         \right|^2   \left(1-\frac{(m_{e_i}-m_{e_j})^2}{m_K^2}\right)    \right. \\ \notag
&\left.
+  \abs{
        \frac{2m_K^2}{(m_{e_i}+m_{e_j})(m_d+m_s)}
        \left(
            \left[\hat{g}_{d_R}^X\right]_{2i} \left[\hat{g}_{d_L}^X\right]_{1j}^* -
            \left[\hat{g}_{d_L}^X\right]_{2i} \left[\hat{g}_{d_R}^X\right]_{1j}^*
         \right)
}^2  
\left(1-\frac{(m_{e_i}+m_{e_j})^2}{m_K^2}\right)
\right] \\ \notag
&\ + (i\leftrightarrow j),
\end{align}
\normalsize
where $m_{e_{i(j)}}$ is the masses of $i$($j$)-th generation charged lepton.
In our numerical analysis, we included contributions
from the Higgs bosons and adjoint scalar $h_\Delta$, $\Delta_8$,
but we have seen that these are always negligible compared with those from the leptoquark
due to the small flavor violating coupling to the charged leptons as discussed in Appendix~\ref{sec-detail}.
We use the same formula for the other leptonic decays of $B$, $B_s$ and $B_c$ mesons
by formally replacing coupling matrices and flavor indices appropriately.
The values of constants used in our numerical analysis is shown in Table~\ref{tab-valmsn},
and the values of observables at the benchmark point are shown in the next section.

\subsubsection{$\mu$-$e$ conversion}

\begin{table}[t]
\centering
\caption{\label{tab-constmue}
Values of vector~\cite{Cirigliano:2009bz}  and scalar~\cite{Junnarkar:2013ac} form factors.
The coefficients $S_N$, $V_N$ are calculated in Ref.~\cite{Kitano:2002mt}.
The capture rates are given in Ref.~\cite{Kitano:2002mt,Suzuki:1987jf}.
}
\begin{tabular}[t]{ccccc} \hline
 $f_{V_p}^u$&  $f_{V_p}^d$ &  $f_{V_n}^u$&  $f_{V_n}^d$ &  $f_{V_p}^s = f_{V_n}^s$ \\ \hline
$2$ & $1$ & $1$ & $2$ & $0$ \\  \hline \hline
 $f_{S_p}^u$&  $f_{S_p}^d$ &  $f_{S_n}^u$&  $f_{S_n}^d$ &  $f_{S_p}^s = f_{S_n}^s$ \\ \hline
$0.0191$ & $0.0363$ & $0.0171$ & $0.0404$ & $0.043$ \\
\hline
& \\
\end{tabular}
\begin{tabular}[t]{c|cccc|c} \hline
Target & $S_p$ & $S_n$ & $V_p$ & $V_n$ & $\Gamma_{\mathrm{capt}}~[10^{6}\cdot s^{-1}]$\\ \hline\hline
Au      & $0.0614$ & $0.0918$ & $0.0974$ & $0.146$   & $13.07$ \\
Al       & $0.0155$ & $0.0167$ & $0.0161$ & $0.0173$ & $0.705$ \\
\hline
\end{tabular}
\end{table}

For the flavor violation involving the electron,
$\mu$-$e$ conversion is also severely constrained particularly in the future experiments.
The conversion rate is given by~\cite{Cirigliano:2009bz}
\begin{align}
\Gamma_{\mathrm{conv}} = 4 m_\mu^5
               \left(
                   \abs{  \sum_{N=p,n} \left(\tilde{C}_{VL}^N V_N+   \left. m_N \tilde{C}_{SL}^N S_N  \right. \right)  }^2  +
                   \abs{ \sum_{N=p,n} \left( \tilde{C}_{VR}^N V_N +   \left. m_N \tilde{C}_{SR}^N S_N \right.\right)   }^2
               \right),
\end{align}
where
\begin{align}
\tilde{C}_{VL}^N = \sum_{q=u,d,s} C^q_{VL} f_{V_N}^q,\quad
\tilde{C}_{SL}^N = \sum_{q=u,d,s} C^q_{SL} f_{S_N}^q + \frac{2}{27} f_G^N \sum_{Q=c,b,t} C^Q_{SL}, \\
\tilde{C}_{VR}^N = \sum_{q=u,d,s} C^q_{VR} f_{V_N}^q,\quad
\tilde{C}_{SR}^N = \sum_{q=u,d,s} C^q_{SR} f_{S_N}^q + \frac{2}{27} f_{G}^N \sum_{Q=c,b,t} C^Q_{SR}.
\end{align}
The values for form factors are shown in Table~\ref{tab-constmue} and
$f_{G}^N = 1- \sum_{q=u,d,s} f_{S_N}^q$.
In our model, the coefficients are given by
\begin{align}
 C^{d_i}_{VL} = \frac{\left[ \hat{g}^X_{d_L}\right]_{i1}^* \left[ \hat{g}^X_{d_L}\right]_{i2}}{2m_X^2},
\quad
 C^{d_i}_{VR} = \frac{\left[ \hat{g}^X_{d_R}\right]_{i1}^* \left[ \hat{g}^X_{d_R}\right]_{i2}}{2m_X^2}, 
\end{align}
and
\begin{align}
 C^{d_i}_{SL} =&\ -\frac{1}{m_{d_i}} \left[
                      \frac{ \left[ \hat{g}^X_{d_L}\right]_{i1}^* \left[ \hat{g}^X_{d_R}\right]_{i2} }{m_X^2}
          +\sum_{S}  \frac{ 1 }{2m_S^2}
                  \left[\hat{Y}_e^S \right]_{12}  \mathrm{Re}\left( \left[ \hat{Y}^S_d \right]_{ii}\right)
                          \right],  \\
 C^{d_i}_{SR} =&\  -\frac{1}{m_{d_i}} \left[
                      \frac{ \left[ \hat{g}^X_{d_R}\right]_{i1}^* \left[ \hat{g}^X_{d_L}\right]_{i2} }{m_X^2}
          +\sum_{S}  \frac{ 1 }{2m_S^2}
                  \left[\hat{Y}_e^S \right]_{21}^*  \mathrm{Re}\left( \left[ \hat{Y}^S_d \right]_{ii}\right)
                          \right],  \\
 C^{u_i}_{SL} =&\  - \frac{1}{m_{u_i}} \sum_{S}  \frac{ 1 }{2m_S^2}
                  \left[\hat{Y}_e^S \right]_{12}  \mathrm{Re}\left( \left[ \hat{Y}^S_u \right]_{ii}\right), \\
 C^{u_i}_{SR} =&\  - \frac{1}{m_{u_i}} \sum_{S}  \frac{ 1 }{2m_S^2}
                  \left[\hat{Y}_e^S \right]_{21}^*  \mathrm{Re}\left( \left[ \hat{Y}^S_u \right]_{ii}\right),
\end{align}
where $S$ runs over all the neutral scalar fields.


\subsection{Benchmark}

We show the values of parameters at a benchmark point whose the input parameters are given by
\begin{align}
 D_{Q_{L,R}} \sim D_{\ell_{L,R}} \sim 5~\mathrm{TeV}, \quad
m_N = 10~\TeV, \quad
 \delta_u = \delta_d = 10^{-4},
\end{align}
and the angles for $V_{Q_{L,R}}$, $V_{\ell_{L,R}}$ are $s_Q = 0.00078$ and $s_\ell = 0.00072$.
The other unitary matrices
in $W_{Q_{L,R}}$, $W_{\ell_{L,R}}$ and $w_{Q_{L,R}}$, $w_{\ell_{L,R}}$ are taken to be identity.
The other parameters are fitted such that the SM fermion masses, CKM and PMNS matrices are explained.
The mass matrices in Eq.~\eqref{eq-parametrization} are given by
\tiny
\begin{align}
&\wt{D}_d =
\begin{pmatrix}
0 & 0 & 0 & 0.000502716 & 0. & 0.  \\
0 & 0 & 0 & 0. & 0.146787 & 0.  \\
0 & 0 & 0 & 0. & 0. & 1.80177  \\
0.00211529 & 0. & 0. & 0 & 0 & 0  \\
0. & 0.0417809 & 0. & 0 & 0 & 0  \\
0. & 0. & 2.97755 & 0 & 0 & 0  \\
\end{pmatrix}, 
\\
& \wt{D}_u =  \\ \notag
&
\begin{pmatrix}
0.00000060\cdot e^{-2.2460i} & 0.00001279\cdot e^{-1.0960i} & 0.00488531\cdot e^{-2.7440i} & 0.00001386\cdot e^{0.0987i} & 0.00015155\cdot e^{0.0092i} & 0.00013351\cdot e^{2.1100i}  \\
0.00000181\cdot e^{-3.1020i} & 0.00002292\cdot e^{-3.1350i} & 0.00697150\cdot e^{-3.0540i} & 0.00007441\cdot e^{3.1210i} & 0.00003288\cdot e^{0.5167i} & 0.00036822\cdot e^{0.0012i}  \\
0.00000197\cdot e^{-3.0910i} & 0.00006067\cdot e^{-3.1380i} & 0.00863796\cdot e^{-3.0440i} & 0.00006500\cdot e^{0.0333i} & 0.00002488\cdot e^{1.1330i} & 0.00051000\cdot e^{0.0012i}  \\
0.00111051\cdot e^{0.0579i} & 0.10784074\cdot e^{3.1410i} & 1.22353000\cdot e^{0.3833i} & 0.00080864\cdot e^{-2.7230i} & 0.00080780\cdot e^{-2.7230i} & 0.00081043\cdot e^{-2.7230i}  \\
0.00057303\cdot e^{3.1160i} & 0.46812929 & 5.64585730\cdot e^{3.1230i} & 0.00371601\cdot e^{-0.0204i} & 0.00371740\cdot e^{-0.0204i} & 0.00373373\cdot e^{-0.0203i}  \\
0.02694553\cdot e^{0.0001i} & 0.01657569 & 191.73898000 & 0.13766851\cdot e^{-3.1420i} & 0.13769853\cdot e^{-3.1420i} & -0.13788805  \\
\end{pmatrix}, 
\end{align}

\begin{align}
 V_{Q_R}^* \wt{D}_{Q_R}^T  =&\
\begin{pmatrix}
3.90023160 & 3.90030370 & 3.90037580  \\
3.90023280 & 3.90030730 & 3.90038170  \\
3.90023400 & 3.90031080 & 3.90038760  \\
5000.29540000 & -0.00912672 & -0.00912690  \\
0.00000000 & 5000.39540000 & -0.00912691  \\
0.00000000 & 0.00000000 & 5000.49540000  \\
\end{pmatrix},
\quad &
 V_{Q_L} \wt{D}_{Q_L}  =&\
\begin{pmatrix}
4999.99540000 & -2.82413960 & -0.00818287  \\
0.00000000 & 3551.78620000 & -2.85575350  \\
0.00000000 & 0.00000000 & 5000.19540000  \\
-3.89999760 & -2.76819210 & -3.89792210  \\
-3.89999880 & -99.99485400 & -3.82138210  \\
-3.90000000 & -3517.92550000 & -2.77154640  \\
\end{pmatrix},
\\
 V_{\ell_R}^* \wt{D}_{\ell_R}^T  =&\
\begin{pmatrix}
5000.89610000 & -0.00777755 & -0.00777770  \\
0.00000000 & 5000.99610000 & -0.00777771  \\
0.00000000 & 0.00000000 & 5001.09610000  \\
-3.60064610 & -3.60071350 & -3.60078080  \\
-3.60064710 & -3.60071630 & -3.60078550  \\
-3.60064800 & -3.60071910 & -3.60079010  \\
\end{pmatrix},
\quad &
V_{\ell_L} \wt{D}_{\ell_L}  =&\
\begin{pmatrix}
3.60043010 & 3.59914710 & 2.48918270  \\
3.60043110 & -140.73388000 & -3517.02360000  \\
3.60043200 & 3.60050310 & 3.60057410  \\
5000.59610000 & 0.09614469 & 2.52787370  \\
0.00000000 & 4998.71670000 & -99.02272300  \\
0.00000000 & 0.00000000 & 3553.69030000  \\
\end{pmatrix},
\end{align},
\normalsize

The charged fermion Dirac masses are given by
\begin{align}
m^e_i =&\ (0.00050112, 0.105789,1.80177,4999.87,5000.6,5000.7,5000.9,5001.,5002.03), \notag \\
m^d_i =&\ (0.0021141,0.04176,2.11584,4999.43,5000.,5000.09,5000.3,5000.41,5001.27), \notag \\
m^u_i =&\ (0.0009744,0.479892,136.433,5000.,5000.2,5000.3,5000.4,5000.5,5001.92).
\end{align}
The neutral fermion Majorana masses are given by
\begin{align}
 m^n_i = (&\ 3.38664\cdot 10^{-13}, 8.68414\cdot 10^{-12},5.01938\cdot10^{-11}, \\ \notag
&\   4934.06,4934.06,5000.55,
 5000.55, 5000.61,5000.61,5000.89,5000.9, \\ \notag
&\   5001.13,5001.14,5070.58,5070.59,
  10000.,10000.,10000).
\end{align}
The W-boson couplings are proportional to
\tiny
\begin{align}
&  \left(U^u_{L}\right)^\dag P_6 U^d_{L}  =  \\ \notag
&\begin{pmatrix}
0.9745 & 0.2245 & 0.0036\cdot e^{-1.20i} & 0.0000 & 0.0000\cdot e^{-0.49i} & 0.0000\cdot e^{-0.30i} & 0.0000\cdot e^{-0.43i} & 0.0000\cdot e^{2.80i} & 0.0000\cdot e^{-3.10i}  \\
0.2244\cdot e^{-3.10i} & 0.9736 & 0.0421 & 0.0000 & 0.0000 & 0.0000 & 0.0000 & 0.0000\cdot e^{-3.10i} & 0.0000\cdot e^{-3.10i}  \\
0.0090\cdot e^{-0.38i} & 0.0413\cdot e^{-3.10i} & 0.9991 & 0.0000\cdot e^{0.02i} & 0.0000\cdot e^{3.10i} & 0.0007\cdot e^{-3.10i} & 0.0000\cdot e^{-3.10i} & 0.0002\cdot e^{-3.10i} & -0.0000  \\
0.0000\cdot e^{-0.42i} & 0.0000\cdot e^{-3.10i} & 0.0000 & 0.0000\cdot e^{1.50i} & 0.9991 & 0.0007\cdot e^{-3.10i} & 0.0298\cdot e^{-3.10i} & 0.0002\cdot e^{-3.10i} & 0.0000\cdot e^{1.40i}  \\
0.0000\cdot e^{-2.80i} & 0.0000\cdot e^{0.73i} & 0.0000\cdot e^{-2.40i} & 0.7621\cdot e^{0.72i} & 0.0000\cdot e^{3.00i} & 0.0000 & 0.0000\cdot e^{-0.33i} & 0.0000\cdot e^{-2.30i} & 0.6468\cdot e^{0.72i}  \\
0.0000\cdot e^{-0.22i} & 0.0000\cdot e^{-3.10i} & 0.0000 & 0.0002\cdot e^{-0.03i} & 0.0290 & 0.0025 & 0.0009\cdot e^{-3.10i} & 0.0005 & 0.0002\cdot e^{-0.03i}  \\
0.0000\cdot e^{-0.41i} & 0.0000\cdot e^{-3.10i} & 0.0000\cdot e^{-0.01i} & 0.0000\cdot e^{1.00i} & 0.0002\cdot e^{-3.10i} & 0.0029 & 0.0000 & 0.0006 & 0.0000\cdot e^{1.00i}  \\
0.0000\cdot e^{-1.30i} & 0.0000\cdot e^{2.10i} & 0.0000\cdot e^{-1.00i} & 0.0227\cdot e^{-1.00i} & 0.0001\cdot e^{-3.10i} & 0.0028\cdot e^{-1.00i} & 0.0000\cdot e^{-0.73i} & 0.0005\cdot e^{-1.00i} & 0.0193\cdot e^{-1.00i}  \\
0.0000\cdot e^{-0.36i} & 0.0000\cdot e^{-3.10i} & 0.0007\cdot e^{0.03i} & 0.0001\cdot e^{-3.10i} & 0.0008\cdot e^{0.01i} & 0.9819\cdot e^{0.03i} & 0.0034\cdot e^{0.03i} & 0.1894\cdot e^{0.03i} & 0.0000\cdot e^{-0.26i}  \\
\end{pmatrix},
\\
& \left(U^n_{L}\right)^\dag P_6 U^e_{L} = \\ \notag
&
\begin{pmatrix}
0.7900\cdot e^{-0.83i} & 0.3877\cdot e^{2.40i} & 0.4748\cdot e^{-0.41i} & 0.0001\cdot e^{2.40i} & 0.0000\cdot e^{-0.38i} & 0.0000\cdot e^{-0.59i} & 0.0000\cdot e^{2.80i} & 0.0000\cdot e^{-0.59i} & 0.0001\cdot e^{-0.77i}  \\
0.5949\cdot e^{1.30i} & 0.5746\cdot e^{1.70i} & 0.5619\cdot e^{-1.70i} & 0.0002\cdot e^{1.70i} & 0.0000\cdot e^{-1.70i} & 0.0000\cdot e^{-1.60i} & 0.0000\cdot e^{1.40i} & 0.0000\cdot e^{-1.60i} & 0.0002\cdot e^{-1.50i}  \\
0.1482\cdot e^{0.13i} & 0.7202\cdot e^{-1.60i} & 0.6774\cdot e^{-1.60i} & 0.0002\cdot e^{-1.60i} & 0.0000\cdot e^{-1.60i} & 0.0000\cdot e^{1.60i} & 0.0000\cdot e^{1.60i} & 0.0000\cdot e^{1.60i} & 0.0002\cdot e^{1.60i}  \\
0.0000\cdot e^{-3.10i} & 0.0135 & 0.0000\cdot e^{-0.12i} & 0.3774 & 0.0045\cdot e^{-2.80i} & 0.0206\cdot e^{-0.02i} & 0.0001\cdot e^{-2.80i} & 0.0020\cdot e^{3.10i} & 0.3280  \\
0.0000\cdot e^{-1.60i} & 0.0135\cdot e^{1.60i} & 0.0000\cdot e^{1.40i} & 0.3774\cdot e^{1.60i} & 0.0045\cdot e^{-1.20i} & 0.0206\cdot e^{1.60i} & 0.0001\cdot e^{-1.20i} & 0.0020\cdot e^{-1.60i} & 0.3280\cdot e^{1.60i}  \\
0.0000\cdot e^{-3.10i} & 0.0000\cdot e^{-0.02i} & 0.0000\cdot e^{-3.10i} & 0.0185\cdot e^{-0.03i} & 0.3424 & 0.5213\cdot e^{-3.10i} & 0.0111 & 0.0509 & 0.0161\cdot e^{-0.03i}  \\
0.0000\cdot e^{-1.50i} & 0.0000\cdot e^{1.60i} & 0.0000\cdot e^{-1.60i} & 0.0185\cdot e^{1.50i} & 0.3473\cdot e^{1.60i} & 0.5200\cdot e^{-1.60i} & 0.0113\cdot e^{1.60i} & 0.0508\cdot e^{1.60i} & 0.0161\cdot e^{1.50i}  \\
0.0000\cdot e^{-3.10i} & 0.0000\cdot e^{0.16i} & -0.0000 & 0.0065\cdot e^{0.27i} & -0.6148 & -0.3247 & -0.0200 & 0.0317 & 0.0057\cdot e^{0.27i}  \\
0.0000\cdot e^{-1.50i} & 0.0000\cdot e^{1.70i} & 0.0000\cdot e^{-1.60i} & 0.0067\cdot e^{1.80i} & 0.6120\cdot e^{-1.60i} & 0.3294\cdot e^{-1.60i} & 0.0199\cdot e^{-1.60i} & 0.0322\cdot e^{1.60i} & 0.0058\cdot e^{1.80i}  \\
0.0000\cdot e^{0.06i} & 0.0000\cdot e^{-3.10i} & 0.0000 & 0.0006\cdot e^{0.10i} & 0.0192\cdot e^{-3.10i} & 0.0203\cdot e^{-3.10i} & 0.0006\cdot e^{-3.10i} & 0.0020 & 0.0005\cdot e^{0.10i}  \\
0.0000\cdot e^{1.70i} & 0.0000\cdot e^{-1.50i} & 0.0000\cdot e^{1.60i} & 0.0002\cdot e^{1.90i} & 0.0205\cdot e^{-1.60i} & 0.0091\cdot e^{-1.60i} & 0.0007\cdot e^{-1.60i} & 0.0009\cdot e^{1.60i} & 0.0002\cdot e^{1.90i}  \\
0.0000\cdot e^{0.04i} & 0.0000\cdot e^{-3.10i} & 0.0000 & 0.0112 & 0.0624 & 0.3418\cdot e^{-3.10i} & 0.0020 & 0.0334 & 0.0097  \\
0.0000\cdot e^{1.60i} & 0.0000\cdot e^{-1.60i} & 0.0000\cdot e^{1.60i} & 0.0111\cdot e^{1.60i} & 0.0619\cdot e^{1.60i} & 0.3397\cdot e^{-1.60i} & 0.0020\cdot e^{1.60i} & 0.0332\cdot e^{1.60i} & 0.0096\cdot e^{1.60i}  \\
0.0000\cdot e^{-3.10i} & 0.0135 & 0.0000\cdot e^{-0.12i} & -0.3767 & 0.0045\cdot e^{0.38i} & 0.0205\cdot e^{3.10i} & 0.0001\cdot e^{0.38i} & 0.0020\cdot e^{-0.02i} & -0.3274  \\
0.0000\cdot e^{-1.60i} & 0.0135\cdot e^{1.60i} & 0.0000\cdot e^{1.40i} & 0.3767\cdot e^{-1.60i} & 0.0045\cdot e^{2.00i} & 0.0205\cdot e^{-1.60i} & 0.0001\cdot e^{1.90i} & 0.0020\cdot e^{1.60i} & 0.3274\cdot e^{-1.60i}  \\
0.0000\cdot e^{1.10i} & 0.0000\cdot e^{-0.22i} & 0.0000\cdot e^{-0.56i} & 0.0000\cdot e^{-0.32i} & 0.0000\cdot e^{1.90i} & 0.0000\cdot e^{-0.76i} & 0.0000\cdot e^{2.00i} & 0.0000\cdot e^{2.40i} & 0.0000\cdot e^{-0.32i}  \\
0.0000\cdot e^{2.70i} & 0.0000\cdot e^{-0.40i} & 0.0000\cdot e^{-0.17i} & 0.0000\cdot e^{-0.27i} & 0.0000\cdot e^{-3.00i} & 0.0000\cdot e^{-0.04i} & 0.0000\cdot e^{-3.00i} & 0.0000\cdot e^{3.10i} & 0.0000\cdot e^{-0.27i}  \\
0.0000\cdot e^{0.23i} & -0.0000 & 0.0000\cdot e^{0.05i} & 0.0000 & 0.0000\cdot e^{-2.70i} & 0.0000\cdot e^{-0.02i} & 0.0000\cdot e^{-2.70i} & 0.0000\cdot e^{3.10i} & 0.0000  \\
\end{pmatrix} \\
& \left(U^u_R\right)^\dag\Pb U^d_{R} =\\
&
\begin{pmatrix}
 0.0000\cdot e^{-0.11i} & 0.0000\cdot e^{-0.23i} & 0.0000\cdot e^{-0.26i} & 0.0000\cdot e^{-3.10i} & 0.0000\cdot e^{2.80i} & 0.0000\cdot e^{2.90i} & 0.0000\cdot e^{3.00i} & 0.0000\cdot e^{2.90i} & 0.0000\cdot e^{-3.10i}  \\
0.0000 & 0.0000 & 0.0000 & 0.0000\cdot e^{-3.10i} & 0.0000\cdot e^{3.10i} & 0.0000\cdot e^{-3.10i} & 0.0000\cdot e^{-3.10i} & 0.0000\cdot e^{-3.10i} & 0.0000\cdot e^{-3.10i}  \\
0.0000 & 0.0000 & 0.0000 & 0.0000\cdot e^{0.05i} & 0.0000\cdot e^{3.10i} & 0.0265\cdot e^{-3.10i} & 0.0001\cdot e^{-3.10i} & 0.0051\cdot e^{-3.10i} & 0.0000\cdot e^{1.50i}  \\
-0.0000 & -0.0000 & 0.0000\cdot e^{0.90i} & 0.0000\cdot e^{1.50i} & 0.9991 & 0.0007\cdot e^{-3.10i} & 0.0298\cdot e^{-3.10i} & 0.0002\cdot e^{-3.10i} & 0.0000\cdot e^{1.40i}  \\
0.0000\cdot e^{-2.40i} & 0.0000\cdot e^{-2.40i} & 0.0000\cdot e^{-2.40i} & 0.7622\cdot e^{0.72i} & 0.0000\cdot e^{3.00i} & 0.0000 & 0.0000\cdot e^{-0.33i} & 0.0000\cdot e^{-2.30i} & 0.6467\cdot e^{0.72i}  \\
0.0000\cdot e^{3.10i} & -0.0000 & -0.0000 & 0.0002\cdot e^{-0.03i} & 0.0290 & 0.0025 & 0.0009\cdot e^{-3.10i} & 0.0005 & 0.0002\cdot e^{-0.03i}  \\
0.0000\cdot e^{-3.00i} & 0.0000\cdot e^{-3.10i} & 0.0000\cdot e^{3.10i} & 0.0000\cdot e^{1.00i} & 0.0002\cdot e^{-3.10i} & 0.0029 & 0.0000 & 0.0006 & 0.0000\cdot e^{1.00i}  \\
0.0000\cdot e^{2.10i} & 0.0000\cdot e^{2.10i} & 0.0000\cdot e^{2.10i} & 0.0227\cdot e^{-1.00i} & 0.0001\cdot e^{-3.10i} & 0.0028\cdot e^{-1.00i} & 0.0000\cdot e^{-0.73i} & 0.0005\cdot e^{-1.00i} & 0.0193\cdot e^{-1.00i}  \\
0.0000\cdot e^{-3.10i} & 0.0000\cdot e^{-3.10i} & 0.0004\cdot e^{-3.10i} & 0.0001\cdot e^{-3.10i} & 0.0008\cdot e^{0.01i} & 0.9815\cdot e^{0.03i} & 0.0034\cdot e^{0.03i} & 0.1893\cdot e^{0.03i} & 0.0000\cdot e^{-0.26i}  \\
\end{pmatrix},
\\
&\left(U^n_{R}\right)^\dag \Pb U^e_{R} =  \\ \notag
&
\begin{pmatrix}
0.0000\cdot e^{-0.64i} & 0.0000\cdot e^{2.50i} & 0.0000\cdot e^{-0.70i} & 0.0000\cdot e^{-0.64i} & 0.0000\cdot e^{1.80i} & 0.0000\cdot e^{2.80i} & 0.0000\cdot e^{1.80i} & 0.0000\cdot e^{-0.29i} & 0.0000\cdot e^{-0.64i}  \\
0.0000\cdot e^{-1.50i} & 0.0000\cdot e^{1.60i} & 0.0000\cdot e^{-1.50i} & 0.0000\cdot e^{-1.50i} & 0.0000\cdot e^{-1.70i} & 0.0000\cdot e^{-2.10i} & 0.0000\cdot e^{-1.70i} & 0.0000\cdot e^{1.10i} & 0.0000\cdot e^{-1.50i}  \\
0.0000\cdot e^{1.70i} & 0.0000\cdot e^{-1.50i} & 0.0000\cdot e^{1.60i} & 0.0000\cdot e^{1.70i} & 0.0000\cdot e^{-1.40i} & 0.0000\cdot e^{1.80i} & 0.0000\cdot e^{-1.40i} & 0.0000\cdot e^{-1.30i} & 0.0000\cdot e^{1.70i}  \\
0.0000 & -0.0000 & 0.0000 & 0.3825 & 0.0046\cdot e^{2.80i} & 0.0209\cdot e^{0.02i} & 0.0002\cdot e^{2.80i} & 0.0021\cdot e^{-3.10i} & 0.3324  \\
0.0000\cdot e^{-1.60i} & 0.0000\cdot e^{1.60i} & 0.0000\cdot e^{-1.60i} & 0.3825\cdot e^{-1.60i} & 0.0046\cdot e^{1.20i} & 0.0209\cdot e^{-1.60i} & 0.0002\cdot e^{1.20i} & 0.0021\cdot e^{1.60i} & 0.3324\cdot e^{-1.60i}  \\
0.0000\cdot e^{0.02i} & 0.0000\cdot e^{-3.10i} & 0.0000 & 0.0185\cdot e^{0.03i} & 0.3424 & -0.5214 & 0.0111 & 0.0509 & 0.0161\cdot e^{0.03i}  \\
0.0000\cdot e^{-1.60i} & 0.0000\cdot e^{1.60i} & 0.0000\cdot e^{-1.60i} & 0.0185\cdot e^{-1.50i} & 0.3473\cdot e^{-1.60i} & 0.5201\cdot e^{1.60i} & 0.0113\cdot e^{-1.60i} & 0.0508\cdot e^{-1.60i} & 0.0161\cdot e^{-1.50i}  \\
0.0000\cdot e^{-0.13i} & 0.0000\cdot e^{3.00i} & 0.0000 & 0.0066\cdot e^{-0.27i} & 0.6148\cdot e^{-3.10i} & 0.3247\cdot e^{-3.10i} & 0.0200\cdot e^{-3.10i} & 0.0317 & 0.0057\cdot e^{-0.27i}  \\
0.0000\cdot e^{-1.70i} & 0.0000\cdot e^{1.50i} & 0.0000\cdot e^{-1.60i} & 0.0067\cdot e^{-1.80i} & 0.6120\cdot e^{1.60i} & 0.3294\cdot e^{1.60i} & 0.0199\cdot e^{1.60i} & 0.0322\cdot e^{-1.60i} & 0.0058\cdot e^{-1.80i}  \\
0.0000\cdot e^{-0.06i} & 0.0000\cdot e^{3.10i} & 0.0000 & 0.0006\cdot e^{-0.10i} & -0.0192 & -0.0203 & -0.0006 & 0.0020 & 0.0005\cdot e^{-0.10i}  \\
0.0000\cdot e^{-1.70i} & 0.0000\cdot e^{1.40i} & 0.0000\cdot e^{-1.60i} & 0.0002\cdot e^{-1.90i} & 0.0205\cdot e^{1.60i} & 0.0091\cdot e^{1.60i} & 0.0007\cdot e^{1.60i} & 0.0009\cdot e^{-1.60i} & 0.0002\cdot e^{-1.90i}  \\
0.0000 & -0.0000 & 0.0000 & 0.0112 & 0.0624 & -0.3418 & 0.0020 & 0.0334 & 0.0097  \\
0.0000\cdot e^{-1.60i} & 0.0000\cdot e^{1.60i} & 0.0000\cdot e^{-1.60i} & 0.0111\cdot e^{-1.60i} & 0.0619\cdot e^{-1.60i} & 0.3397\cdot e^{1.60i} & 0.0020\cdot e^{-1.60i} & 0.0332\cdot e^{-1.60i} & 0.0096\cdot e^{-1.60i}  \\
-0.0000 & 0.0000 & -0.0000 & 0.3716\cdot e^{-3.10i} & 0.0044\cdot e^{-0.38i} & 0.0202\cdot e^{-3.10i} & 0.0001\cdot e^{-0.38i} & 0.0020\cdot e^{0.02i} & 0.3228\cdot e^{-3.10i}  \\
0.0000\cdot e^{1.60i} & 0.0000\cdot e^{-1.60i} & 0.0000\cdot e^{1.60i} & 0.3716\cdot e^{1.60i} & 0.0044\cdot e^{-2.00i} & 0.0202\cdot e^{1.60i} & 0.0001\cdot e^{-1.90i} & 0.0020\cdot e^{-1.60i} & 0.3228\cdot e^{1.60i}  \\
0.0000\cdot e^{-0.34i} & 0.0000\cdot e^{2.80i} & 0.0000\cdot e^{-0.34i} & 0.0000\cdot e^{-0.34i} & 0.0000\cdot e^{1.50i} & 0.0000\cdot e^{-0.69i} & 0.0000\cdot e^{1.50i} & 0.0000\cdot e^{2.50i} & 0.0000\cdot e^{-0.34i}  \\
0.0000\cdot e^{-0.22i} & 0.0000\cdot e^{2.90i} & 0.0000\cdot e^{-0.22i} & 0.0000\cdot e^{-0.22i} & 0.0000\cdot e^{3.00i} & 0.0000\cdot e^{-0.15i} & 0.0000\cdot e^{3.00i} & 0.0000\cdot e^{3.00i} & 0.0000\cdot e^{-0.22i}  \\
0.0000 & -0.0000 & 0.0000 & 0.0000 & 0.0000\cdot e^{2.70i} & 0.0000\cdot e^{0.02i} & 0.0000\cdot e^{2.70i} & 0.0000\cdot e^{-3.10i} & 0.0000  \\
 \end{pmatrix}.
\end{align}
\normalsize
The upper-left $3\times 3$ block of $U^\dag_{u_L} P_6 U_{d_L}$ and
$U^\dag_{n_L} P_6 U_{e_L}$ correspond to the CKM matrix and hermitian conjugate of the PMNS matrix,
respectively.
We also note that the W-boson coupling to the SM fermions in the right-handed current are negligible.
The leptoquark couplings with $g_4 = 1$ are given by
\tiny
\begin{align}
& \hat{g}^{X}_{d_L} =  \\ \notag
&
\begin{pmatrix}
0.0000 & 0.0000 & 0.0000 & 0.0000 & 0.7067 & 0.0000 & 0.0230 & 0.0000\cdot e^{-3.10i} & 0.0000  \\
0.0000 & 0.0283 & 0.0001 & 0.0211 & 0.0000\cdot e^{-3.10i} & 0.7026 & 0.0000 & 0.0686\cdot e^{-3.10i} & 0.0184  \\
0.0000\cdot e^{-3.10i} & 0.7065 & 0.0012 & 0.0055\cdot e^{-3.10i} & 0.0000\cdot e^{-3.10i} & 0.0279\cdot e^{-3.10i} & 0.0000\cdot e^{-3.10i} & 0.0027 & 0.0052\cdot e^{-3.10i}  \\
0.0000 & 0.0007 & -0.5391 & 0.3009 & 0.0000 & -0.0000 & -0.0000 & -0.0001 & -0.3447  \\
-0.7068 & -0.0000 & 0.0000 & 0.0001 & -0.0006 & 0.0000 & 0.0211 & 0.0000 & 0.0001  \\
-0.0002 & -0.0064 & -0.0012 & -0.5235 & 0.0001 & 0.0146 & -0.0027 & -0.1360 & -0.4551  \\
0.0211 & -0.0000 & -0.0000 & -0.0018 & -0.0230 & -0.0001 & 0.7064 & -0.0015 & -0.0016  \\
0.0000 & -0.0012 & -0.0002 & -0.1009 & -0.0000 & 0.0728 & 0.0010 & 0.6905 & -0.0879  \\
-0.0000 & 0.0010 & -0.4576 & -0.3531 & 0.0000 & 0.0000 & 0.0000 & 0.0001 & 0.4074  \\
\end{pmatrix},
\\
& \hat{g}^{X}_{d_R} =  \\ \notag
&
\begin{pmatrix}
0.0000\cdot e^{-3.10i} & 0.0000\cdot e^{-3.10i} & 0.0000\cdot e^{-3.10i} & 0.0000 & 0.0230\cdot e^{-3.10i} & 0.0000\cdot e^{-3.10i} & 0.7067 & 0.0001\cdot e^{-3.10i} & 0.0000\cdot e^{-3.10i}  \\
0.0000\cdot e^{-3.10i} & 0.0000\cdot e^{-3.10i} & 0.0000\cdot e^{-3.10i} & 0.0000 & 0.0000 & 0.0687\cdot e^{-3.10i} & 0.0001\cdot e^{-3.10i} & 0.7038\cdot e^{-3.10i} & 0.0001\cdot e^{-3.10i}  \\
0.0000\cdot e^{-3.10i} & 0.0000\cdot e^{-3.10i} & 0.0000\cdot e^{-3.10i} & 0.4638 & 0.0000 & 0.0003 & 0.0001\cdot e^{-3.10i} & 0.0000 & 0.5338\cdot e^{-3.10i}  \\
0.0000 & -0.0001 & 0.4575 & 0.4070 & -0.0000 & 0.0006 & 0.0000 & -0.0001 & 0.3536  \\
0.0211 & 0.0003 & 0.0000 & 0.0000 & -0.7064 & 0.0002 & -0.0230 & -0.0000 & -0.0000  \\
-0.0046 & 0.1339 & 0.0001 & 0.0007 & -0.0003 & -0.6910 & -0.0000 & 0.0675 & 0.0002  \\
0.7067 & 0.0116 & 0.0000 & 0.0000 & 0.0211 & -0.0024 & 0.0007 & 0.0002 & -0.0000  \\
0.0109 & -0.6942 & -0.0001 & 0.0001 & 0.0000 & -0.1333 & -0.0000 & 0.0131 & 0.0001  \\
-0.0000 & 0.0001 & -0.5392 & 0.3453 & 0.0000 & 0.0004 & -0.0000 & -0.0000 & 0.3001  \\
\end{pmatrix},
\\
& \left(\hat{g}^{X}_{u_L}\right)^\dag =  \\ \notag
&
\begin{pmatrix}
0.0029\cdot e^{-2.00i} & 0.0221\cdot e^{-2.40i} & 0.2727\cdot e^{-2.40i} & 0.5584\cdot e^{-2.30i} & 0.3358\cdot e^{2.80i} & 0.0163\cdot e^{-2.30i} & 0.0001\cdot e^{1.10i} & 0.0171\cdot e^{-1.60i} & 0.0013\cdot e^{0.84i}  \\
0.0044\cdot e^{-1.40i} & 0.0329\cdot e^{-1.70i} & 0.4043\cdot e^{-1.70i} & 0.4204\cdot e^{1.80i} & 0.3975\cdot e^{-2.20i} & 0.0121\cdot e^{1.80i} & 0.0001\cdot e^{-1.60i} & 0.0227\cdot e^{-0.56i} & 0.0032\cdot e^{1.40i}  \\
0.0055\cdot e^{1.90i} & 0.0414\cdot e^{1.50i} & 0.5083\cdot e^{1.50i} & 0.1048\cdot e^{3.00i} & 0.4784\cdot e^{-2.30i} & 0.0030\cdot e^{3.10i} & 0.0000\cdot e^{-1.90i} & 0.0006\cdot e^{-0.25i} & 0.0051\cdot e^{-1.60i}  \\
0.0039\cdot e^{0.32i} & 0.0291 & 0.0042 & 0.0004\cdot e^{-0.12i} & 0.0111\cdot e^{-0.72i} & 0.0010\cdot e^{-3.10i} & 0.0011\cdot e^{-3.10i} & 0.3539\cdot e^{-2.10i} & 0.3518\cdot e^{3.10i}  \\
0.0039\cdot e^{-1.30i} & 0.0291\cdot e^{-1.60i} & 0.0042\cdot e^{-1.60i} & 0.0004\cdot e^{-1.70i} & 0.0111\cdot e^{-2.30i} & 0.0010\cdot e^{1.60i} & 0.0011\cdot e^{1.60i} & 0.3539\cdot e^{2.60i} & 0.3518\cdot e^{1.50i}  \\
0.1532 & -0.4131 & 0.0318\cdot e^{0.01i} & 0.0000 & 0.0000\cdot e^{-0.99i} & 0.0010 & 0.2319 & 0.0002\cdot e^{-2.10i} & 0.0328\cdot e^{3.10i}  \\
0.1567\cdot e^{-1.60i} & 0.4130\cdot e^{1.60i} & 0.0317\cdot e^{-1.60i} & 0.0001\cdot e^{1.60i} & 0.0000\cdot e^{-2.60i} & 0.0046\cdot e^{-1.60i} & 0.2297\cdot e^{-1.60i} & 0.0002\cdot e^{2.60i} & 0.0327\cdot e^{1.60i}  \\
-0.4755 & -0.1260 & 0.0151\cdot e^{-0.11i} & 0.0005 & 0.0000\cdot e^{3.00i} & 0.0166\cdot e^{-3.10i} & 0.0855 & 0.0001\cdot e^{-2.30i} & 0.0154\cdot e^{3.00i}  \\
0.4743\cdot e^{1.60i} & 0.1297\cdot e^{1.60i} & 0.0154\cdot e^{-1.70i} & 0.0004\cdot e^{-1.60i} & 0.0000\cdot e^{1.50i} & 0.0147\cdot e^{1.60i} & 0.0865\cdot e^{-1.60i} & 0.0001\cdot e^{2.40i} & 0.0157\cdot e^{1.40i}  \\
-0.0165 & -0.0109 & 0.0010\cdot e^{-0.06i} & 0.0145\cdot e^{-3.10i} & 0.0001\cdot e^{2.40i} & 0.4993 & 0.0113\cdot e^{-3.10i} & 0.0001\cdot e^{-2.20i} & 0.0024\cdot e^{3.10i}  \\
0.0156\cdot e^{1.60i} & 0.0030\cdot e^{1.60i} & 0.0004\cdot e^{-1.70i} & 0.0145\cdot e^{1.60i} & 0.0001\cdot e^{0.82i} & 0.4995\cdot e^{-1.60i} & 0.0051\cdot e^{1.60i} & 0.0001\cdot e^{2.50i} & 0.0017\cdot e^{1.50i}  \\
0.0113\cdot e^{3.10i} & -0.2451 & 0.0199 & 0.0004 & 0.0000\cdot e^{-0.17i} & 0.0157\cdot e^{-3.10i} & 0.4345\cdot e^{-3.10i} & 0.0000\cdot e^{0.97i} & 0.0188\cdot e^{3.10i}  \\
0.0114\cdot e^{1.60i} & 0.2436\cdot e^{1.60i} & 0.0198\cdot e^{-1.60i} & 0.0001\cdot e^{-1.60i} & 0.0000\cdot e^{-1.70i} & 0.0064\cdot e^{1.60i} & 0.4356\cdot e^{1.60i} & 0.0000\cdot e^{-0.60i} & 0.0187\cdot e^{1.50i}  \\
0.0037\cdot e^{-2.80i} & -0.0275 & 0.0148 & 0.0004\cdot e^{-3.10i} & 0.0100\cdot e^{-0.72i} & 0.0009 & 0.0009 & 0.3526\cdot e^{-2.10i} & 0.3530\cdot e^{-0.03i}  \\
0.0037\cdot e^{1.90i} & 0.0275\cdot e^{1.60i} & 0.0148\cdot e^{-1.60i} & 0.0004\cdot e^{1.70i} & 0.0100\cdot e^{-2.30i} & 0.0009\cdot e^{-1.60i} & 0.0009\cdot e^{-1.60i} & 0.3526\cdot e^{2.60i} & 0.3529\cdot e^{-1.60i}  \\
0.0000\cdot e^{0.36i} & 0.0000\cdot e^{0.34i} & 0.0000\cdot e^{0.21i} & 0.0000\cdot e^{2.00i} & 0.0000\cdot e^{3.00i} & 0.0000\cdot e^{3.00i} & 0.0000\cdot e^{0.39i} & 0.0000\cdot e^{-1.40i} & 0.0000\cdot e^{-2.90i}  \\
0.0000\cdot e^{0.53i} & 0.0000\cdot e^{0.24i} & 0.0000\cdot e^{0.40i} & 0.0000\cdot e^{0.48i} & 0.0000\cdot e^{2.60i} & 0.0000\cdot e^{3.10i} & 0.0000\cdot e^{0.21i} & 0.0000\cdot e^{-2.00i} & 0.0000\cdot e^{-2.90i}  \\
0.0000\cdot e^{0.32i} & 0.0000 & -0.0000 & -0.0000 & 0.0000\cdot e^{2.40i} & 0.0006 & 0.0006\cdot e^{-3.10i} & 0.0006\cdot e^{1.00i} & 0.0000\cdot e^{3.10i}  \\
\end{pmatrix},
\\
& \left(\hat{g}^{X}_{u_L}\right)^\dag =  \\ \notag
&
\begin{pmatrix}
0.0000\cdot e^{2.50i} & 0.0000\cdot e^{2.50i} & 0.0000\cdot e^{2.50i} & 0.0000\cdot e^{-0.92i} & 0.0000\cdot e^{1.30i} & 0.0000\cdot e^{2.20i} & 0.0000\cdot e^{1.50i} & 0.0000\cdot e^{-0.07i} & 0.0000\cdot e^{-1.10i}  \\
0.0000\cdot e^{1.60i} & 0.0000\cdot e^{1.60i} & 0.0000\cdot e^{1.60i} & 0.0000\cdot e^{-1.40i} & 0.0000\cdot e^{0.83i} & 0.0000\cdot e^{1.70i} & 0.0000\cdot e^{-1.60i} & 0.0000\cdot e^{-0.59i} & 0.0000\cdot e^{1.50i}  \\
0.0000\cdot e^{-1.60i} & 0.0000\cdot e^{-1.60i} & 0.0000\cdot e^{-1.60i} & 0.0000\cdot e^{1.60i} & 0.0000\cdot e^{0.85i} & 0.0000\cdot e^{-1.60i} & 0.0000\cdot e^{1.60i} & 0.0000\cdot e^{-0.57i} & 0.0000\cdot e^{1.50i}  \\
0.0000\cdot e^{-0.01i} & 0.0000 & 0.3486 & 0.0033\cdot e^{-0.38i} & 0.3582\cdot e^{-0.72i} & 0.0003\cdot e^{-3.00i} & 0.0005\cdot e^{-0.04i} & 0.0101\cdot e^{1.00i} & 0.0051\cdot e^{-3.10i}  \\
0.0000\cdot e^{-1.60i} & 0.0000\cdot e^{-1.60i} & 0.3486\cdot e^{-1.60i} & 0.0032\cdot e^{-2.00i} & 0.3582\cdot e^{-2.30i} & 0.0000\cdot e^{2.90i} & 0.0002\cdot e^{-1.70i} & 0.0105\cdot e^{-0.57i} & 0.0051\cdot e^{1.60i}  \\
0.0011\cdot e^{-3.10i} & 0.2320 & -0.0100 & 0.2425\cdot e^{-3.10i} & 0.0170\cdot e^{-0.69i} & 0.0064\cdot e^{-3.10i} & 0.0015 & 0.0012\cdot e^{0.99i} & 0.3700\cdot e^{-0.03i}  \\
0.0047\cdot e^{1.60i} & 0.2298\cdot e^{-1.60i} & 0.0099\cdot e^{1.60i} & 0.2460\cdot e^{1.60i} & 0.0170\cdot e^{-2.30i} & 0.0063\cdot e^{1.60i} & 0.0013\cdot e^{-1.60i} & 0.0014\cdot e^{-0.58i} & 0.3691\cdot e^{-1.60i}  \\
0.0166 & 0.0855 & -0.0062 & 0.4345 & 0.0059\cdot e^{-1.00i} & 0.0130 & 0.0008 & 0.0006\cdot e^{1.00i} & 0.2310\cdot e^{-0.03i}  \\
0.0147\cdot e^{-1.60i} & 0.0865\cdot e^{-1.60i} & 0.0063\cdot e^{1.60i} & 0.4326\cdot e^{-1.60i} & 0.0061\cdot e^{-2.60i} & 0.0131\cdot e^{-1.60i} & 0.0007\cdot e^{-1.60i} & 0.0007\cdot e^{-0.57i} & 0.2344\cdot e^{-1.60i}  \\
-0.4995 & -0.0112 & -0.0004 & 0.0136 & 0.0005\cdot e^{-0.83i} & 0.0012 & 0.0007\cdot e^{-3.10i} & 0.0008\cdot e^{1.00i} & 0.0144\cdot e^{-0.03i}  \\
0.4997\cdot e^{1.60i} & 0.0051\cdot e^{1.60i} & 0.0002\cdot e^{1.60i} & 0.0145\cdot e^{-1.60i} & 0.0002\cdot e^{-2.60i} & 0.0007\cdot e^{-1.60i} & 0.0003\cdot e^{1.60i} & 0.0003\cdot e^{-0.57i} & 0.0065\cdot e^{-1.60i}  \\
0.0156 & -0.4344 & -0.0065 & 0.0444\cdot e^{-3.10i} & 0.0102\cdot e^{-0.72i} & 0.0000\cdot e^{-3.10i} & 0.0001 & 0.0016\cdot e^{1.00i} & 0.2427\cdot e^{-0.03i}  \\
0.0063\cdot e^{-1.60i} & 0.4355\cdot e^{1.60i} & 0.0065\cdot e^{1.60i} & 0.0440\cdot e^{1.60i} & 0.0101\cdot e^{-2.30i} & 0.0004\cdot e^{1.60i} & 0.0005\cdot e^{-1.60i} & 0.0012\cdot e^{-0.57i} & 0.2412\cdot e^{-1.60i}  \\
-0.0000 & -0.0000 & 0.3581 & 0.0031\cdot e^{2.80i} & 0.3479\cdot e^{2.40i} & 0.0007\cdot e^{3.10i} & 0.0006\cdot e^{0.03i} & 0.0109\cdot e^{-2.10i} & 0.0237\cdot e^{-0.02i}  \\
0.0000\cdot e^{1.60i} & 0.0000\cdot e^{1.60i} & 0.3582\cdot e^{-1.60i} & 0.0031\cdot e^{1.20i} & 0.3479\cdot e^{0.85i} & 0.0003\cdot e^{1.50i} & 0.0003\cdot e^{-1.50i} & 0.0105\cdot e^{2.60i} & 0.0237\cdot e^{-1.60i}  \\
0.0000\cdot e^{-0.27i} & 0.0000\cdot e^{-0.45i} & 0.0000\cdot e^{-2.70i} & -0.0160 & 0.0090\cdot e^{-1.20i} & 0.5537 & 0.3132\cdot e^{0.47i} & 0.3081\cdot e^{-2.60i} & 0.0017\cdot e^{-2.70i}  \\
0.0000\cdot e^{-0.32i} & 0.0000\cdot e^{-0.19i} & 0.0000\cdot e^{0.17i} & 0.0047\cdot e^{1.60i} & 0.0145\cdot e^{-0.88i} & 0.1627\cdot e^{-1.50i} & 0.4850\cdot e^{-3.00i} & 0.4879\cdot e^{-2.30i} & 0.0028\cdot e^{0.12i}  \\
-0.0012 & -0.0012 & -0.0012 & 0.0119 & 0.0123\cdot e^{-0.72i} & 0.4080\cdot e^{-3.10i} & 0.4083 & 0.4081\cdot e^{-2.10i} & 0.0010\cdot e^{-0.03i}  \\
\end{pmatrix},
\end{align}
\normalsize
The Yukawa couplings $h_{qq^\prime}$ defined in Eq.~\eqref{eq-hqq} are given by
\begin{align}
h_{dd} = 0.00000183\cdot e^{-0.01i}, \quad
h_{uu} = 0.00000182\cdot e^{-0.01i}, \quad
h_{du}  = 0.00000183\cdot e^{-0.01i},
\end{align}
where the Yukawa coupling in the gauge basis, $h$ is set to be identity matrix.

Finally we summarize the prediction of observables at this benchmark point, lower and upper limits in Table~\ref{tab-benchobs}.
We take $\tan\beta = 50$ and $m_H = 3.0$ TeV to calculate the neutral meson mixing.
The benchmark point explains the $\RK$ anomaly.
On the other hand it is hard to explain the $\RD$ anomaly.
The all experimental constraints listed in the table are satisfied and the tuning level defined in Eq.~(\ref{eq:tuning}) is about $0.1\%$.
\begin{table}[t]
 \centering
\caption{\label{tab-benchobs}
The values of predictions at this benchmark point, lower and upper limit are shown. 
The value of the fine-tuning parameter is 
$\Delta_{\mathrm{FT}}^{-1} = 1.39\times 10^{3}$. 
}
\begin{tabular}[t]{c|cccc} \hline
observable & prediction & lower limit & upper limit& Ref. \\ \hline \hline
$\mathrm{Re}\Delta C_9$ ($\RK$ anomaly)&-0.505&-0.59&-0.41 & \cite{Alguero:2019ptt} \\
$\Delta C_V$ ($\RD$ anomaly)&$2.75\times 10^{-8}$&$5.2\times 10^{-2}$&0.124&\cite{Iguro:2020keo}\\
$\mathrm{BR}(B  \to e\nu )$&$8.74\times 10^{-12}$&0&$9.8\times10^{-7}$ & \cite{Zyla:2020zbs}\\
$\mathrm{BR}(B  \to \mu\nu )$&$3.88\times 10^{-7}$& $2.90\times10^{-7}$ & $1.07\times10^{-6}$ & \cite{Zyla:2020zbs}\\
$\mathrm{BR}(B  \to \tau\nu)$&$8.82\times 10^{-5}$&$8.5\times 10^{-5}$&$1.33\times 10^{-4}$&\cite{Zyla:2020zbs}\\
$\mathrm{BR}(B_c\to e\nu )$&$2.20\times 10^{-9}$&0&0.6&\cite{Blanke:2018yud}\\
$\mathrm{BR}(B_c\to \mu\nu )$&$9.79\times 10^{-5}$&0&0.6&\cite{Blanke:2018yud}\\
$\mathrm{BR}(B_c\to \tau\nu)$&$2.39\times 10^{-2}$&0&0.6&\cite{Blanke:2018yud}\\
$\mathrm{BR}(K_L\to e\mu )$&$3.04\times 10^{-13}$&0&$4.7\times 10^{-12}$&\cite{Zyla:2020zbs}\\
$\mathrm{BR}(B_d  \to e\tau)$&$1.27\times 10^{-18}$&0&$2.8\times 10^{-5}$&\cite{Zyla:2020zbs}\\
$\mathrm{BR}(B_d  \to \mu\tau)$&$4.70\times 10^{-13}$&0&$2.2\times 10^{-5}$&\cite{Zyla:2020zbs}\\
$\mathrm{BR}(B_d  \to \mu  e   )$&$5.46\times 10^{-13}$&0&$2.8\times 10^{-9}$&\cite{Zyla:2020zbs}\\
$\mathrm{BR}(B_d  \to ee    )$&$2.47\times 10^{-15}$&0&$8.3\times10^{-8}$&\cite{Zyla:2020zbs}\\
$\mathrm{BR}(B_d \to \mu\mu)$&$1.00\times 10^{-10}$&$0$&$3.9\times 10^{-10}$&\cite{Zyla:2020zbs}\\
$\mathrm{BR}(B_d   \to \tau\tau)$&$2.34\times 10^{-8}$&0&$2.1\times10^{-3}$&\cite{Zyla:2020zbs}\\
$\mathrm{BR}(B_s\to e\tau)$&$2.83\times 10^{-18}$&-&-&\cite{Zyla:2020zbs}\\
$\mathrm{BR}(B_s\to \mu\tau)$&$7.51\times 10^{-13}$&0&$4.2\times 10^{-5}$&\cite{Zyla:2020zbs}\\
$\mathrm{BR}(B_s\to \mu  e  )$&$7.99\times 10^{-13}$&0&$5.4\times 10^{-9}$&\cite{Zyla:2020zbs}\\
$\mathrm{BR}(B_s\to ee    )$&$7.48\times 10^{-14}$&0&$2.8\times 10^{-7}$&\cite{Zyla:2020zbs}\\
$\mathrm{BR}(B_s\to \mu\mu )$&$3.81\times 10^{-9}$&$2.2\times 10^{-9}$&$3.8\times 10^{-9}$&\cite{Zyla:2020zbs}\\
$\mathrm{BR}(B_s\to \tau\tau)$&$7.16\times 10^{-7}$&0&$6.8\times 10^{-3}$&\cite{Zyla:2020zbs}\\
$\mathrm{BR}(\mu\to e)^{\mathrm{Al}}$&$7.67\times 10^{-16}$&0&$6\times 10^{-17}$&\cite{Bertl:2006up}\\
$\mathrm{BR}(\mu\to e)^{\mathrm{Au}}$&$9.37\times 10^{-16}$&0&$3\times 10^{-13}$&\cite{Kuno:2013mha}(prospect)\\
$|C_{B_d}|$&0.94&0.83&1.27&\cite{Bona:2005vz,Bona:2007vi} \\
$\mathrm{Arg} C_{B_d}$&-0.75&-5.6&1.6&\cite{Bona:2005vz,Bona:2007vi} \\
$|C_{B_s}|$&1.03&0.93&1.29&\cite{Bona:2005vz,Bona:2007vi} \\
$\mathrm{Arg} C_{B_s}$&$-1.71\times 10^{-2}$&-1.36&2.2&\cite{Bona:2005vz,Bona:2007vi} \\
$\mathrm{Im} C_{K}$&0.92&0.88&1.36&\cite{Bona:2005vz,Bona:2007vi} \\
$\mathrm{BR}(\mu\to e\gamma)$&$3.69\times 10^{-14}$&0&$4.2\times 10^{-13}$&\cite{TheMEG:2016wtm}\\
$\mathrm{BR}(\tau\to e\gamma)$&$4.38\times 10^{-15}$&0&$3.3\times 10^{-8}$&\cite{Zyla:2020zbs}\\
$\mathrm{BR}(\tau\to \mu\gamma)$&$5.09\times 10^{-9}$&0&$4.4\times 10^{-8}$&\cite{Zyla:2020zbs}\\
\hline
\end{tabular}
\end{table}

\clearpage
{\small
\bibliographystyle{JHEP}
\bibliography{reference_vectorlike,PSleptoquark,PSleptoquark_iguro210120}

}

\end{document}